\documentclass[preprint2]{aastex}
\usepackage{natbib}
\usepackage{graphicx}

\usepackage{txfonts}
\usepackage{enumerate}  %enumerate with capital roman numbers
\usepackage{amssymb}
\usepackage{lscape}
\usepackage{rotating}
\usepackage{afterpage}

\bibpunct{(}{)}{;}{a}{}{,} % to follow the A&A style

\newcommand{\filteri}{\textit{F814W}~}
\newcommand{\filterb}{\textit{F435W}~}
\newcommand{\filterh}{\textit{F160W}~}
\newcommand{\mi}{$M_{F814W}$~}
\newcommand{\mb}{$M_{F435W}$~}
\newcommand{\mbi}{\mbox{$M_{F435W}$-$M_{F814W}$}~}

\newcommand{\reff}{$r_{\rm{eff}}$~}
\newcommand{\reffknot}{$r^{\rm{knot}}_{\rm{eff}}$~}
\newcommand{\reffpsf}{$r^{\rm{PSF}}_{\rm{eff}}$~}

\newcommand{\lir}{$L_{\rm{IR}}$~}
\newcommand{\ha}{H$_ \alpha$~}
\newcommand{\hb}{H$_ \beta$~}
\newcommand{\hst}{\textit{HST}~}

\newcommand{\msun}{M$_{\odot}$~}
\newcommand{\lsun}{$L_{\odot}$~}
\newcommand{\hii}{H{\scriptsize{II}}~}
\newcommand{\ld}{$D_{\rm{L}}$~}
\newcommand{\av}{$A_V$~}
\newcommand{\twospace}{$\!\!$}
\newcommand{\onespace}{$\!$}

\newcommand{\captionfonts}{\small}
\makeatletter  % Allow the use of @ in command names
\long\def\@makecaption#1#2{%
  \vskip\abovecaptionskip
  \sbox\@tempboxa{{\captionfonts #1: #2}}%
  \ifdim \wd\@tempboxa >\hsize
    {\captionfonts #1: #2\par}
  \else
    \hbox to\hsize{\hfil\box\@tempboxa\hfil}%
  \fi
  \vskip\belowcaptionskip}
\makeatother   % Cancel the effect of \makeatletter

\shorttitle{}
\shortauthors{Miralles-Caballero et al.}

\begin{document}
\title{Characterization of optically-selected star forming knots in (U)LIRGs}
\author{Daniel Miralles-Caballero\altaffilmark{1,2}}
\email{dmiralles@damir.iem.csic.es}
\author{Luis Colina\altaffilmark{1,2}}
\author{Santiago Arribas\altaffilmark{1,2}}
\author{Pierre-Alain Duc\altaffilmark{3}}

%\altaffiltext{$\star$}{This work is based on observations made with the Spitzer Space Telescope, which is operated %by the Jet Propulsion Laboratory, California Institute of Technology under NASA contract 1407}

\altaffiltext{1}{Instituto de Estructura de la Materia (IEM/CSIC), C/ Serrano 121, 28006 Madrid, Spain }
\altaffiltext{2}{Now at Departamento de Astrof{\'{\i}}sica, Centro de Astrobiolog{\'{\i}}a, CSIC-INTA, Ctra. de Torrej\'on Ajalvir km 4, Torrej\'on de Ardoz, 28850 Madrid, Spain}
\altaffiltext{3}{Laboratoire AIM, CEA, CNRS – Universit{\'e} Paris Diderot, Irfu/Service d’Astrophysique CEA-Saclay,
	91191 Gif sur Yvette Cedex, France}

\begin{abstract}

We present a comprehensive characterization of the general properties (luminosity functions, mass, size, ages, etc) of optically selected compact stellar objects (knots) in a representative sample of 32 low-z Luminous and Ultraluminouos Infrared Galaxies, (U)LIRGs. It is important to understand the formation and evolution of these properties in such systems, which represent the most extreme cases of starbursts in the low-z Universe. We have made use of high angular resolution ACS images from the Hubble Space Telescope in \filterb ($\sim$ \textit{B}) and \filteri ($\sim$ \textit{I}) bands. The galaxies in the sample represent different interaction phases (first contact, pre-merger, merger and post-merger) and cover a wide luminosity range (11.46 $\leq$ log (\lir/\lsun\twospace) $\leq$ 12.54). With a median size of 32 pc, most of the nearly 3000 knots detected consists of complexes of star clusters. Some of the knots ($\sim$15\%) are so blue that their colors indicate a young (i.e., $<$ 30 Myr) and almost extinction-free population. There is a clear correlation of the mass of these blue knots with their radius, where $M \propto R^{1.91\pm0.14}$, similar to that found in complexes of clusters in M51 and in giant molecular clouds.  This suggests that the star formation within the knots is proportional to the gas density at any given radius. This relation does not depend significantly on either the infrared luminosity of the system or on the interaction phase. The star formation of all the knots is characterized by luminosity functions (LFs) of the knots with slopes close to 2. Nevertheless, we see a marginally significant indication that the LF evolves with the interaction process, becoming steeper from early to advanced merger phases. Due to size-of-sample effects we are probably sampling knots in ULIRGs intrinsically more luminous (by a factor of about four) than in less luminous systems. They also have sizes and are likely to have masses characteristic of clumps in galaxies at z$\gtrsim$1. Knots in post-mergers are on average larger ($\times$ 1.3-2), more luminous (2 mag) in the \textit{I} band, and 0.5 mag redder than those in systems in earlier phases. Two scenarios are briefly discussed: (1) the likely presence of relatively high extinction in the most advanced mergers; (2) the dissolution of the less massive clusters and/or their coalescence into more massive, evolved superclusters. 
\end{abstract}


\section{Introduction}

Star clusters tend to form where strong star formation occurs and, specially, in starbursts triggered by galaxy interactions and mergers (\citealt{Schweizer98}). During the last two decades young and massive compact clusters  have been extensively studied in different environments including starburst galaxies~\citep{Meurer95a}, barred galaxies~\citep{Barth95}, spiral disks~\citep{Larsen99a}, interacting galaxies (\citealt{Whitmore99,Whitmore07};~\citealt{Bik03};~\citealt{Weilbacher00,Weilbacher03};~\citealt{Knierman03};~\citealt{Peterson09};~\citealt{Mullan11}) and compact groups of galaxies (\citealt{Iglesias-Paramo01};~\citealt{Gallagher10};~\citealt{Konstantopoulos10}). 

With masses of $10^{4}-10^{6}$ \msun and a median size of 3.5 pc (in terms of \reff), the light of these clusters follow the same luminosity function (LF) in the optical in a wide range of galactic environments, a power-law ($\psi$(L)dL$\propto$L$^{-\alpha}$dL) with an index of $\alpha$ = 2. Yet, there is some controversy, since the range is rather large depending on the study (\mbox{1.8 $\lesssim~\alpha~\lesssim$ 2.8}). Recent studies suggest that the LF can be curved, showing a systematic decrease of the slope from roughly 1.8 at low luminosities to roughly 2.8 at high luminosities (\citealt{Gieles10}). Such truncation can be reproduced by modeling LFs using an underlying cluster IMF with a Schechter function, shape of the mass function (MF) claimed to be observed in star-forming environments (e.g.,~\citealt{Haas08,Larsen09,Gieles09}). On the other hand, there are works that still favor the universality of the LF and the MF (e.g.,~\citealt{Whitmore07,Whitmore10};~\citealt{Fall09};~\citealt{Chandar11}).

These massive clusters do not form normally in isolation but tend to be clustered themselves (\citealt{Zhang01,Larsen04}). Star-forming complexes represent the largest units of star formation in a galaxy (see the review by~\citealt{Efremov95}). They are typically kpc-scale regions encompassing several clusters. Their properties have also been investigated in some nearby interacting galaxies, such as in M51 (\citealt{Bastian05a}) and in NGC 4038/4039 (\citealt{Whitmore99}), in order to understand the hierarchy of the star formation in embedded groupings. These works have confirmed that the properties of complexes of clusters resemble those of giant molecular clouds (GMCs) from which they are formed. In particular, the mass-radius relation, namely \mbox{$M_{GMC} \propto R_{GMC}^2$}, reflects the state of virial equilibrium in these clouds (\citealt{Solomon87}) and holds down to the scale of cloud clumps a few parsecs in radius (\citealt{Williams95}). Complexes of young clusters in M51 have a similar correlation, showing the imprint from the parent GMC (\citealt{Bastian05a}). Likewise, young massive clusters with masses above 10$^6$ \msun follow the same relation (\citealt{Kissler-Patig06}). However, young clusters with lower mass does not (\citealt{Larsen04,Bastian05b,Kissler-Patig06}). A star formation efficiency that depends on the binding energy (predicted by~\citealt{Elmegreen97}) of the progenitor GMC could destroy such a relation during the formation of young clusters. Another possible explanation for this lack of a mass-radius relation in young clusters is that dynamical encounters between young clusters (and gas clouds) add energy into the forming clusters, thereby increasing their radii (\citealt{Bastian05a}).

Although much effort has been made to understand the star formation and evolutionary processes in interacting systems, little is known for Luminous (LIRGs) and  Ultraluminous (ULIRGs) Infrared Galaxies, which represent the most extreme cases of starbursts and interactions in our nearby Universe. LIRGs and ULIRGs are objects with infrared luminosities of $10^{11} L_{\odot} \leq L_{bol} \sim L_{IR}[8 - 1000 \mu m]$\footnote{For simplicity, we will identify the infrared luminosity as \mbox{\lir(\lsun) $\equiv$ log ($L_{IR}[8 - 1000 \mu m]$) }.} $< 10^{12} L_{\odot}$ and  $10^{12} L_{\odot} \leq L_{IR}[8 - 1000 \mu m] < 10^{13} L_{\odot}$, respectively \citep{Sanders96}. The main source of this luminosity is thought to be the starburst activity, although an AGN may also be present, and even be the dominant energy source in a small percentage of the most luminous systems \citep{Genzel98a,Farrah03,Yuan10}. They are rich in gas and dust, and more than 50\% of the LIRGs and most ULIRGs show signs of being involved in a major interaction/merger  \citep[e.g][]{Surace98,Surace00,Cui01,Farrah01,Bushouse02,Evans02,Veilleux06}. 
   
(U)LIRGs are therefore natural laboratories for probing how star formation is affected by major rearrangements in the structure and kinematics of galactic disks. Establishing the general properties (luminosity functions, mass, size, ages, etc) of the compact stellar objects in (U)LIRGs as a function of luminosity and interaction phase would provide relevant information in order to understand the mechanisms that govern the star formation and evolution 
in these systems. It is not known either whether the processes that (U)LIRGs undergo produce stellar objects similar or very different in terms of size or luminosity than in less luminous \mbox{(non-)} interacting galaxies.

In a cosmological context, (U)LIRGs become very relevant as tracers of the infrared phase (i.e., the dust-enshrouded star-forming phase), occurring during the early stages in the formation of massive galaxies. The so-called sub-millimeter galaxies, detected at \mbox{z $>$ 2}, are similar to the ULIRGs observed in the local Universe, in the sense that they are merging systems with extremely high rates of star formation \citep{Chapman03,Frayer03,Engel10}. Besides, contrary to what it happens in the local Universe, (U)LIRGs are major contributors to the star formation rate density at z$\sim$1-2~\citep{Perez-Gonzalez05}. Star-forming galaxies become increasingly irregular at higher redshifts, with a blue clumpy structure, asymmetry and lack of central concentration (\citealt{Abraham96,Vandenbergh96,Im99}). It has been claimed that clumps in clumpy galaxies represent star-forming complexes intrinsically more massive by one or two orders of magnitudes than similar-size complexes in local galaxies (\citealt{Efremov95,Elmegreen09,Gallagher10}). Although most of the clumpy galaxies, observed at \mbox{z = 0.7-2}, do not show signs of interactions, given the extreme star formation properties in local (U)LIRGs it is relevant to study the properties of their clumpy structure in order to establish the similarities to and differences from high-z clumpy galaxies.

Previous studies of compact stellar structures in (U)LIRGs were focused only on small samples, mostly with low angular resolution ground-based imaging (\citealt{Surace98};~\citealt{Surace00}), or on detailed multi-frequency studies of compact stellar objects in individual galaxies (\citealt{Surace00};~\citealt{Diaz-Santos07}). Thus, no study has been done so far on a representative sample of luminous infrared galaxies covering the different phases of the interaction process as well as the entire LIRG and ULIRG luminosity range. 

This paper presents the first attempt at obtaining an homogeneous and statistically significant study of the photometric properties (magnitudes, colors and sizes) of optically-selected compact stellar objects found in these systems as a function of $L_{IR}$, morphology (i.e., interaction phase) and radial distance to the nucleus of the galaxy. These stellar objects, some of them identified as YMCs candidates for the closest galaxies (i.e., at \ld$\lesssim$60 Mpc) and practically all of them identified as cluster complexes further away, will be referred to as ''knots''. The luminosity function of the knots is also evaluated and compared with that obtained in other studies. This study also represents the starting point of a more quantitative analysis based on the direct comparison of these measurements with state-of-the-art simulations of galaxy encounters with linear resolutions comparable to those presented here (e.g., simulations by~\citealt{Bournaud08a}). The results of this analysis, in particular the evolution of the luminosity functions of the knots, will be presented in a forthcoming paper (Miralles-Caballero et al. 2011, in preparation).

The paper is organized as follows. We first describe the sample and classify the systems according to their morphology in section~\ref{sec:sample}. The data are presented in section~\ref{sec:data}. We then explain the analysis methodology, including the source selection, photometric measurements, completeness tests, etc. In section~\ref{sec:results}  we present the main photometric results, including the properties of the objects measured as a function of \lir and the interaction phase of the system.  Age and mass derivation for the young objects is also discussed. Finally, the summary and conclusions are given in section~\ref{sec:summary}. Throughout this paper we use $\Omega_{\Lambda}$ = 0.73, $\Omega_{M}$ = 0.27 and $H_{0}$ = 73 $km s^{-1} Mpc^{-1}$.

\section{Sample and properties}
\label{sec:sample}
\subsection{Sample selection}

Galaxies were selected from the flux-limited IRAS Revisited Galaxy Sample (RBGS,~\citealt{Sanders03}) and~\cite{Sanders88b}, using the following additional criteria: to sample the wide range of IR luminosities in (U)LIRGs, to cover all types of nuclear activity, to span different phases of interaction and to optimize the linear scales by selecting low-z galaxies. 

For all the selected systems, archival high angular resolution \hst blue and red optical images are available. The \textit{I} band, less affected by extinction, can roughly indicate the extent of the system and the tidal features emerging from it, whereas the \textit{B} band can give us an idea of the young population. Two of the systems were not considered for this study, since they are so close ($\lesssim$ 40 Mpc) that individual stars belonging to them start to contaminate, and selection effects can be encountered for these galaxies. For instance, the faintest objects detected in IRAS 10257-4338 by~\cite{Knierman03} with the WFPC2 can be individual stars (though they expect them to be few). Since the ACS images are more sensitive at similar exposure times, more individual stars can be detected. In addition, the galaxies in the sample were observed with IFS (VIMOS, INTEGRAL and PMAS), with the aim at identifying Tidal Dwarf Galaxy candidates\footnote{These data will be presented in a second paper.}. 

\begin{figure}[!t]
\hspace{0.5cm}\includegraphics[angle=90,width=0.95\columnwidth]{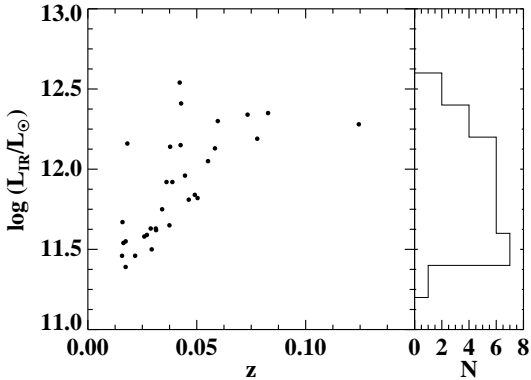}
      \caption{Distribution of the 32 systems of the sample in the luminosity-redshift plane. The sample consists of low-z (z$\lesssim$0.1) systems and covers a factor of  $\sim$ 10 in infrared luminosity.\label{fig:LirVsZ}
              }
      
\end{figure}

Our sample comprises 32 low-z systems, 20 LIRGs and 12 ULIRGs. Individual properties of the systems are presented in Table~\ref{table:sample}. It is not complete in either volume, flux or luminosity. However, this sample is essentially representative of the local (U)LIRG systems (see table~\ref{table:sample} and Figure~\ref{fig:LirVsZ}): (1) it co\-vers the luminosity range 11.39 $\leq$ \lir $\leq $ 12.54 and the redshift range 0.016 $\leq$ z $\leq$ 0.124 (from  65 to about 550 Mpc), with a median value of 0.037; (2) it spans all types of nuclear activity, with different excitation mechanisms such as LINER (i.e., shocks, strong winds, weak AGN), HII (star formation) and Seyfert-like (presence of an AGN); (3) and all the different morphologies usually identified in these systems are also sampled (see section~\ref{sub:Morphology} and Figure~\ref{fig:composite34}).  Therefore, the sample is appropriate for the purpose of this work.

\begin{deluxetable}{lccccccccc}
\tabletypesize{\scriptsize}
\tablewidth{0pt}
\tablecaption{Sample of (U)LIRGs\label{table:sample}}

\tablehead{
\colhead{IRAS}	&	\colhead{Other name}	&	\colhead{R. A. (J2000)}	&	\colhead{DEC (J2000)}	&	\colhead{$D_{L}$}	&	\colhead{$L_{IR}$}	&	\colhead{Morphology}	&	\colhead{Interaction}	&	\colhead{Spectral Class}	&	\colhead{Ref.}	\\
\colhead{name} &		&	\colhead{(hh:mm:ss)}	&	\colhead{(deg:mm:ss)}	&	\colhead{(Mpc)}	&		&		&	\colhead{phase}	&	\colhead{(Optical)}	& \\
\colhead{(1)} &	\colhead{(2)}	&	\colhead{(3)}	&	\colhead{(4)}	&	\colhead{(5)}	&	\colhead{(6)}	&	\colhead{(7)}	&	\colhead{(8)}	&	\colhead{(9)}	&	\colhead{(10)}	
}
\startdata
00506+7248	&		&	00:54:04.91 	&	+73:05:05.5 	&	65.3	&	11.46	&	IP	&	III	&	HII	&	[1]	\\
02512+1446	&		&	02:54:01.99 	&	+14:58:26.2 	&	131.3	&	11.63	&	IP	&	III	&	HII	&	[2]	\\
04315-0840	&	NGC 1614 	&	04:33:59.99 	&	-08:34:42.8 	&	66.1	&	11.67	&	SN	&	IV (2)	&	HII	&	[3]	\\
05189-2524	&		&	05:21:01.42 	&	-25:21:48.3	&	180.6	&	12.15	&	SN	&	V (2)	&	Sy2	&	[4],[5]	\\
06076-2139	&		&	06:09:45.98 	&	-21:40:25.6 	&	158.3	&	11.65	&	IP	&	III (1)	&	Sy/L-HII	&	[6]	\\
06259-4708	&		&	06:27:22.53 	&	-47:10:45.4 	&	164.2	&	11.92	&	MI	&	I-II (1)	&	HII-HII	&	[7]	\\
07027-6011	&		&	07:03:27.46 	&	-60:16:00.2 	&	131.8	&	11.62	&	IP	&	I-II (0)	&	Sy2	&	[7]	\\
08355-4944	&	 	&	08:37:01.98 	&	-49:54:27.5 	&	108.5	&	11.58	&	SN	&	IV (2)	&	HII	&	[8]	\\
08520-6850	&		&	08:52:28.86 	&	-69:02:00.3 	&	197.1	&	11.81	&	IP	&	I-II (1)	&	-	&	-	\\
08572+3915	&		&	09:00:25.62 	&	+39:03:54.1 	&	250.6	&	12.13	&	IP	&	III	&	L/HII	&	 [9], [10]	\\
09022-3615	&		&	09:04:12.26 	&	-36:26:58.1 	&	256.4	&	12.30	&	SN	&	IV (2)	&	-	&	-	\\
F10038-3338	&		&	10:06:05.04 	&	-33:53:14.8 	&	143.8	&	11.75	&	DN	&	IV (2)	&	-	&	-	\\
10173+0828	&		&	10:20:00.10 	&	+08:13:33.3 	&	209.4	&	11.84	&	SN	&	V	&	-	&	-	\\
12112+0305	&		&	12:13:45.82 	&	+02:48:38.0 	&	318.3	&	12.34	&	IP	&	III	&	L-L	&	[4], [5], [9]	\\
12116-5615	&		&	12:14:21.77 	&	-56:32:27.6 	&	113.7	&	11.59	&	SN	&	V (2)	&	HII	&	[7]	\\
12540+5708	&	Mrk 231	&	12:56:13.73 	&	+56:52:29.6 	&	178.9	&	12.54	&	SN	&	V (2)	&	Sy1	&	[2], [5]	\\
13001-2339	&		&	13:02:52.49 	&	-23:55:19.6 	&	90.7	&	11.46	&	DN	&	IV (2)	&	L	&	[11]	\\
13428+5608	&	Mrk 273	&	13:44:41.93 	&	+55:53:12.3 	&	159.8	&	12.14	&	DN	&	IV	&	L-Sy2	&	[2]	\\
13536+1836	&	Mrk 463	&	13:56:02.80	&	18:22:17.20	&	215.0	&	11.82	&	DN	&	III	&	Sy1-Sy2	&	[12]	\\
14348-1447	&		&	14:37:40.36 	&	-15:00:29.4 	&	361.6	&	12.35	&	DN	&	III	&	L	&	 [5], [9]	\\
15206+3342	&		&	15:22:38.00	&	+33:31:35.9	&	559.6	&	12.28	&	SN	&	V	&	HII	&	[5], [13]	\\
15250+3609	&		&	15:26:59.05 	&	+35:58:38.0	&	236.3	&	12.05	&	SN	&	IV	&	L	&	 [5]	\\
15327+2340	&	Arp 220	&	15:34:54.63 	&	+23:29:40.5 	&	75.5	&	12.16	&	DN	&	IV 	&	L	&	[13]	\\
16104+5235	&	NGC 6090	&	16:11:40.10 	&	+52:27:21.5 	&	123.1	&	11.50	&	IP	&	III	&	HII	&	[2]	\\
F17138-1017	&		&	17:16:35.64 	&	-10:20:34.9 	&	72.2	&	11.39	&	SN	&	V (2)	&	HII	&	[14]	\\
17208-0014	&		&	17:23:19.14 	&	-00:17:22.5 	&	181.7	&	12.41	&	SN	&	V	&	L	&	[5], [15]	\\
F18093-5744	&	IC 4686	&	18:13:38.66 	&	-57:43:53.7 	&	72.2	&	11.55	&	MI	&	III (1)	&	HII-HII	&	[7]	\\
18329+5950	&	NGC 6670	&	18:33:37.06 	&	+59:53:19.3	&	121.2	&	11.63	&	IP	&	III	&	-	&	-	\\
20550+1656	&		&	20:57:23.51 	&	+17:07:34.6 	&	152.5	&	11.92	&	IP	&	III	&	HII	&	[2]	\\
22491-1808	&		&	22:51:45.79 	&	-17:52:22.5 	&	338.6	&	12.19	&	DN	&	III (1)	&	HII	&	[5]	\\
23007+0836	&		&	23:03:16.92 	&	+08:53:06.4 	&	67.9	&	11.54	&	IP	&	I-II	&	Sy1-Sy2/L	&	 [2]	\\
23128-5919	&		&	23:15:46.46 	&	-59:04:01.9 	&	189.6	&	11.96	&	DN	&	III (1)	&	Sy/L-HII	&	 [5], [9]	\\
\enddata

\tablecomments{Col (1): object designation in the IRAS Point and Faint Source catalogs. Col (2): other name. Cols (3) and (4): right ascension (hours, minutes and seconds) and declination (degrees, arcminutes, arcseconds) taken from the from the NASA Extragalactic Database (NED). Col (5): luminosity distances derived assuming a $\Lambda$CDM cosmology with H$_{0}$ = 73 $km s^{-1} Mpc^{-1}$, $\Omega_{\Lambda}$=0.73 and $\Omega_{M}$=0.27, and using the Eduard L. Wright Cosmology calculator, which is based in the prescription given by~\cite{Wright06}. Col (6): logarithm of the infrared luminosity, $L_{\rm{IR}}=L(8-1000μm)$, in units of solar bolometric luminosity, computed following \citealt{Sanders96} ($L_{\rm{IR}}=4\pi D_{\rm{L}}^{2} F_{\rm{IR}}$, where $F_{\rm{IR}}=1.8 x 10^{-14} \times (13.48f_{12}+5.16f_{25}+2.58f_{60}+f_{100}) [\rm{Wm}^{-2}]$, being $f_{12}, f_{25}, f_{60}$ and $f_{100}$ the IRAS flux densities in Jy at 12, 25, 60 and 100 $\mu m$). The IRAS flux densities were obtained from the IRAS Point Source \& IRAS Faint Source catalogs \citep{Moshir93,Joint94}. Col (7): MI stands for Multiple Interacting galaxies, IP interacting pair, DN double nucleus and SN single nucleus. Col (8):  morphological classification used in this study, where I-II stands for first approach, III for pre-merger, IV for post-merger and V for relaxed system (see text). Between brackets the classification by~\cite{Rodriguez-Zaurin10} is given, when available. In  their study, class 0 corresponds to classes I-II, class 1 to class III and class 2 to classes IV-V in this study, respectively. Note a disagreement in IRAS 06259-4708 and IRAS 08520-6850 due to the slight different of the definitions of the early interaction phases in both studies (see text). Col (9): spectral class in the optical. A slash between two different classes indicates that no clear classification is known between both, whereas a dash separates the class of each nucleus in the system when known. L stands for LINER and Sy refers to a Seyfert activity. Col (10): references to the spectral class-- [1]~\cite{Alonso-Herrero09}, [2]~\cite{Wu98}, [3]~\cite{Kotilainen01}, [4]~\cite{Risaliti06}, [5]~\cite{Nardini08}, [6]~\cite{Arribas08}, [7]~\cite{Kewley01b}, [8]~\cite{Cohen92}, [9]~\cite{Evans02}, [10]~\cite{Arribas00}, [11]~\cite{Zenner93}, [12]~\cite{Garcia-Marin07}, [13]~\cite{Kim98}, [14]~\cite{Corbett03}, [15]~\cite{Arribas03}.}
\end{deluxetable}

\subsection{Morphology}
\label{sub:Morphology}

With a sample of 32 systems we have the opportunity to investigate the properties of their compact stellar objects as a function of the morphology of the galaxies. Besides, setting a rough estimate of the timing of the merger according to its morphological class, we will compare the properties of these objects with those of compact stellar objects measured in high re\-so\-lu\-tion interacting galaxy simulations by~\cite{Bournaud08a} in a forthcoming paper (Miralles-Caballero et al. 2011, in preparation). The galaxies exhibit a variety of morphologies, from a wide pair with a projected nuclear distance of 54 kpc, to close nuclei with a projected radial distance of \mbox{d$\lesssim$5 kpc}, and to a single nucleus with a distorted stellar envelope. We derived a morphology class for each system of the sample by using the red (\filteri filter in ACS and WFPC2 instruments) \hst images. A mer\-ging classification scheme similar to that given by~\cite{Veilleux02} has been con\-si\-de\-red. Though, due to our limited number of galaxies some grou\-ping has been made, so as to have enough statistics in all phases. For instance, phases I and II in~\cite{Veilleux02} have been grouped for this study. The sample has been divided in four groups according to their projected morphology and trying to match well identified temporal snapshots in the evolution of the interaction according to the models (see images in Figure~\ref{fig:composite34}):

 \begin{figure*}
   \centering
 \vspace{-3cm}
   \includegraphics[width=\textwidth]{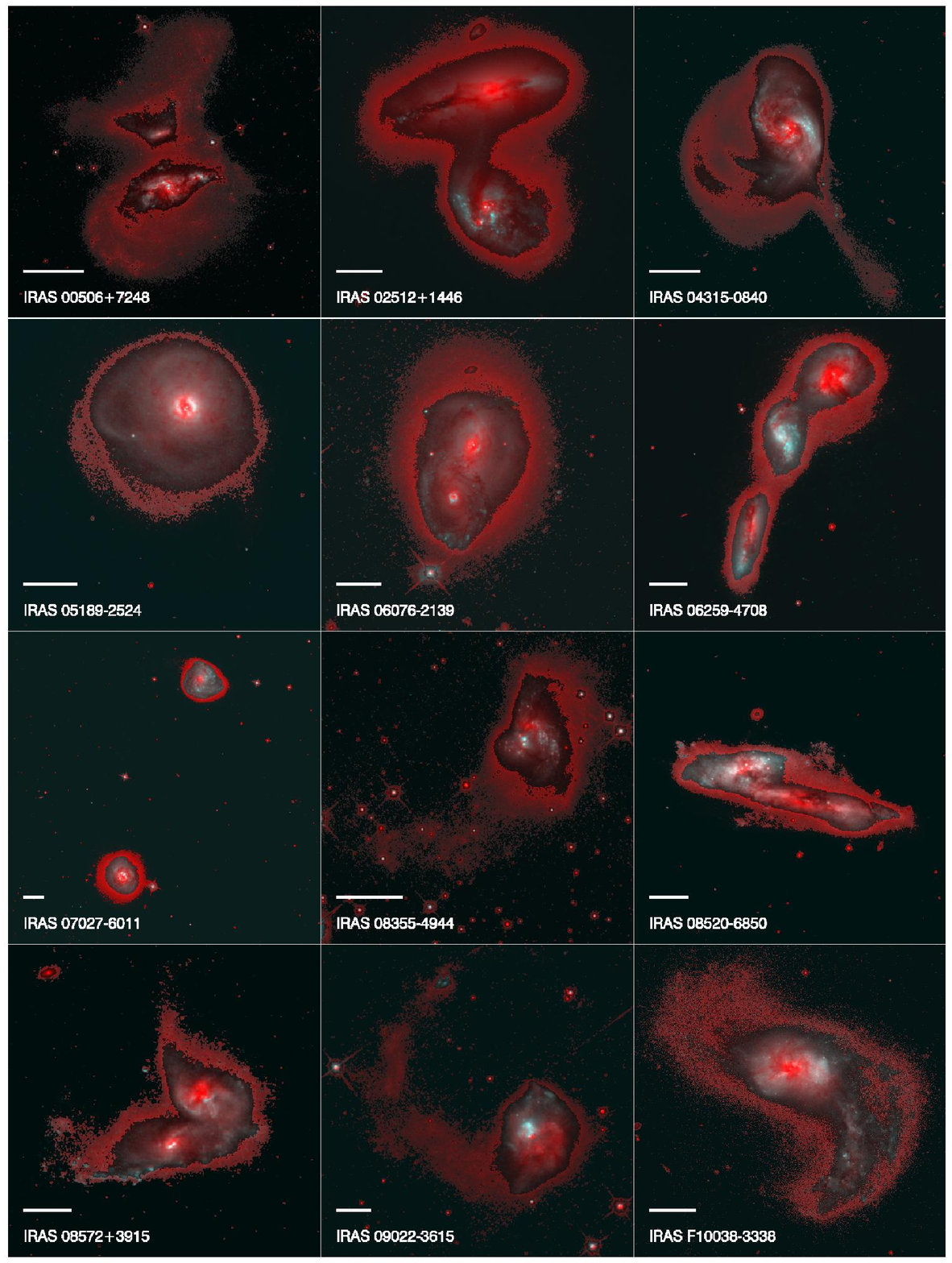} 
\vspace{-1cm}
\caption{Systems under study. False color composite image of the complete sample using \filteri (red) and \filterb (blue) \hst\twospace-ACS images. We have saturated the images below a given surface brightness, typically 5-10 times the global background deviation. The result of this, the diffuse red light, shows the lowest surface brightness features (tails, plums and shells).  The white horizontal line indicates a scale of 5kpc. Blue knots along the tails are clearly visible in some systems. North points up and East to the left.\label{fig:composite34}}
    \end{figure*}
  \begin{figure*}
   \centering
 \vspace{-3cm}
   \includegraphics[width=\textwidth]{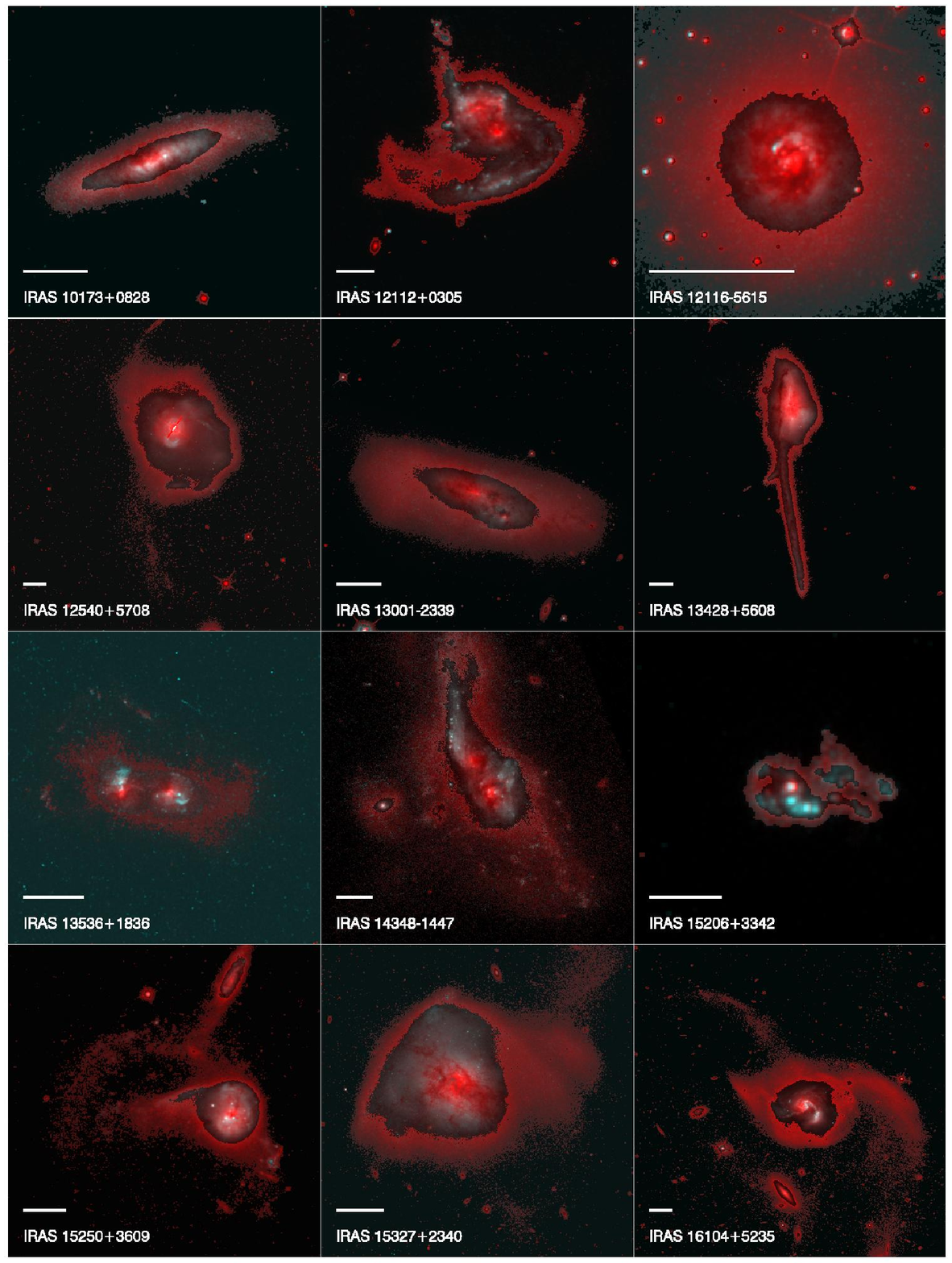} 
\addtocounter{figure}{-1}   
   \caption{- Continued}
    \end{figure*}
  \begin{figure*}
   \centering
 \vspace{-4cm}
   \includegraphics[trim = 0cm 0cm 0cm 0cm,width=\textwidth]{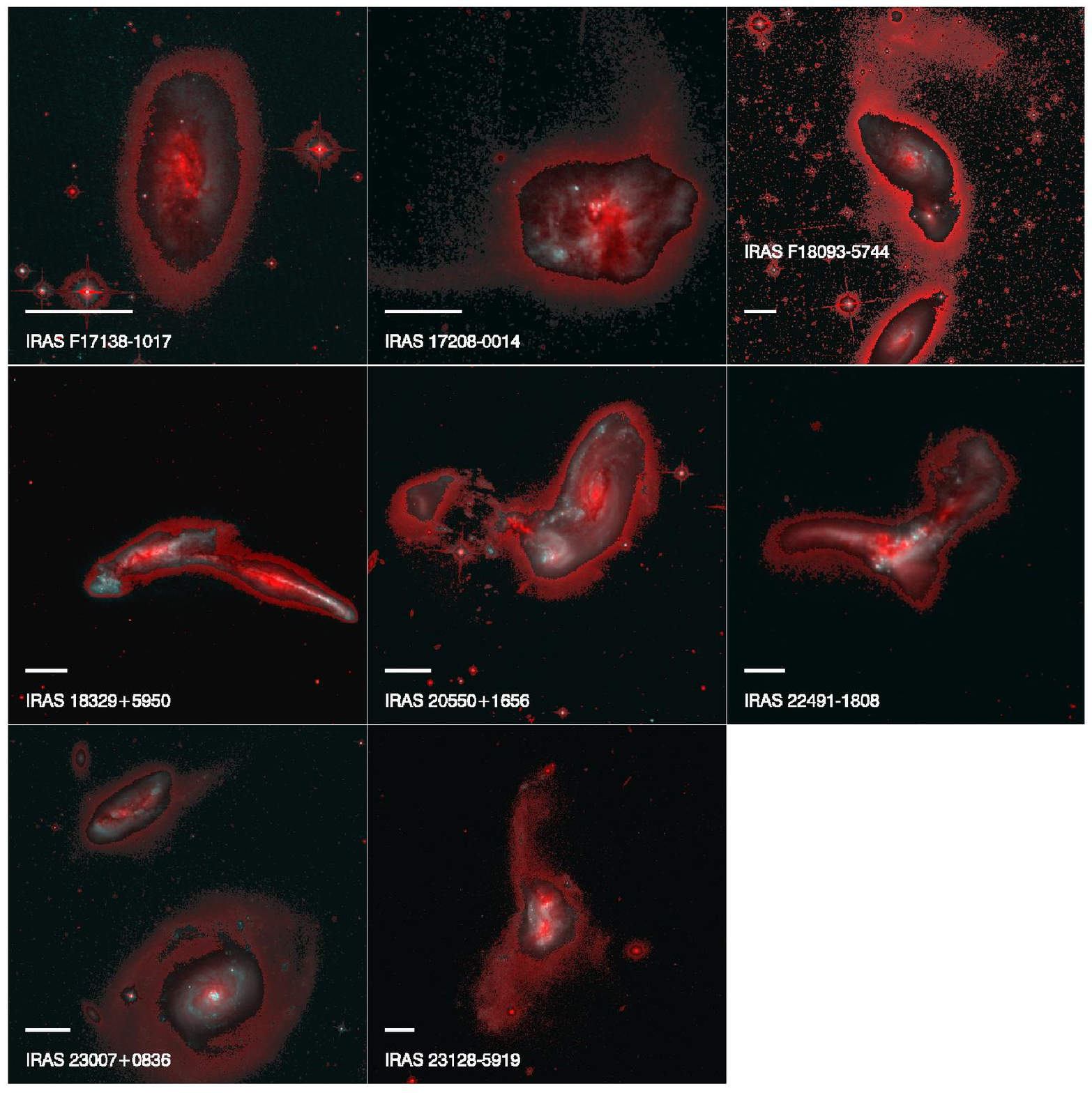} 
\addtocounter{figure}{-1}   
 \vspace{-6cm}
   \caption{- Continued}
    \end{figure*}
  
\begin{enumerate}[]
 \item  ~I-II. \textit{First Approach}.-- In this early phase, the first passage has not been completed. It is prior to the first close passage or during that first passage. The galaxy disks remain relatively unperturbed and bars and tidal tails have not yet formed. In some cases the disks can overlap, but no projected morphological disruption is seen on either galaxy. Some examples of systems in this phase are IRAS 06259-4708 and IRAS 07027-6011. \\
\item  ~III. \textit{Pre-Merger}. --  Two identifiable nuclei, with very well identified tidal tails and bridges, characterizes this phase.  Both disks are still re\-cog\-ni\-za\-ble. This occurs 200-400 Myr after the first approach, according to the similar morphologies observed in~\cite{Bournaud08a} models. Some examples representing this phase are IRAS 08572+3915 and IRAS 12112+0305. \\
\item  ~IV. \textit{Merger}. -- We have only included phase IVa of~\cite{Veilleux02}, since the systems in phase IVb have morphologies more similar to those in phase V than in IVa. In this phase both nuclei have apparently coalesced. We consider that the nuclei have coalesced as soon as their separation is less than 1.5 kpc, since numerical simulations have shown that by the time the two nuclei have reached a separation of $\leq$ 1 kpc, the stellar system has basically achieved equilibrium~\citep{Mihos99,Naab06}. Long tails and shell structures are seen in this phase. Both disks are no longer recognizable, but a common internal structure has formed. This phase is seen 500-700 Myr after the first approach, according to the models mentioned. IRAS 04315-0840 and IRAS F10038-3338 are representative galaxies of this phase. It has to be taken into account that we are considering projected distances. Hence, those systems with double nuclei allocated to this phase could well belong to the previous phase if the real distance of the nuclei is considerably greater.  \\
\item  ~V. \textit{Post-Merger}. -- We have also included phase IVb of~\cite{Veilleux02}. In this phase there are no prominent and bright tidal tails or bridges, since their surface brightness starts to fall below the detection limit, though some shell structures around the nucleus can still be detected. These isolated nuclei can have disturbed stellar envelopes, similar to those in the previous phase, indirectly indicating the past interaction. Typical ages of \mbox{t $\gtrsim$ 1 Gyr} after the first passage characterizes the timescale of this phase, according to the models mentioned. IRAS 05189-2524 and IRAS 12116-5615 are good examples of systems in this phase. 
\end{enumerate}

Since this classification has been performed only by evaluating the apparent morphology of the systems under study, it is always subject to improvements based on data with a higher angular resolution and sensitivity. Therefore, misclassification could occur in a few systems. In addition, for certain studies, slightly different and simpler classifications have been used previously (\citealt{Rodriguez-Zaurin10}; see also Table 1). Although they have
the advantage of reducing subjectivity and therefore uncertainty, their classification is too coarse for
detail comparison with model predictions. While the classification of a few objects can differ, as is the case for IRAS 06259-4807 and IRAS 08520-6850 in common with Rodr\'{i}guez-Zaur\'{i}n's sample (the definitions of the early phases in both studies are slightly different), the agreement between both classifications is excellent. Using this classification, 4 systems are assigned to category I-II, 13 to category III, 8 to category IV and 7 to category V (see table~\ref{table:sample}).

\section{The data}
\label{sec:data}

\begin{deluxetable}{lcccccccc}
\tabletypesize{\footnotesize}
\tablewidth{0pt}
\tablecaption{Available \hst data: observing log details \label{table:dataset}}

\tablehead{
\colhead{IRAS}	&	\colhead{\hst opti-}	&	\colhead{PI and}	&	\colhead{Obs.}	&	\colhead{Exp.}	&	\colhead{\hst NIR}	&	\colhead{PI and}	&	\colhead{Obs.}	&	\colhead{Exp.}	\\
\colhead{name} &	\colhead{cal data}	&	\colhead{proposal ID}	&	\colhead{date}	&	\colhead{time (s)}	&	\colhead{data}	&	\colhead{proposal ID}	&	\colhead{date}	& \colhead{time (s)}\\
\colhead{(1)} &	\colhead{(2)}	&	\colhead{(3)}	&	\colhead{(4)}	&	\colhead{(5)}	&	\colhead{(6)}	&	\colhead{(7)}	&	\colhead{(8)}	&	\colhead{(9)}}
\startdata

00506+7248	&	ACS/WFC	&	Evans, 10592	&	04/09/05	&	800, 1500	&	NICMOS	&	Alonso-Herrero, 10169	&	09/10/04	&	311	\\
02512+1446	&	ACS/WFC	&	Evans, 10592	&	22/07/06	&	720, 1260	&	-	&	-	&	-	&	-	\\
04315-0840	&	ACS/WFC	&	Evans, 10592	&	14/08/06	&	720, 1260	&	NICMOS	&	Rieke, 7218	&	07/02/98	&	191	\\
05189-2524	&	ACS/WFC	&	Evans, 10592	&	13/08/06	&	730, 1275	&	NICMOS	&	Veilleux, 9875	&	28/08/04	&	2560	\\
06076-2139	&	ACS/WFC	&	Evans, 10592	&	14/11/05	&	720, 1260	&	-	&	-	&	-	&	-	\\
06259-4708	&	ACS/WFC	&	Evans, 10592	&	19/04/06	&	780, 1350	&	-	&	-	&	-	&	-	\\
07027-6011	&	ACS/WFC	&	Evans, 10592	&	19/09/05	&	830, 1425	&	-	&	-	&	-	&	-	\\
08355-4944	&	ACS/WFC	&	Evans, 10592	&	08/09/05	&	780, 1350	&	-	&	-	&	-	&	-	\\
08520-6850	&	ACS/WFC	&	Evans, 10592	&	15/04/06	&	870, 1485	&	-	&	-	&	-	&	-	\\
08572+3915	&	ACS/WFC	&	Evans, 10592	&	09/12/05	&	750, 1305	&	NICMOS	&	Maiolino, 9726	&	17/03/04	&	160	\\
09022-3615	&	ACS/WFC	&	Evans, 10592	&	07/06/06	&	750, 1305	&	-	&	-	&	-	&	-	\\
F10038-3338	&	ACS/WFC	&	Evans, 10592	&	06/11/05	&	740, 1290	&	-	&	-	&	-	&	-	\\
10173+0828	&	ACS/WFC	&	Evans, 10592	&	01/12/05	&	720, 1260	&	-	&	-	&	-	&	-	\\
12112+0305	&	ACS/WFC	&	Evans, 10592	&	20/02/06	&	720, 1260	&	NICMOS	&	Scoville, 7219	&	15/11/97	&	192	\\
12116-5615	&	ACS/WFC	&	Evans, 10592	&	07/09/05	&	830, 1425	&	NICMOS	&	Surace, 11235	&	01/07/07	&	2495	\\
12540+5708	&	ACS/WFC	&	Evans, 10592	&	11/05/06	&	830, 1425	&	NICMOS	&	Veilleux, 9875	&	09/09/04	&	2560	\\
13001-2339	&	ACS/WFC	&	Evans, 10592	&	26/02/06	&	720, 1260	&	-	&	-	&	-	&	-	\\
13428+5608	&	ACS/WFC	&	Evans, 10592	&	17/11/05	&	820, 1425	&	NICMOS	&	Maiolino, 9726	&	11/05/04	&	599	\\
13536+1836	&	WFPC2/PC	&	Sanders, 5982	&	27/11/95	&	360, 1030	&	NICMOS	&	Low, 7213	&	27/12/97	&	480	\\
14348-1447	&	ACS/WFC	&	Evans, 10592	&	06/04/06	&	720, 1260	&	NICMOS	&	Scoville, 7219	&	31/12/97	&	480	\\
15206+3342	&	WFPC2/PC	&	Sanders, 5982	&	05/09/95	&	343, 750	&	-	&	-	&	-	&	-	\\
15250+3609	&	ACS/WFC	&	Evans, 10592	&	25/01/06	&	750, 1305	&	NICMOS	&	Scoville, 7219	&	20/11/97	&	223	\\
15327+2340	&	ACS/WFC	&	Evans, 10592	&	06/01/06	&	720, 1260	&	NICMOS	&	Maiolino, 9726	&	10/01/04	&	599	\\
16104+5235	&	ACS/WFC	&	Evans, 10592	&	18/09/05	&	800, 1380	&	NICMOS	&	Maiolino, 9726	&	01/12/03	&	599	\\
F17138-1017	&	ACS/WFC	&	Evans, 10592	&	31/03/06	&	720, 1260	&	NICMOS	&	Alonso-Herrero, 10169	&	23/09/04	&	240	\\
17208-0014	&	ACS/WFC	&	Evans, 10592	&	04/04/06	&	720, 1260	&	NICMOS	&	Scoville, 7219	&	26/10/97	&	223	\\
F18093-5744	&	ACS/WFC	&	Evans, 10592	&	08/04/06	&	830, 1425	&	NICMOS	&	Alonso-Herrero, 10169	&	26/09/04	&	240	\\
18329+5950	&	ACS/WFC	&	Evans, 10592	&	31/10/05	&	830, 1425	&	NICMOS	&	Surace, 11235	&	07/03/09	&	2495	\\
20550+1656	&	ACS/WFC	&	Evans, 10592	&	15/04/06	&	720, 1260	&	-	&	-	&	-	&	-	\\
22491-1808	&	ACS/WFC	&	Evans, 10592	&	04/05/06	&	720, 1260	&	NICMOS	&	Scoville, 7219	&	21/11/97	&	480	\\
23007+0836	&	ACS/WFC	&	Evans, 10592	&	12/06/06	&	720, 1260	&	NICMOS	&	Scoville, 7219	&	10/11/97	&	351	\\
23128-5919	&	ACS/WFC	&	Evans, 10592	&	22/03/06	&	830, 1425	&	NICMOS	&	Maiolino, 9726	&	13/10/03	&	599	\\
\enddata
\tablecomments{Col (1): object designation in the IRAS Point and Faint Source catalogs. Col (2): optical instruments used on board \hst. WFC refers to the Wide Field Channel of the ACS camera. Sources with the instrument WFPC2 were focused on the PC chip. Col (3): last name of the principal investigator, followed by the proposal ID of the optical observation. Col (4): day when the optical observation was taken. Col (5) total exposure times for each galaxy, with filter \filteri (left) and filter \filterb (right). For observations with the WFPC2 the blue filter corresponds to \textit{F439W}. Col (6): NIR instruments used on board \hst to observe the retrieved \filterh images. For all these observations the camera used was NIC2. Col (7): same as (3) for the infrared data. Col (8): same as (4) for the infrared data. Col (9): same as (5) for the infrared data, with filter \filterh\twospace. }
\end{deluxetable}

The results presented in this paper are based on high angular resolution archival \hst images. The available dataset for each galaxy at the moment of the analysis is presented in table~\ref{table:dataset}. 

Advanced Camera for Surveys (ACS) broad-band images were taken for thirty systems from the Hubble Legacy Archive, with the filters \filteri and \filterb\twospace. The former is equivalent to the ground-based Johnson-Cousins \textit{I}, whereas the latter differs from the ground-based Johnson-Cousins \textit{B} between 7 and 12\% in flux~\citep{Sirianni05}. The field of view (FoV) is 210'' x 210''. Total integration times were taken from 720 to 870s with filter \filteri and from 1260 to 1500s with filter \filterb\twospace. For IRAS 13536+1836 and IRAS 15206+3342, Wide Field Planetary Camera 2 (WFPC2) images were taken, with the filters \filteri and \textit{F439W}, being equi\-va\-lent to ground-based Johnson \textit{I} and \textit{B}, respectively~\citep{Origlia00}. Here, the systems fall in the PC chip, giving a FoV of about 40'' x 40''. Whenever both ACS and WFPC2 images were available in both filters, ACS images were preferred for its superior sensitivity and larger FoV. The pixel size in both instruments is about 0.05''. From now on, we will refer to the blue filter as \filterb for ACS and WFPC2 for simplicity. Complementary Near Infrared Camera and Multi Object Spectrometer (NICMOS) broad-band images with filter \filterh ($\sim$ \textit{H}) were also available for twenty galaxies, which were helpful to locate the nuclei of the systems. However, they have not been included in the photometric study since they have small FoV (20'' x 20''), not covering the outskirts of the systems or interacting tail-bridge structures. 

Images were reduced on the fly (\hst pipeline), with the highest quality available reference files at the time of retrieval. The calibrated ACS \filterb images had only few cosmic rays remaining, since 3 images were observed and then combined. However, all the ACS \filteri images were severely contaminated with cosmic rays in a $\sim$ 6 arcsec width band across the entire FoV. Two exposures were taken and when they did not overlap the \hst pipeline was not able to discriminate the cosmic rays. In most cases that band went through part of the system, covering about 10-20\% of its extension. Cosmic ray rejection was then carried out using the IRAF task \textit{credit} with the aid of the blue images since they were barely contaminated. Most of the systems have observations that cover the entire FoV with only this instrument and the two filters, so the aid of multi-band images was limited to few systems. 

\section{Analysis techniques}
\label{sec:analysis}
\subsection{Source detection and photometry}
Compact regions with high surface brightness (knots hereafter) that are usually bluer than the underlying galaxy  have been identified for each galaxy. They have been detected inside a box that encloses each system including the diffuse emission observed in the red band. They appear similar to the bright blue knots found in interacting galaxies like in NCG 7252~\citep{Whitmore93}, NGC 4038/39~\citep{Whitmore95}, some ULIRGs~\citep{Surace98}, M51~\citep{Bik03,Lee05} and Arp284~\citep{Peterson09}. These knots were identified on the basis of having a flux above the 5$\sigma$ detection level of the local background. Due to the irregular and distorted morphologies for most of the systems, we performed a two step analysis to identify the compact regions: we first smoothed the image, averaging in a box of 30$\times$30 pixels, and subtracted this smoothed image from the original one in order to make a first rough local background subtraction; we then ran SExtractor~\citep{Bertin96} on the resulting image to detect the knots. Due to the steepness of the light profile of the local background in the inner regions and its larger deviation with respect to outer regions, an appropriate background subtraction was not always possible and some inner knots were never detected. Thus, visual inspection was needed afterward to include some inner knots and also to eliminate some spurious detections. 

Once the knots were detected, all photometric measurements were done using the tasks PHOT and POLYPHOT within the IRAF~\citep{Tody93} environment. We first identified the point-like sources against the extended ones, since the  photometry of the former requires aperture corrections. We fitted the knots with a Moffat profile, and those with a \mbox{FWHM $\lesssim$ 1.5$\times$FHWM} the profile of the stars in the field were considered as point-like knots. Note that although they actually consist of point-like and slightly resolved objects, they are considered as point-like objects here just to assign an aperture radius, since both require aperture corrections when the photometry is done. Once their size is derived via psf fitting (see section~\ref{sec:sizes}) we differentiate between resolved and unresolved knots.  

Aperture photometry with a 3 pixel aperture radius (0.15'') was performed for the point-like knots. For the resolved knots, we used aperture radii between 5 and 7 pixels (0.25-0.35''), depending on when the radial profile reached background levels. The 7-pixel apertures were large enough to include practically all the light from the knot but small enough to avoid confusion with other sources and to avoid adding more noise to the measurements.  In some cases, the irregular shape of the knots demanded the use of polygonal apertures. In some cases, especially for knots belonging to the most distant galaxies, the polygonal apertures were larger than 7 pixels. Estimates of the underlying background flux were made by using the mode of flux values measured in an 8-13 pixel annulus centered on the computed centroid with POLYPHOT, and eliminating the 10\% of the extreme values at each side of the flux background distribution (e.g., if there is a nearby knot it adds spurious flux at the bright end of the distribution). Most of the knots are apart from each other by more than 7 pixels (0.35''), hence confusion does not seem to be a serious problem, though in the inner regions of few systems (IRAS 04315-0840, IRAS 15206+3342 and IRAS 20550+1656) there is some overcrowding. 

Photometric calibrations were performed following~\cite{Sirianni05} for the ACS images and \linebreak \cite{Holtzman95} for the WFPC2 images (hereafter calibration reference papers). Magnitudes were then computed in VEGA system. Correction for charge transfer efficiency (CTE) was needed in the WFPC2 photometry but was not applied to the ACS photometry, since tests in some images shown that the effect was less than 1\% for the faintest regions identified.  Aperture corrections were considered for the point-like knots. To that end, we performed the photometry for isolated stars in the field (from a few to about a hundred, depending on the image) in the same way as we did for the point-like knots, then a second time but using a 10 pixel radius. The ratio of both values gives us an aperture correction up to 10 pixels. From 10 pixels to an \textit{infinite} aperture radius the values in the calibration reference papers were taken, since the PSF profile from 10 pixels on is stable enough regardless of the observing conditions. Typical aperture corrections were about 0.3 and 0.2 mag for the \filteri and \filterb filters, respectively. Using the deviation of the different aperture corrections computed with stars in the field we estimated the uncertainty due to the aperture corrections to be of the order of 0.05-0.1 mag in both filters. Typical photometric uncertainties for all the knots lie between 0.05 and 0.15 mag, depending on the brightness of the knot. According to the completeness test performed in section~\ref{sec:completeness}, the systematics affect both filters nearly equally, thus the typical uncertainty in color is between 0.10 and 0.20 mag. 

We corrected all the magnitudes for reddening due to our galaxy taking the values directly from the NASA/IPAC Extragalactic Database (NED), computed following~\cite{Schlegel98}. Apart from this, no internal reddening corrections have been applied to the magnitudes reported in this paper. 

Initially, we probably detected old globular clusters, super star clusters, star complexes, \hii regions, foreground stars, background objects, etc. Some of them were rejected after the photometry was made. Foreground stars were mostly easily identified by their high brightness and red colors and their status as point-like objects. The fields in systems at low galactic la\-ti\-tu\-des (IRAS 08355-4944, IRAS 09022-3615 and IRAS 12116-5615) are so crowded with foreground stars that some may still  contaminate the pho\-to\-me\-tric sample knots after rejection, especially if a faint foreground star lies within the inner regions of the galaxy. No cluster/knot with color \mbox{\mbi$\gtrsim$ 2.2} fits in stellar population models for an extinction-free starburst at redshift z$\sim$0. Since we do not expect much extinction in the outer parts of the galaxies (usually where we do not see diffuse emission from the galaxy in the red band), all the red knots outside the diffuse emission of the systems are likely to be background objects and were also rejected. After these rejections a total of 2961 knots were considered under study. 

The nuclei of the systems were identified using the \textit{H} band images and with the aid of other works that have detected them in radio (e.g.,~\citealt{Dinshaw99}). The identification of the true nucleus is of relevance in the derivation of the projected distances of the knots with respect to the closest nucleus, to properly study the spatial distribution of these knots. 

\subsection{Determination of sizes}
\label{sec:sizes}

\subsubsection{Effective radius}

The half-light radius of the knots was derived by fitting radial Gaussians, technique that has been successfully used in other works (e.g.,~\citealt{Surace98,Whitmore93}). The measured radial profile of a knot corresponds to the convolved profile of the PSF with the intrinsic profile of the knot.  We then fitted 2D-Gaussians to each knot and to foreground stars in the image (to compute the PSF profile). We can estimate the $\sigma$ for the intrinsic profile of the knot as $\sigma$=$\sqrt{\sigma_{m}^2-\sigma_{PSF}^2}$, where $\sigma_{m}$ corresponds to the $\sigma$ of the measured profile (average of $\sigma$ in the x and y direction) and $\sigma_{PSF}$ to that of the PSF profile.  An effective radius, \reff\twospace, was derived as the half value of the FWHM derived from the computed $\sigma$ (FWHM=2.354$\sigma$). 

Note that \reff defined in this study does not strictly correspond to the half-light radius, since the light profile of the globular clusters and young massive clusters detected in the Milky way and in other galaxies is not necessarily Gaussian. However, given the fact that the distance of 80\% of the systems is larger than 100 Mpc we do not expect to measure resolved sizes as small as the typical size of young massive clusters (typically, \mbox{\reff $<$ 20 pc};~\citealt{Whitmore99}), but associations or cluster complexes. Sizes of globular clusters in the Milky Way are even smaller. Therefore, it is not worth using  in this study more robust methods (e.g.,~\citealt{Larsen99b}) that better characterize the light profile of the source, and where the derived \reff is closer to the half-light radius . The main interest here is the relative change in the sizes of the knots among the different systems. 

\begin{figure}[!t]
   \centering
\includegraphics[angle=90,width=0.95\columnwidth]{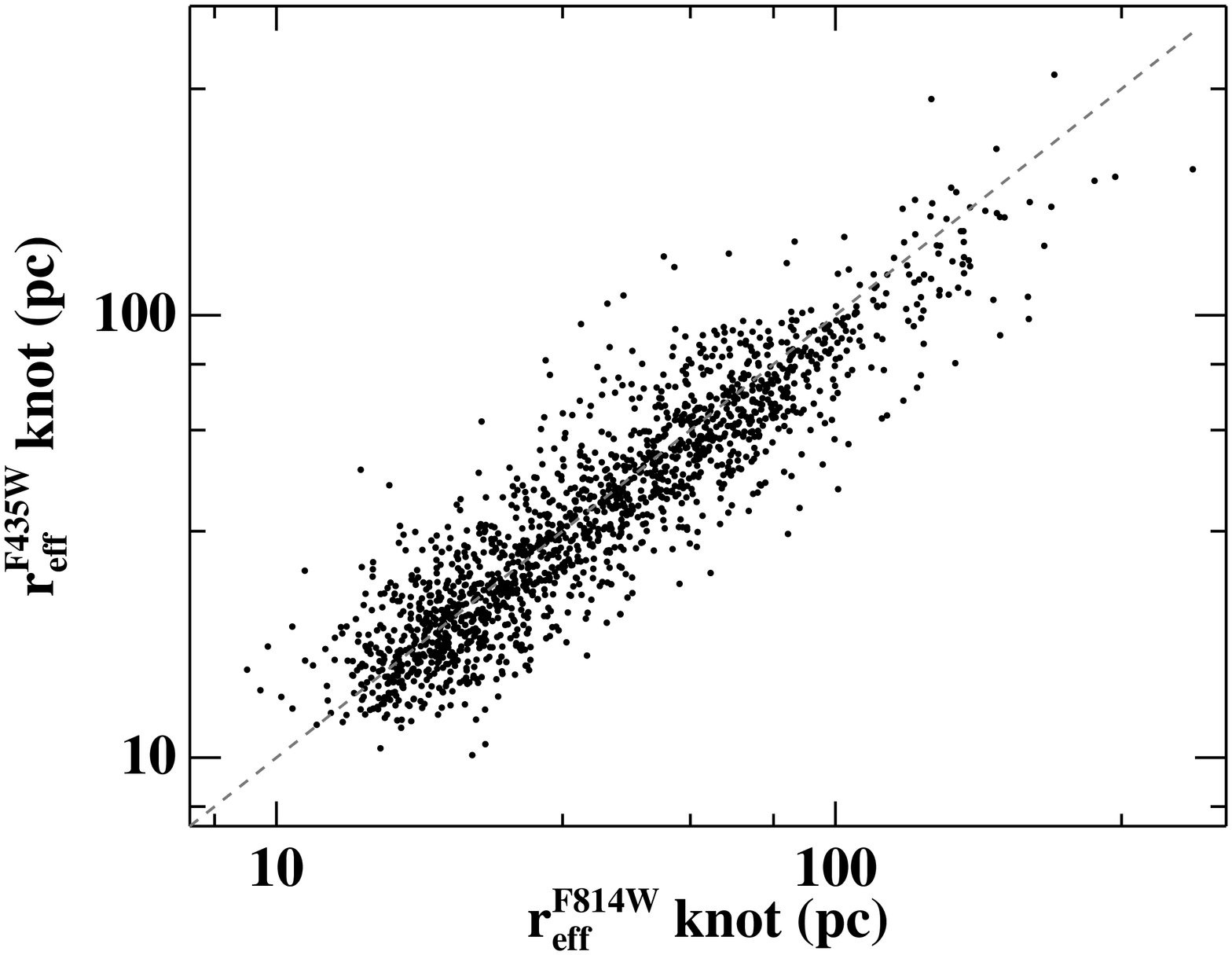} 
    \caption{Relation between effective radius for all the knots measured in images with filter \filteri and \filterb\twospace. Both axes are in log scale. The line indicates a ratio of unity.\label{fig:reff3}}
              
 \end{figure}

The rms scatter around the 1:1 line in Fig~\ref{fig:reff3} is about 30\%, from 3 pc to 90 pc for knots with sizes of 10 and 300 pc, respectively. This should give a reasonable estimate of the uncertainty in the measurement of \reff\twospace. Differential extinction, which is wavelength dependent, is also responsible for part of the scatter. Then, under similar local background conditions, knots will look a bit more extended when observed with the \filteri filter. Another source of uncertainty comes from the young population. In fact, faint young blue knots easily detectable in the \filterb image but with low S/N ratio in the \filteri image will look more extended in the blue ones. In any case, this measurement is unlikely to be affected by the nebular emission-lines (i.e., \ha\twospace, \hb or oxygen lines), since in the vast majority of the systems these lines do not significantly contaminate the \hst filters used in this study.

In order to estimate the minimum size we can achieve by using this technique we estimated which would be the Gaussian that, convolved with that of the PSF profile, would give a total FWHM = FWHM$_{PSF}$ + 3{\scriptsize{STD}}, where  {\scriptsize{STD}} gives the standard deviation of the FWHM distribution measured for all the stars that were used to compute the PSF profile. For the closest systems we are typically limited to about 10 pc and for the furthest systems, 40 pc.

We could estimate \reff using Gaussian fits for about 88\% of the knots in the sample. The profile for the rest of the knots could not be fitted by Gaussians, generally due to their more extended, diffuse and irregular shape, to very steep variations of the local background and/or to low S/N of the faintest knots. 

\subsubsection{Total size}
\label{sub:tot_size}

The total size of the knots was obtained from the \filteri images, since this filter is less affected by extinction than the other photometric filter in this study and therefore is more convenient to measure the radial extent of the light of the knots. We derived the edge of the knots at the point where the flux, as a function of radius, equals the local background flux. In order to determine this point, we assumed that the knots are circular and measured the surface brightness in concentric rings, as done in ~\cite{Bastian05a}.
%We also tried other used by \cite{Maiz01} and \cite{Bastian05a} with closer systems, which defines the edge of the complex at the point where the color, as a function of radius, becomes constant. However, the knot does not generally show a clear color dependence on its position and an the performance of an automatic script to derive the size of the knots has not been possible.}

The size of the knots has been derived then by fitting the radial profile of the light with a power-law of the form Flux \mbox{$\propto$ r$^{-\beta}$}. This function follows the data quite well and corresponds to the projected density profile of a knot with \mbox{$\rho \propto$ r$^{-(\beta+1)}$}, where $\rho$ is in units of \msun pc$^{-3}$ (\citealt{Bastian05a}). For slightly-resolved knots the PSF may be dominating the size measured. We then estimated the half-light radius of the PSF and applied the ratio \mbox{\reffknot/ \reffpsf} to the derived size.

\begin{figure}[!t]
   \centering
\includegraphics[angle=90,width=0.95\columnwidth]{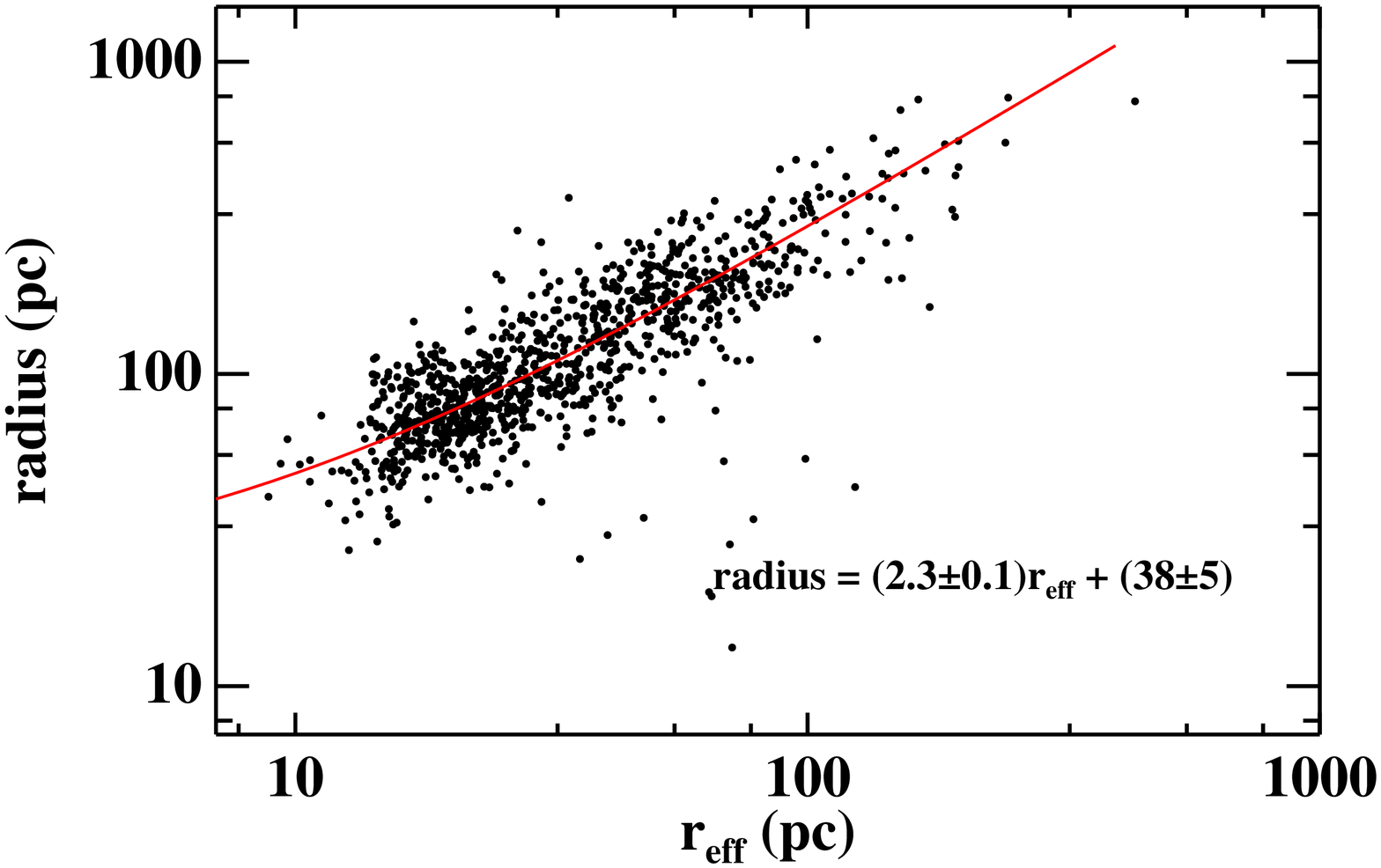} 
   \caption{Relation between effective radius and total radius of the knots for which we could estimate the total size value with an uncertainty of less than 50\%. Both axes are in log scale. The fit for the data is shown at the bottom-right corner.\label{fig:size_reff}}
              
 \end{figure}

Due to the complex nature of (U)LIRGs, a large fraction of them involved in an interaction process, the light profile of the local background of the galaxies is usually very steep and irregular. Thus, the application of this method is somewhat limited. Besides, the size of the unresolved knots cannot be derived. We could measure reliable sizes, with an uncertainty of less than 50\%, for a third of the detected knots. However, we could estimate the total size of the remaining knots (as long as they are resolved) by using the relation found between \reff and the total size (see Fig.~\ref{fig:size_reff}). In any case, given these high uncertainties and that the half-light radius is an observable which is more generally used, the study of the size of the knots in this paper will be focused on \reff\twospace. The value of the total size of the knots will be used for the specific purpose of characterizing the mass-radius relation of the knots.

\subsection{Completeness tests}
\label{sec:completeness}

We have run a series of artificial star tests in order to assess the completeness and measurement quality of the data set. Some considerations were taken before applying the tests:
\begin{enumerate}[-]
 \item Since each system has a different local background level we had to run a test per image. The same happens if we consider different filters. \\
 \item In the central regions of the galaxies the local background level is higher and steeper than in the outer regions. Hence, different completeness levels are expected as the galactocentric distance increases. To that end we defined two surface brightness regions based on the median value across the system: the inner region, corresponding to high surface brightness levels, and the outer region, corresponding to low surface brightness levels. Using this approach, we ran completeness tests separately for each region and afterwards we weighted each computed value by the number of knots within each region. Some works have used more varying levels of background (e.g., 7 in~\citealt{Whitmore99}, 3 in~\citealt{Haas08}), but given the angular size of most of the systems in this study it is not worth defining such number of levels in this study. \\
 \item  We have also taken into account the size of the knots, since most of them are actually slightly resolved. To include this effect we also ran two tests separately. First, we added artificial stars with a PSF computed using foreground stars. Additionally, we artificially broadened that PSF by convolving it with a Gaussian that gives a \mbox{FWHM $\sim$ 1.5 $\times$ FWHM} of the initial PSF. We have then taken the average value of the completeness limit in each case. 
\end{enumerate}

The tests were run by adding artificial stars from apparent magnitudes of 20 to 30, in bins of 0.2 mag. We followed several steps to run a completeness for each of the 50 magnitude bins defined. We first added 1000 artificial stars per bin in a random position inside a field that covers the entire system. Poisson noise was added and only stars which were more than 10 pixels apart were selected to avoid contamination between them. Next, we determined in which of the two defined regions with different surface brightness (inner and outer) each artificial star was located. Finally, artificial stars were recovered by per\-for\-ming the photometry in the same way as we did for the knots.  

\begin{figure*}
   \centering
   %F814W
   \includegraphics[width=0.27\textwidth]{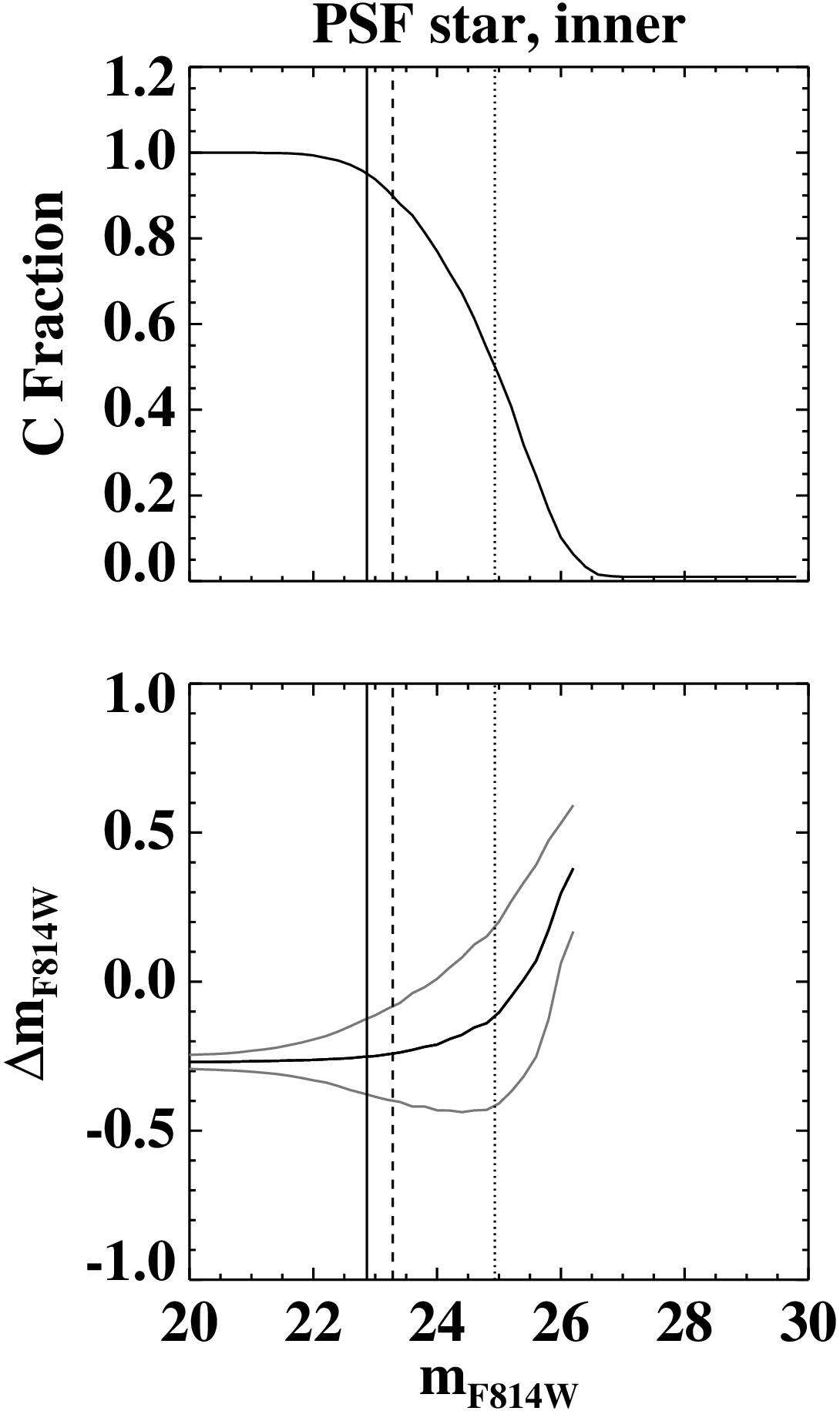}
    \hspace{-0.9cm}
   \includegraphics[width=0.27\textwidth]{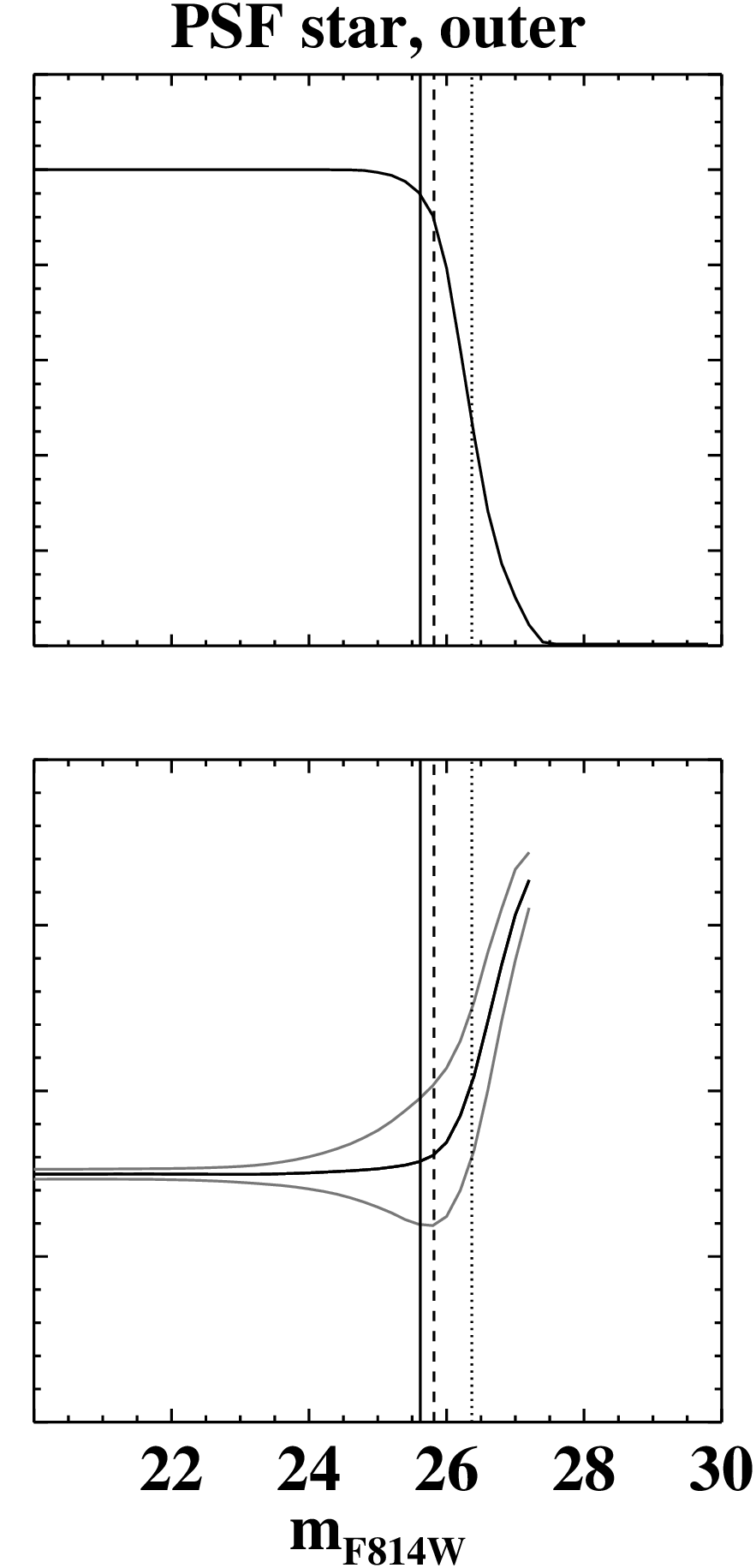}
    \hspace{-0.9cm}
   \includegraphics[width=0.27\textwidth]{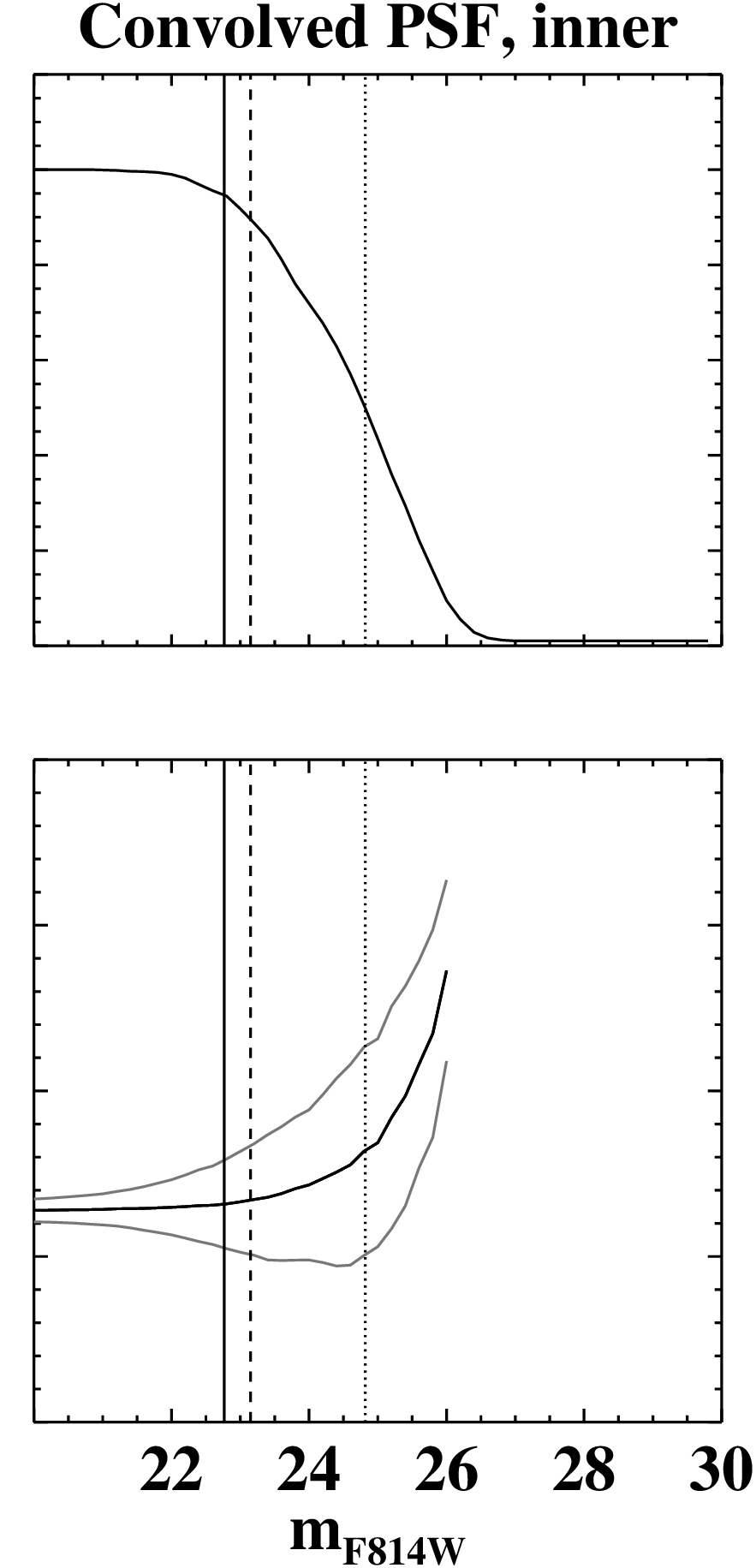}
    \hspace{-0.9cm}
  \vspace{3cm}
   \includegraphics[width=0.27\textwidth]{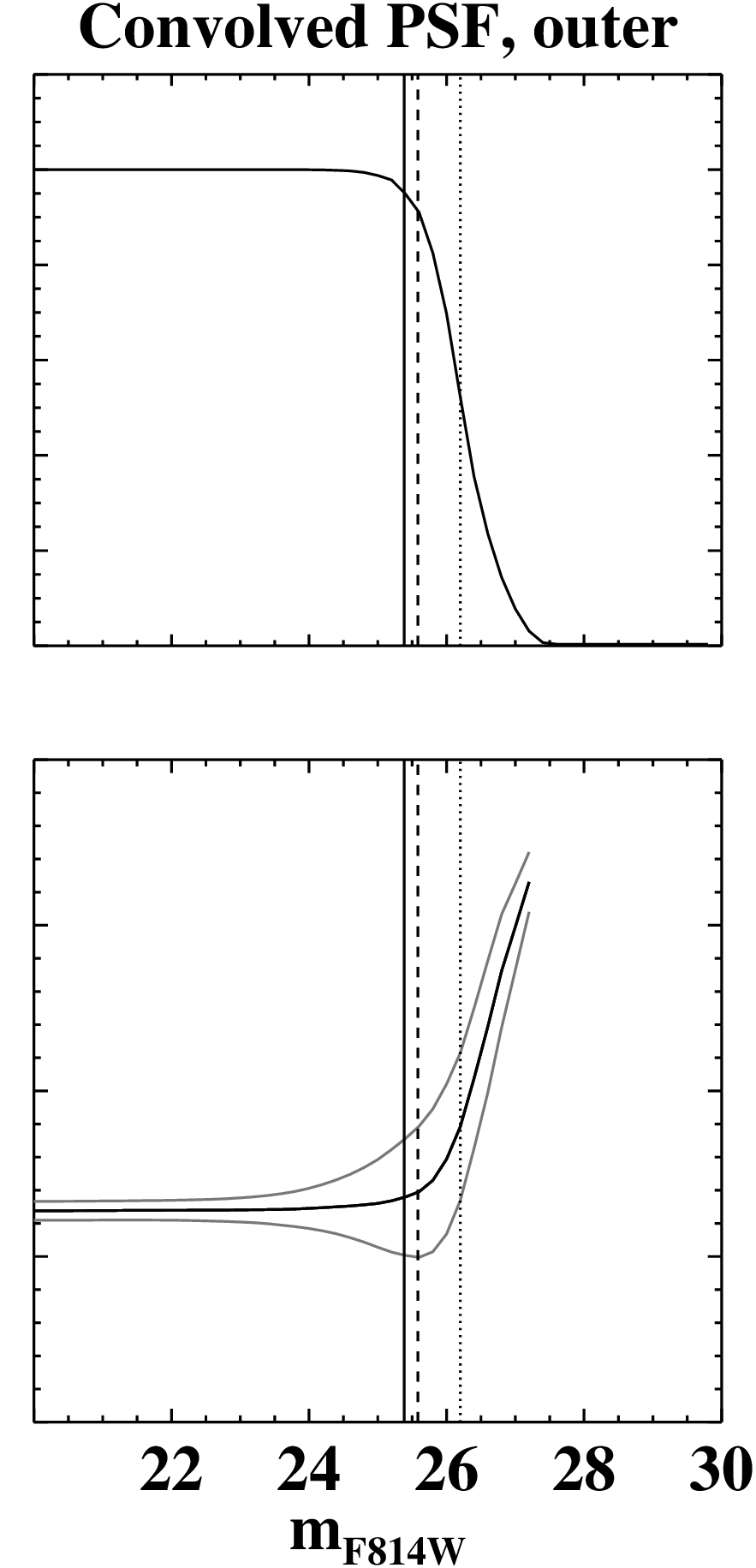}
   %F435W

   \includegraphics[width=0.27\textwidth]{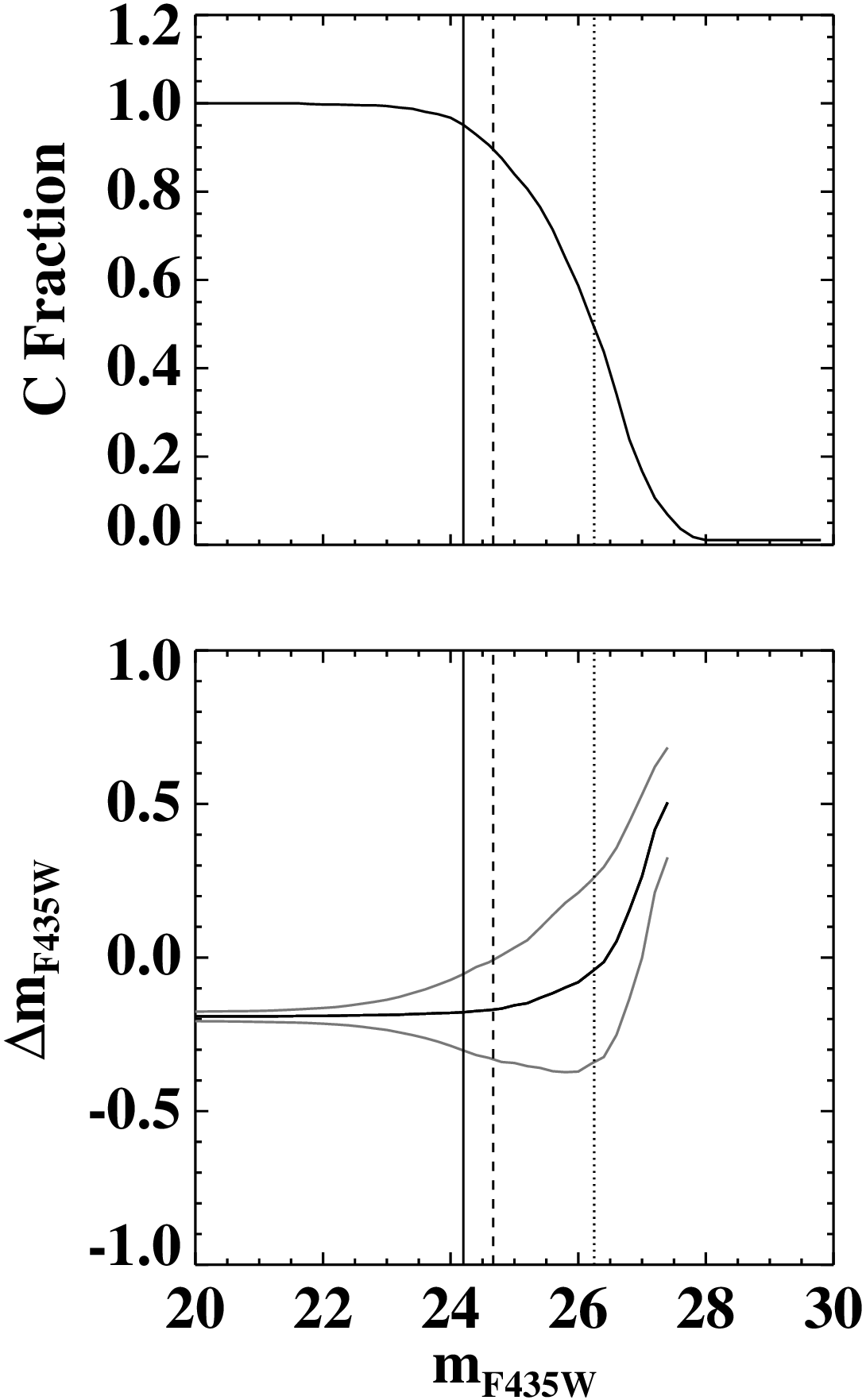}
    \hspace{-0.9cm}
   \includegraphics[width=0.27\textwidth]{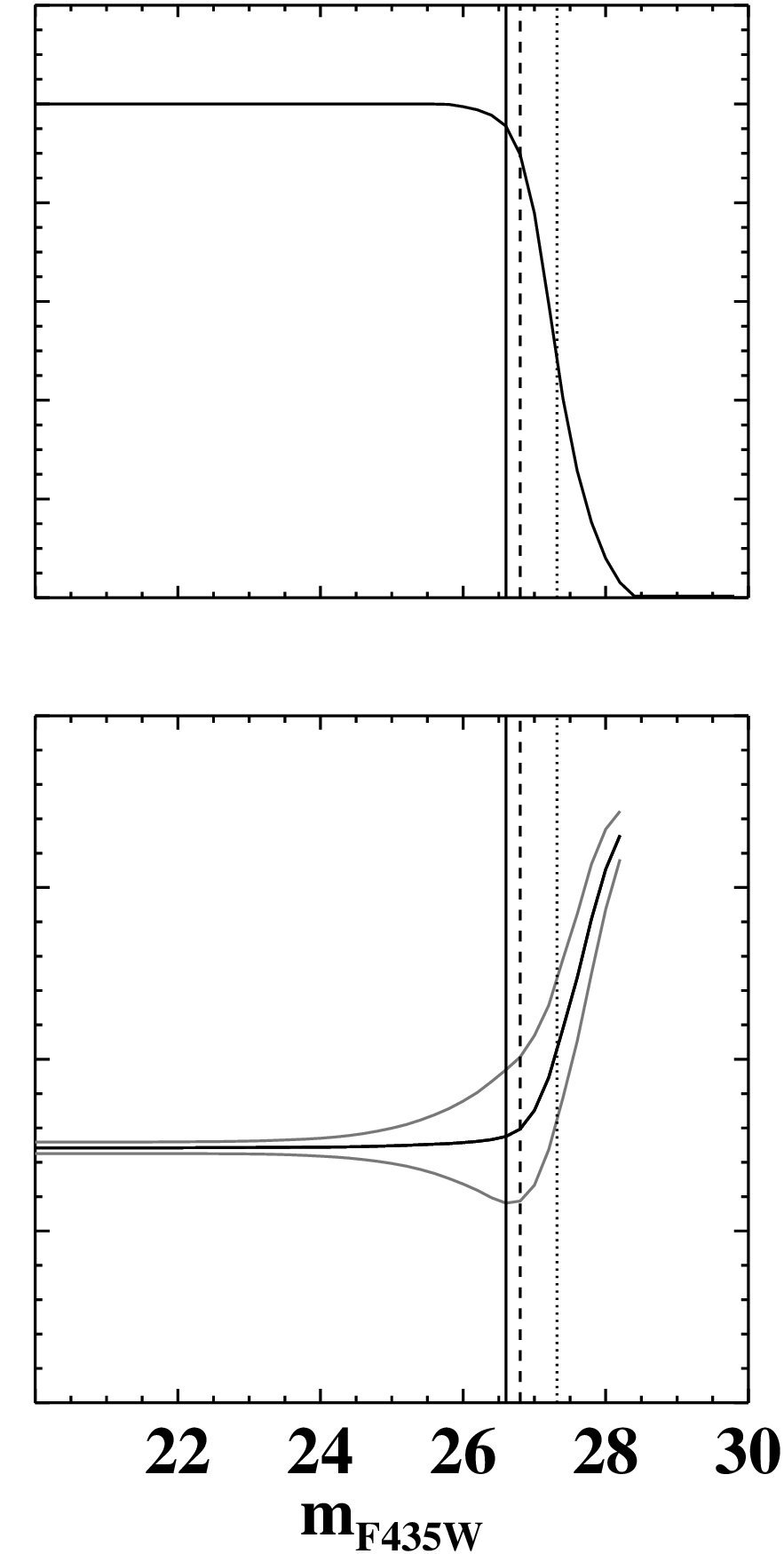}
    \hspace{-0.9cm}
   \includegraphics[width=0.27\textwidth]{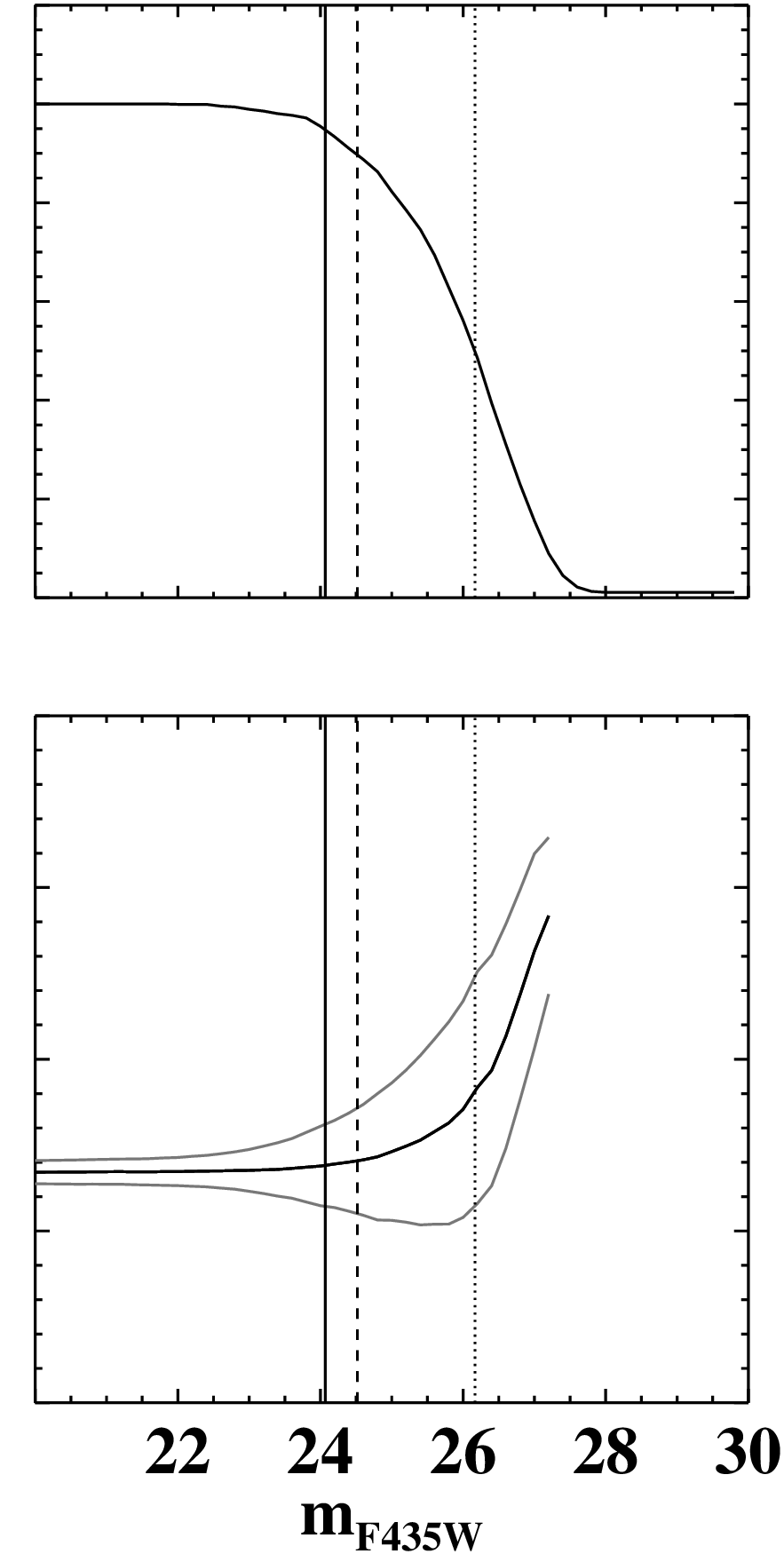}
    \hspace{-0.9cm}
   \includegraphics[width=0.27\textwidth]{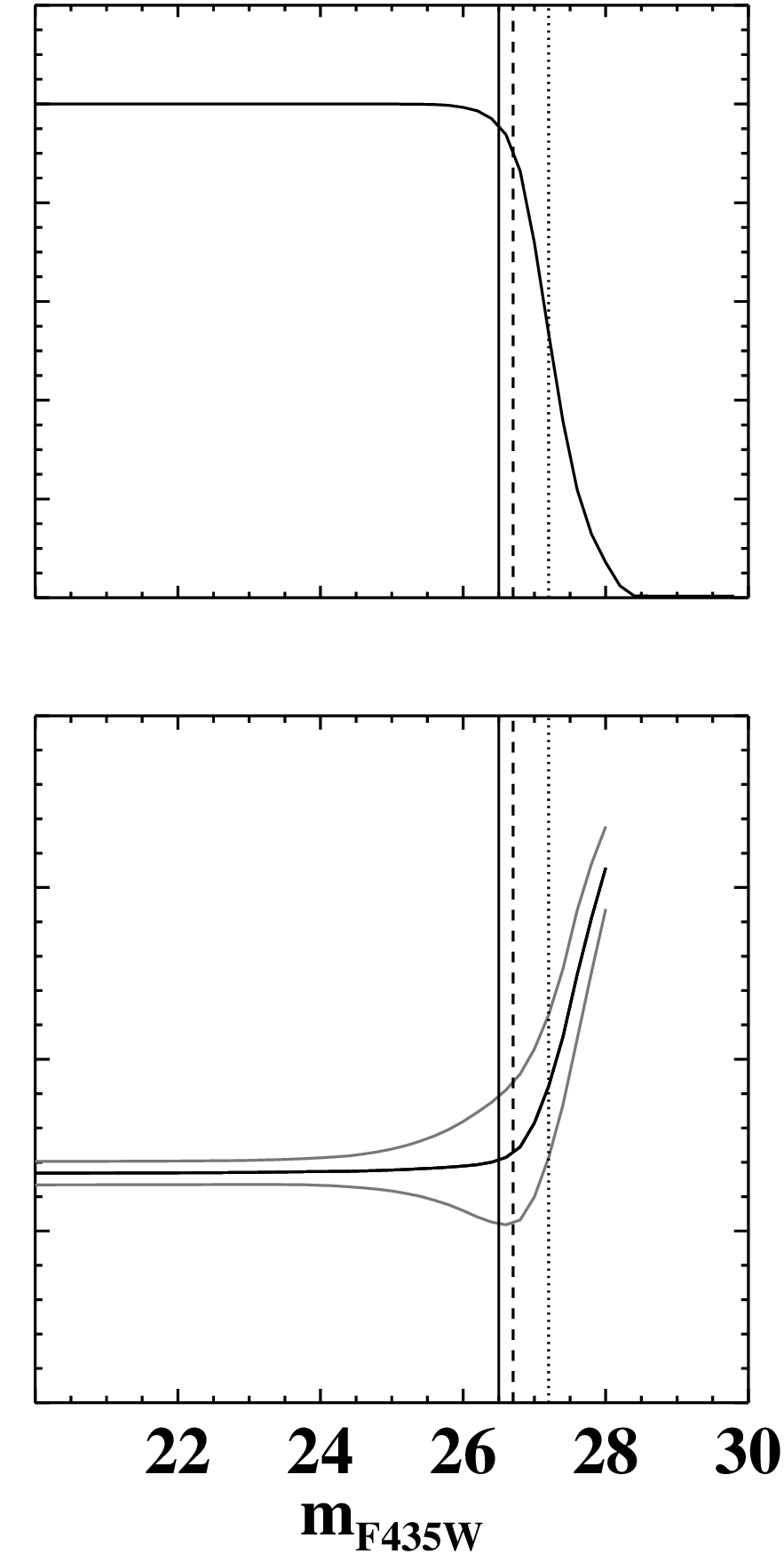}
  \vspace{0.5cm}

   \caption{Result of the completeness test for IRAS 04315-0840. In each case, from left to right, the curve is shown for the inner region using the PSF of foreground stars, the outer region with the same PSF, and then the same using the convolved PSF. From top to bottom, completeness fraction for filter \filteri\twospace, input-output of the recovered stars in black and 1$\sigma$ value added and subtracted in gray are shown, and the same for filter \filterb\twospace. The solid vertical line indicates a completeness level of 95\%, the dashed line a level of 90\% and the pointed one a level of 50\%.\label{fig:completeness}}
              
 \end{figure*}

Since the artificial stars were placed in random positions, some could reside in the same position as a real cluster/star/knot in the original image, increasing the measured flux. In order to avoid this, we rejected those cases where a recovered magnitude differed from the input magnitude by more than a certain value. We have taken the value of 0.75 mag because, as seen in Figure~\ref{fig:completeness}, the systematic uncertainty for faint magnitudes is of the order of 0.4 mag, thus a 2$\sigma$ rejection level seems reasonable. False-positive detections with that filtering for the brightest artificial stars are not expected, since few real bright objects are encountered in the images. The completeness at each magnitude level is computed as the number ratio of recovered  over added artificial stars. All the process was done 10 times and an average completeness value at each magnitude level was computed. An example of the result for the images of IRAS 04315-0840 is shown in Figure~\ref{fig:completeness}.

The mean difference between input and output magnitudes of the artificial stars ($\Delta m_{F814W}$ and $\Delta m_{F435W}$ in Figure~\ref{fig:completeness}) gives an estimate of the aperture correction, which is about -0.3 mag for the \filteri filter and a value close to -0.25 mag for the \filterb filter. This is consistent with the aperture corrections applied in the photometry of the sources. Its standard deviation also represents a robust estimate of the actual photometric error to be associated with each magnitude bin.  This error is obviously larger than that computed by Poisson statistics only, especially toward the fainter magnitudes, since the former consider both the systematic and random uncertainties (see gray curve in Figure~\ref{fig:completeness}). Therefore, a more realistic uncertainty value for the knots with magnitudes near the 50\% completeness limit would move up to 0.3-0.4 mag. 

The systems are 90\% complete within the intervals $m_{F814W}$=23-25 and $m_{F435W}$=25-26.5, but due to the colors of the regions the \filterb completeness limit usually shifts to brighter magnitudes. We do not see a significant difference in the completeness limits when the PSF is convolved for the \filterb filter (less than 0.05 magnitudes), but for the red filter the value of $\Delta m_{F814W}$ is increased by 0.1 mag when the PSF is convolved. Some difference is expected, since as the PSF widens we recover less flux by using the same aperture.

\subsection{Determination of the luminosity function}
\label{sec:LF}
The luminosity function (hereafter LF) quantifies the number of objects per luminosity bin. It is usually characterized in studies of stars in a cluster, star clusters and galaxies. The initial mass function and the processes that stars, clusters and galaxies undergo once they are formed defines the shape of the LF. While the globular cluster LF generally appears to be well fitted with a Gaussian function  (e.g.,~\citealt{Whitmore97}), for the young knots and clusters observed in merging systems and other environments, it is well described as a power-law distribution: $dN \propto L^{-\alpha}_{\lambda} dL_{\lambda}$ or the equivalent form using magnitudes, $dN \propto 10^{\beta  M_{\lambda}} dM_{\lambda}$,  $L_{\lambda}$ and $M_{\lambda}$ \mbox{being} the luminosity and magnitude respectively and N being the number of clusters. The slopes in both equations are related as $\alpha = 2.5  \beta + 1$. The slope generally lies between 1.8 and 2.4 in the optical for systems between 30 and 85 Mpc~\citep{Whitmore95,Schweizer96,Miller97,Surace98}.

As we have seen in section~\ref{sec:sizes}, the knots is this study consist generally of cluster complexes, especially for galaxies located further than 100 Mpc. Thus, the LFs of the knots in this study must not be confused with those in studies of individual star clusters. In section~\ref{sec:distance_effect} we explore how the shape of the LF changes due to blending caused by the lack of angular resolution to resolve individual entities.

The shape of the LF has  usually been determined by fitting an equal sized bin distribution. Instead, we used a method described and tested in~\cite{Maiz05}, based on \linebreak \cite{Dagostino86}. It is based on using bins variable in width, such that every bin contains approximately an equal number of knots. Hence, the same statistical weight is assigned to each bin and biases are minimized.~\cite{Haas08} have shown that this method is more accurate than those based on fitting an equal sized bin distribution. The number of bins ($N_{bins}$) is different for all samples and is related to the total number of objects in the samples. Based on the prescription in~\cite{Maiz05} we computed the number of bins using the expression: $N_{bins} = 2  N_{C}^{2/5} + 15$, where $N_{C}$ corresponds to the total number of knots. 

The fit was performed on the whole range from the 90\% completeness limit to the brightest knot in the sample, thus the slope of the LF might be sensitive to an incompleteness of about 10\%. Using a higher completeness limit cuts off a large number of the knots. By correcting for incompleteness at the 90 and 95\% limits the LF would
become steeper (we underestimate the slope of the LF if we do not correct for incompleteness). The typical value of the underestimate of the slope of the LF in~\cite{Haas08} is $\Delta \alpha \lesssim$ 0.1 for the ACS \filterb and \filteri filters. We have computed a similar value, so we adopt it for this study. 

The uncertainties associated with the fit of the LF (shown in Tables and figures) should be considered lower limits since they do not take into account other sources of error. In fact, if the brightest bins are dropped in the fit the values obtained for the slope in all cases agree to within 0.08 dex. Moving the magnitude limit higher by 0.5 mag changes the slope generally by up to 0.05 dex. Finally, we have also explored how the slope varies when performing a linear regression fit to  the cumulative luminosity distribution (see the third case in the Appendix in~\citealt{Haas08}). The variations here are generally within 0.06 dex. Hence, combining all these effects, a more realistic uncertainty for the slopes of the LF that we have obtained would be about 0.1 dex.

\subsection{Distance dependence of the photometric properties}
\label{sec:distance_effect}

\begin{figure*}
\vspace{-2cm}
   \includegraphics[angle=90,trim = 0cm 0.5cm 1cm -1.5cm,clip=true,width=0.95\textwidth]{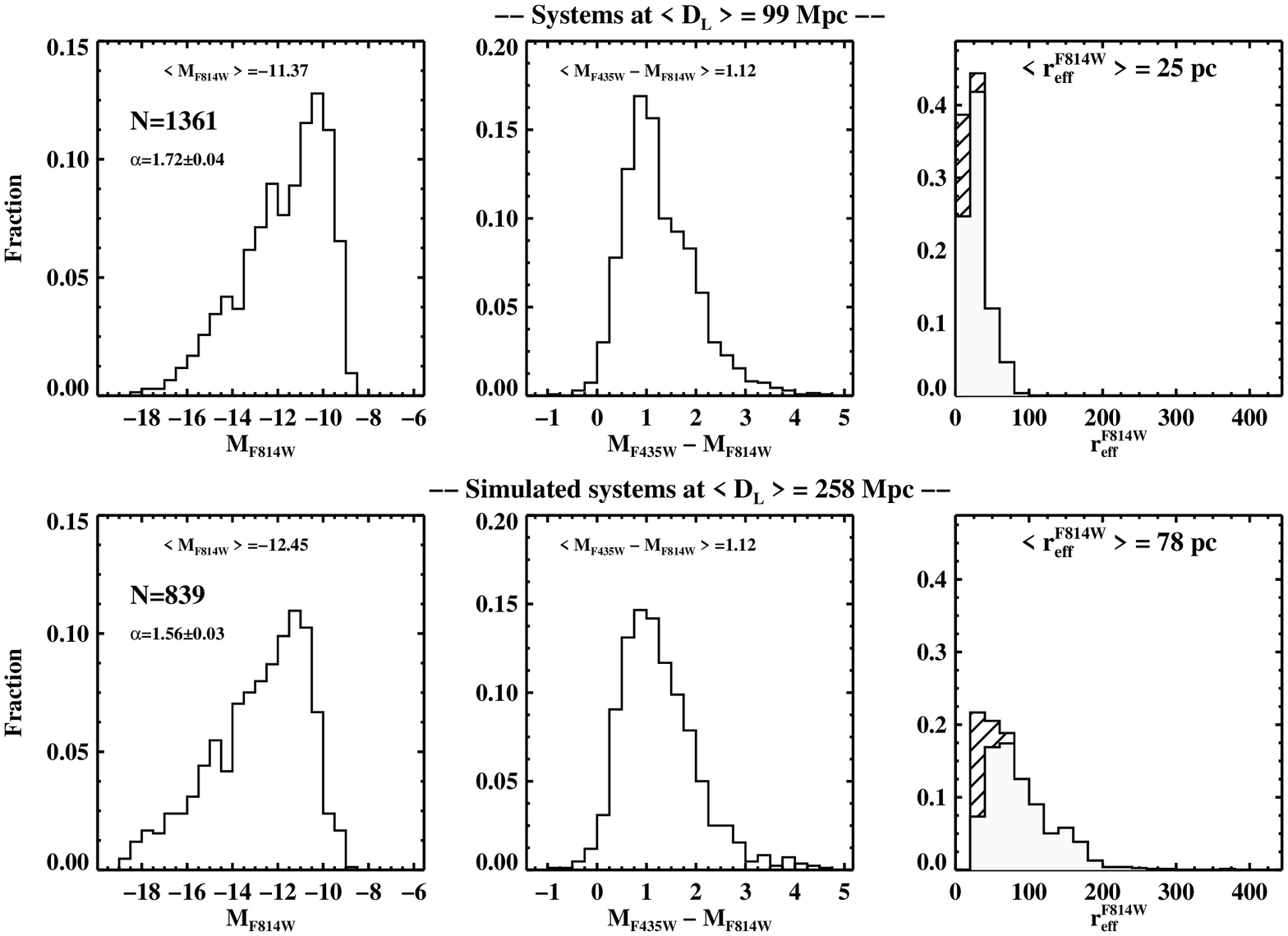}
   \caption{Photometric properties for real (top) and simulated (bottom) data. The median values (between brackets $<~>$ throughout the paper) of the magnitude, color and \reff distributions are shown. The slope of the LF for the magnitude distributions is also presented ($\alpha$). Throughout the paper N represents the number of detected knots in any given distribution.\label{fig:simulation}}
              
\end{figure*}

In the analysis we also consider a selection effect that we can encounter when studying the photometric properties of the knots in (U)LIRGs located at different distances. The classical Malmquist bias is caused by the fact that systematically brighter objects are observed as distance (and volume) increases, as a result of a combination of the selection and the intrinsic scatter of absolute magnitudes. 

Our sample is affected by this bias, since the systems are located at different distances (from 65 to 560 Mpc). As the system stand further, the faintest knots become undetected and we can only see the bright ones. Besides, for the same angular resolution the linear resolution decreases linearly with distance, and therefore the knots that we detect at large distances can be associations of knots instead of individual knots as those observed in less distant systems. Thus, larger sizes are computed. Though this is not a complete sample, Figure~\ref{fig:LirVsZ} shows that in general more distant \mbox{galaxies} have higher infrared luminosity. The same tendency is seen if the RBGS~\citep{Sanders03} are taken. This effect has to be subtracted in order to observe reliable intrinsic differences between systems at different distances.

We have divided the sample in three \lir intervals, to study the dependence of the photometric properties with the total infrared luminosity of the system (see section~\ref{sub:lir}). To asses how the photometric properties depend on the distance of the system a simulation has been performed. The median distance of the low luminosity interval (\lir$<$ 11.65) corresponds to 99 Mpc, whereas that of ULIRGs (the high luminosity interval) is 258 Mpc. Hence, for this sample ULIRGs are located a factor of 2.6 further away than the low luminous LIRGs. We have simulated the galaxies within the first \lir interval as if they were a factor of 2.6 further, obtaining the same median distance as in ULIRGs. To that end, the \hst images have been convolved with a Gaussian at the resolution of the PSF at the new distance of each galaxy and re-binned in order to have the same ''pixel size'' at that distance. Using foreground stars we have computed for each image the factor needed in order to preserve the flux, since the flux of the point-like objects must be preserved. Then photometry has been carried out as in the original images, with circular and in some cases polygonal apertures. 

The results of the simulation for the \filteri filter are presented in Figure~\ref{fig:simulation}. The magnitudes and the sizes increase \mbox{by about 1.1} mag (a factor of 2.8 in flux) and by a factor of 3, respectively. The number of detected objects drops by about a factor of 1.6. Therefore, in the simulated image, individual knots are associations of knots from the original one. As a result, not only do the sizes become larger, but its distribution also flattens. The median values of \reff distribution in the simulation are 25 and 78 pc for the real and simulated knots, respectively. 

The flattening of the LF (from $\alpha$ = 1.72 to 1.56) is also expected, since as the knots are artificially grouped due to angular resolution effects the bright tail of the LF starts to get more populated. Another expected result of the simulation is that the color distribution does not change significantly. 

If we consider the spatial distribution of the knots (inner and outer knots, with projected distances of less and more than 2.5 kpc to the nucleus, respectively), some changes are also observed; the ratio of inner to outer knots shifts from 0.7 to 0.5. Due to the higher surface brightness and steeper background profile in the inner regions, when convolving the images to get lower linear resolution, a large fraction of inner knots are diluted by the local background, in contrast to a lower dilution degree in the outermost regions. Besides, more grouping is expected in the innermost regions, where there is more crowding, hence also losing also a larger number of inner knots when compared with those located in the outer regions. 

\begin{figure}[!t]
   \includegraphics[width=1.05\columnwidth]{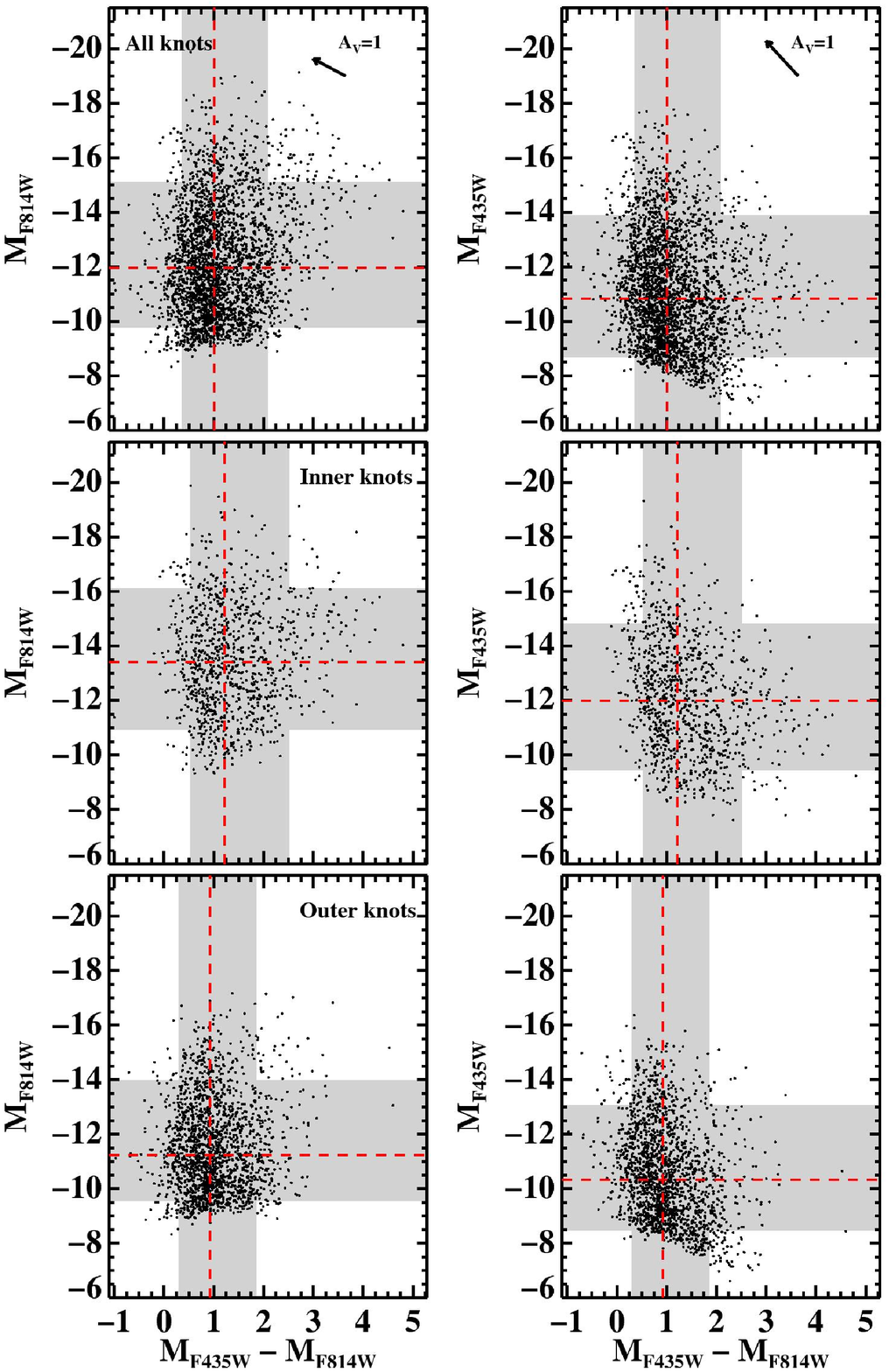}
   \caption{Color-magnitude diagrams for the identified knots. Top: all knots. The red dashed line indicates the median of the distribution and the gray band covers 80 \% of the total number of knots in each plot, rejecting the 10 \% at either side of the distribution. No reddening corrections have been applied to the magnitudes given. A de-reddening vector of A$_{V}$=1 mag is shown in the top right corner. Middle: the same diagrams for the knots in this study within 2.5 kpc. Bottom: same as before but for knots further than 2.5 kpc from the closest nucleus.\label{fig:mag_col}}
              
 \end{figure}

\section{Results and discussion}
\label{sec:results}
\subsection{General properties of the knots}
\label{sub:general_properties}

Owing to the high angular resolution and sensitivity of the ACS camera, and to the definition of our large sample, covering a wider \lir  range than previous studies focused only on ULIRGs ~\citep{Surace98,Surace00}, we have detected close to 3000 knots. This is more than a factor of ten larger than in previous investigations and allows us to carry out a statistical study exploring the different physical properties of the knots over the entire LIRG and ULIRG luminosity range.  We have then performed the photometry in both \filterb and \filteri filters for these knots. For this study we consider the knots with positive detection in both filters. The objects identified as the nuclei are not studied here, since many of them are believed to be contaminated by an AGN (see table~\ref{table:sample} and the references regarding the spectral class).

\subsubsection{Magnitudes and colors. Average values}

The knots detected in our sample of luminous infrared galaxies have observed \filteri and \filterb absolute magnitudes (uncorrected for internal extinction) in the  \mbox{-20 $\lesssim$ \mi $\lesssim$ -9} and \mbox{-19.5 $\lesssim$ \mb $\lesssim$ -7} range, respectively, while colors cover the -1 $\lesssim$ \mbi $\lesssim$ 5 range (Figure~\ref{fig:mag_col}).  The corresponding median magnitudes are \mbox{$<$ \mi $>$=-11.96} and \mbox{$<$ \mb $>$=-10.84}. The median color of the knots  corresponds to \mbox{$<$ \mbi $>$ = 1.0}, which clearly indicates the presence of a young stellar population (about half of the knots have colors bluer than that value). By contrast, the mean color of the diffuse local background light, which traces the old stellar population in the host galaxies, is 1.72, consistent with the typical colors measured in spirals (\mbox{B-I=1.80};~\citealt{Lu93}). 

A small fraction of the knots (2\% of the total) show extreme red colors (\mbi\twospace=3 mag or redder), likely tracing regions with very high internal obscuration, usually located in the more central regions ~\citep{Alonso-Herrero06,Garcia-Marin09b}. If the stellar population in these knots was dominated by young stars (i.e., t$\sim$10 Myr), the extinction could be as high as \mbox{A$_V$=5-7} mag. Heavily obscured young clusters with \av\onespace=5-10 mag have also been detected in the optical in less luminous interacting systems like the Antennae (e.g., clusters S1\_1 and 2000\_1, whose extinction  was computed with the aid of infrared spectroscopy;~\citealt{Mengel08}). 

\begin{figure*}
   \centering
\vspace{-3cm}
   \includegraphics[width=\textwidth]{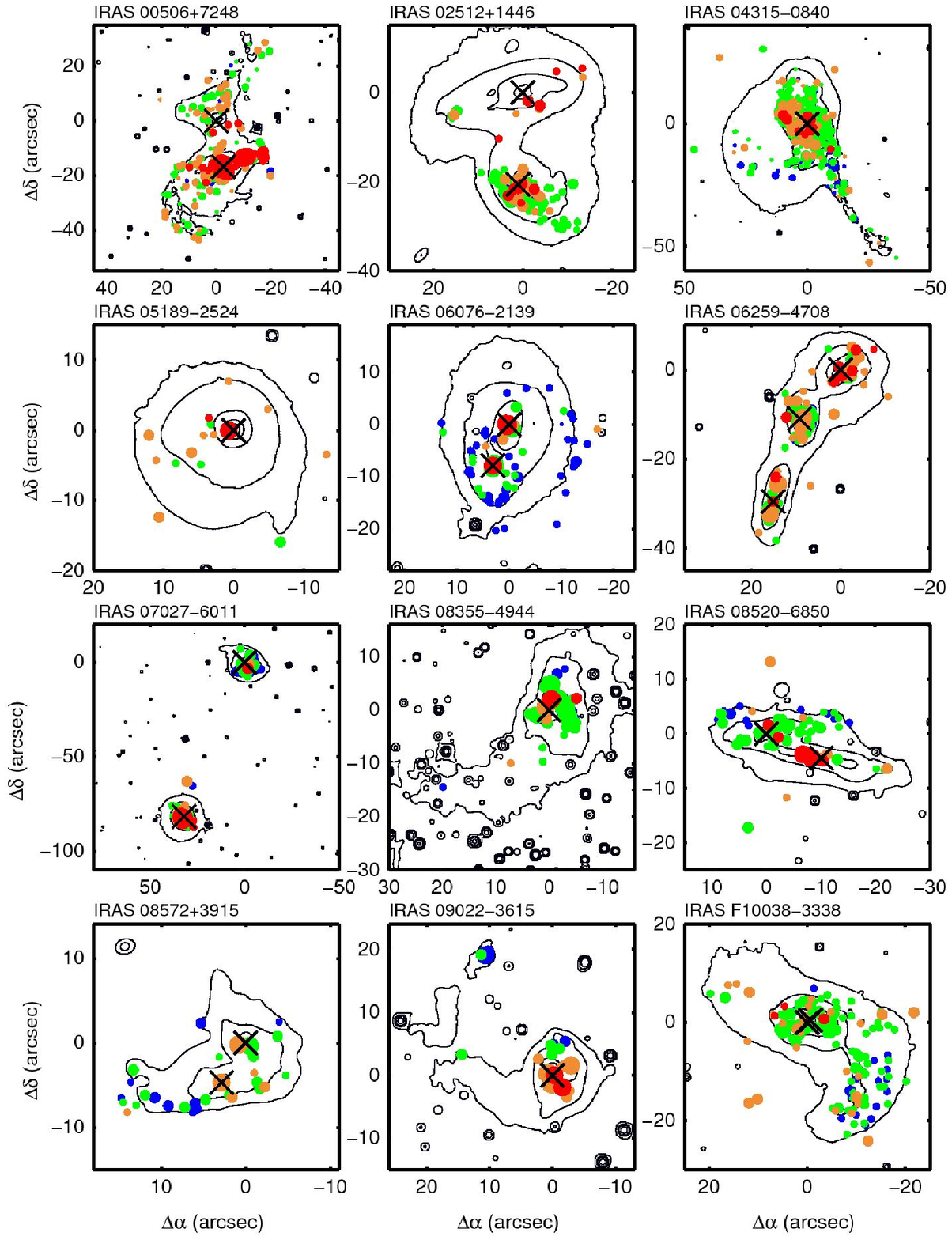}
\vspace{-1cm}
   \caption{Spatial distribution of the clusters measured on the isophotal map of each system. Each \filteri image has been smoothed to diminish pixel-to-pixel noise and then avoid spurious contours. The field of view is the same as in Figure~\ref{fig:composite34}. The symbol 'X' marks the knot identified as the nucleus . From bluer to redder colors, knots are blue, green, orange or red, and the sizes plotted depend on their \mi~(see the legend at the end of the figure). North points up and East to the left.\label{fig:spatial_dist}}
              
    \end{figure*}

  \begin{figure*}
   \centering
   \vspace{-3cm}
   \includegraphics[width=\textwidth]{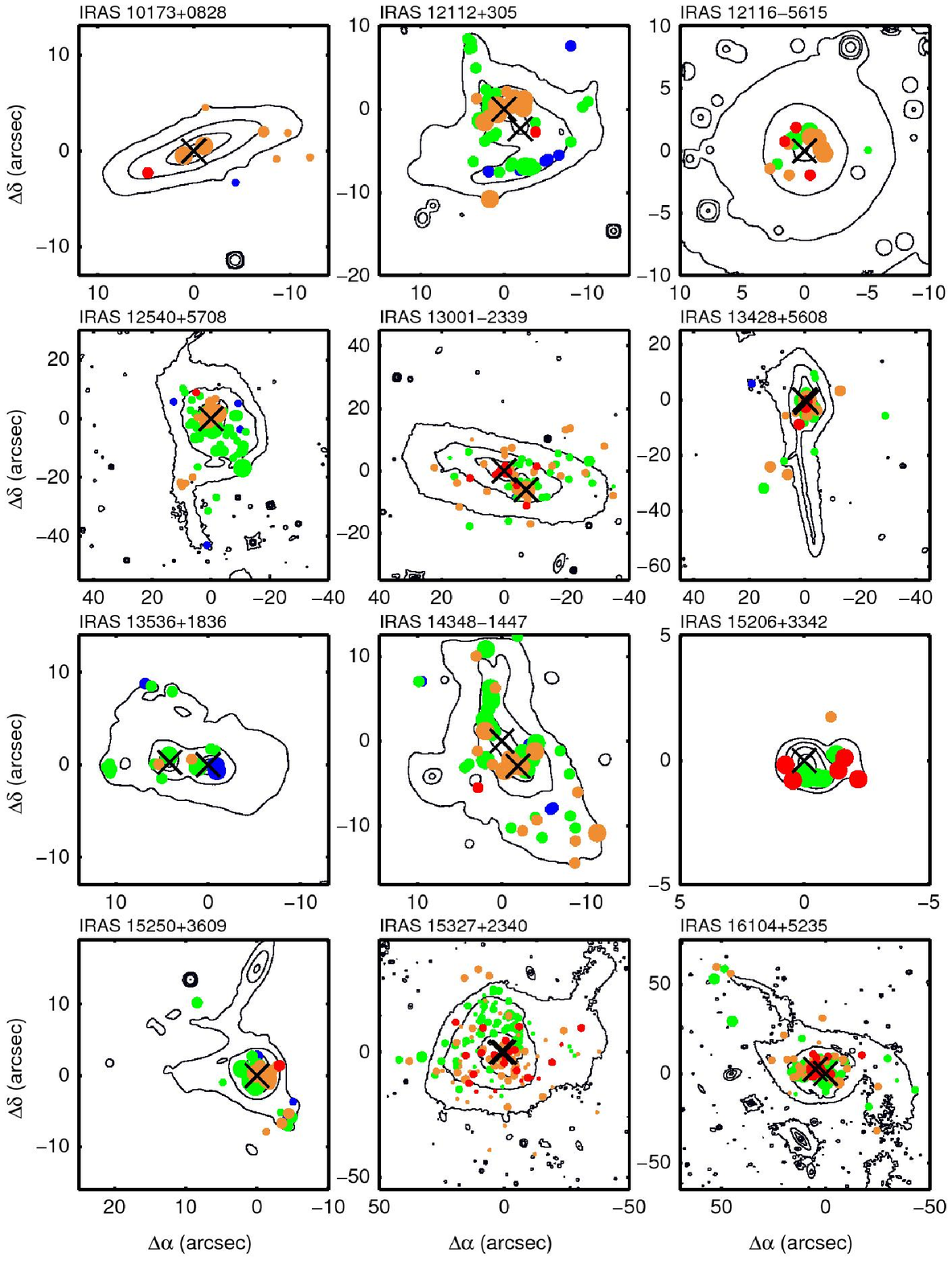}
    \vspace{-1cm}
\addtocounter{figure}{-1}   
   \caption{- Continued}
    \end{figure*}
  \begin{figure*}
   \centering
   \vspace{-2cm}
   \includegraphics[trim = 0cm 7cm 0cm 4cm,clip=true,width=\textwidth]{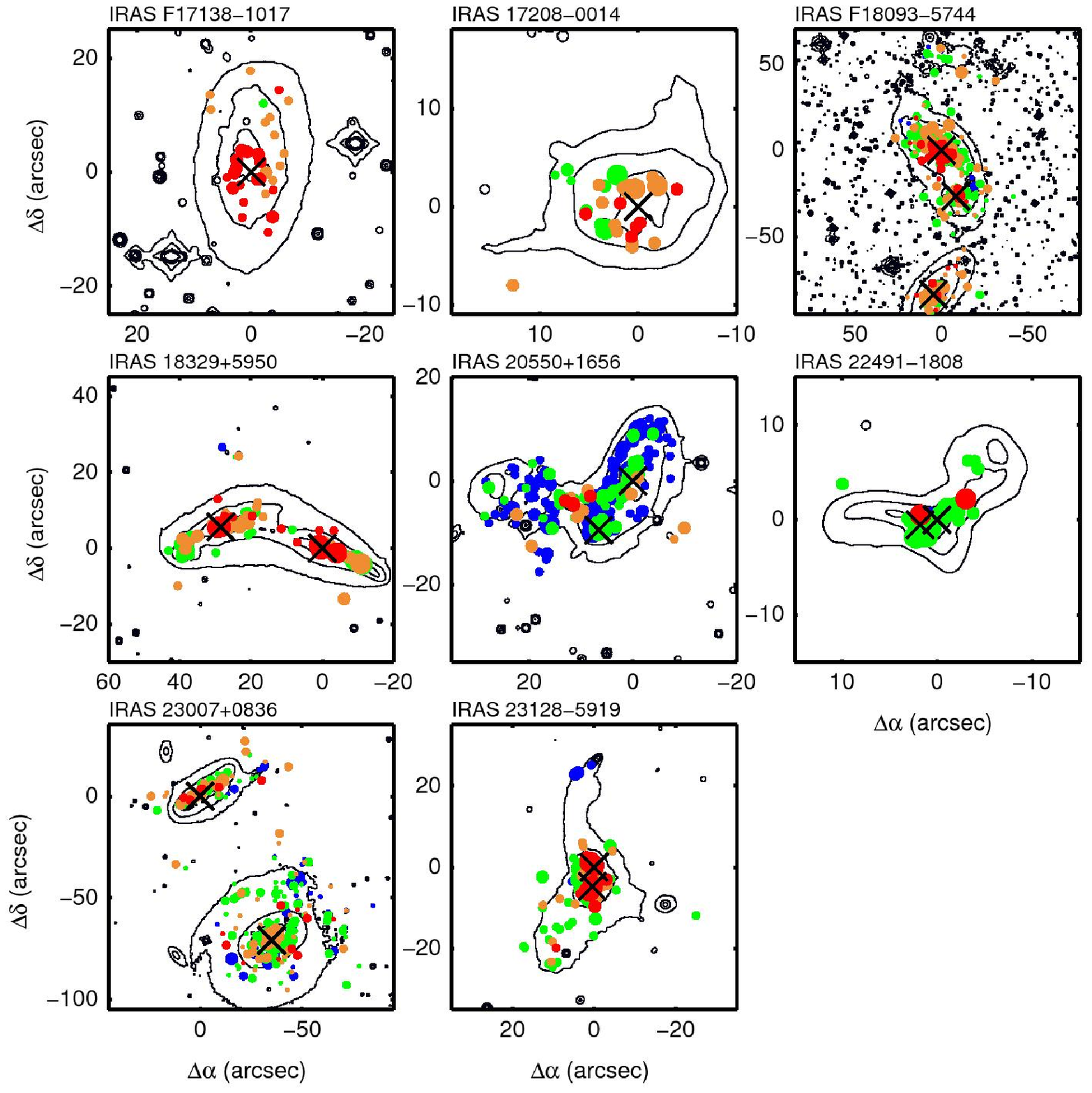}
\addtocounter{figure}{-1}   
   \caption{- Continued}
\vspace{-6.5cm}
  \hspace{8.5cm}
   \includegraphics[angle=90,width=0.35\textwidth]{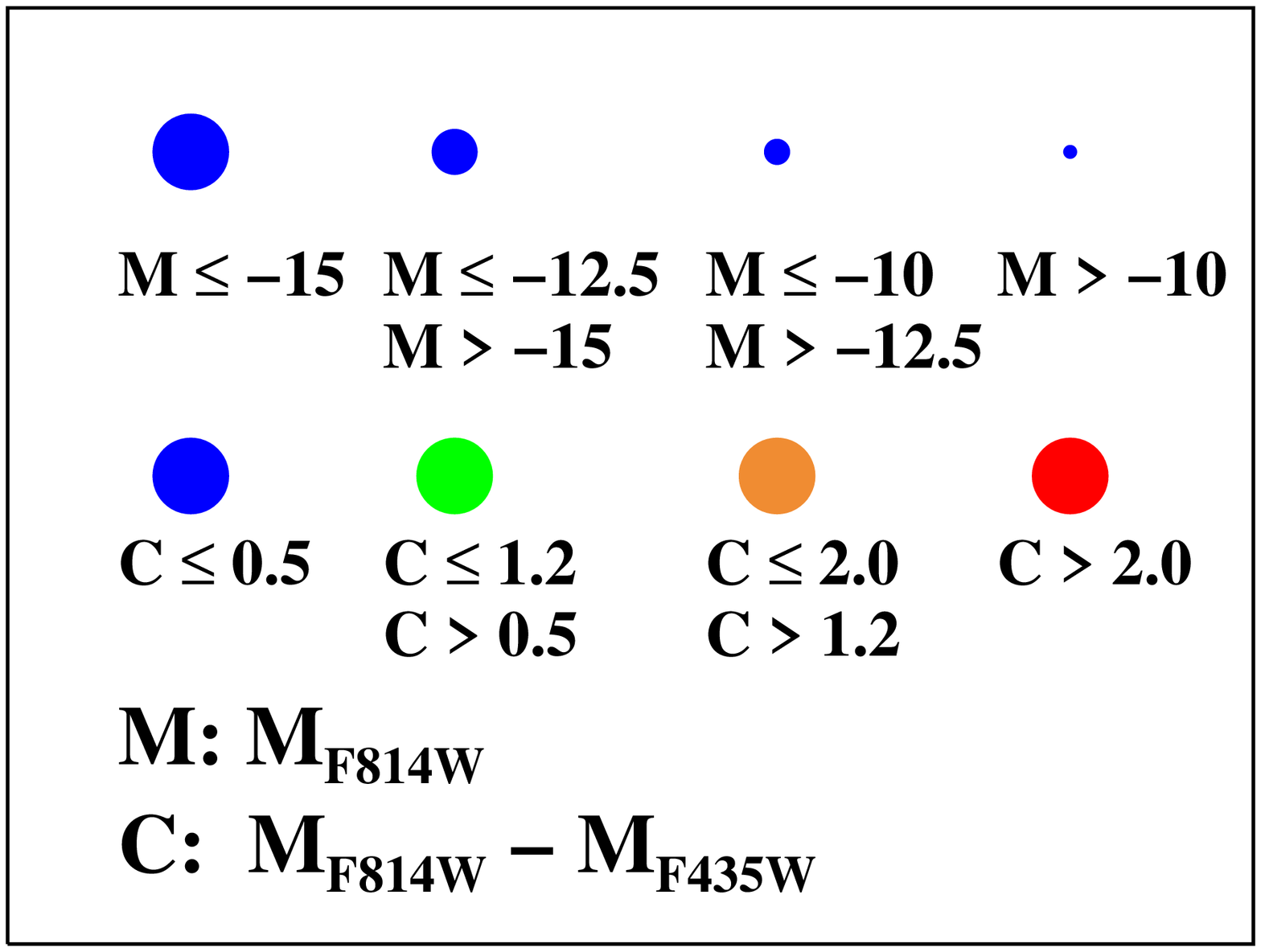}
\end{figure*}

\subsubsection{Magnitudes and colors. Radial distribution} 
\label{sub:mag_col_rad}
In order to understand the star formation in luminous infrared galaxies it is important to investigate whether the observed properties of the knots (magnitudes, colors and sizes) do show dependence on the galactocentric distance. The spatial distribution of the knots is shown in 
Figure~\ref{fig:spatial_dist}, where their sizes are related to their absolute magnitude and their color to the photometric color \mbox{\mbi\twospace}. About 1/3 of the knots lie within a projected galactocentric radius of 2.5 kpc while this fraction increases up to about 2/3 within a radius of 5 kpc. Thus, a substantial fraction of the knots are located close to the nucleus at distances of less than 5 kpc. Also, as can be seen in the figure, in general they represent the population with redder colors, likely indicative of higher internal extinction, although an older population can not be excluded. Yet, there is also an external population of knots (10.8\% of the total) at distances of more than 10 kpc and up to 40 kpc, located along the tidal tails and at the tip of these tails. Most of these external knots are blue (\mbox{\mbi$<$1}), and some (5.7\%) are very luminous (\mbox{\mb$<$-12.5}), suggesting the presence of young massive (M$>$10$^5$\msun in young stars) objects in the outer parts of these systems. Examples of these knots can be seen at the \mbox{northern} tip in IRAS 23128-5919 and in IRAS 09022-3675, along the northern tail in IRAS 14348-1447, etc. 

Throughout the paper, we identify the inner sample of knots as those with projected distances of less than 2.5 kpc to the closest nucleus, and the outer sample as those with distances of at least 2.5 kpc. Both distributions have a significantly different magnitude and color distribution (see Figure~\ref{fig:mag_col}), the inner knots being about 2 mag brighter in both filters and 0.3 mag redder than the outer knots. These two distributions are different, and do not come from the same parent distribution according to the Kolmogorov-Smirnov test (KS, using the idl routine KSTWO;~\citealt{Press92}). The inner knots also have a larger color dispersion with a tail toward red colors. This is likely the effect of having higher and patchier internal extinction in the innermost regions, in agreement with recent spectroscopic studies of these systems~\citep{Alonso-Herrero06,Garcia-Marin09b}. Additionally, the brighter blue magnitudes in the inner regions suggest that there must be more young star formation there than in the outer field.

\subsubsection{Effective radius of the knots}
\label{sub:sizes_general}

As mentioned in section~\ref{sec:sizes}, the angular resolution limit (and therefore upper sizes for unresolved knots) varies from around 10 to 40 pc, depending on the distance to the galaxy. The detected knots do show a wide range of sizes ranging from unresolved (12\%) to a few with very extended sizes (a radius of 200 pc, and up to 400 pc). The median size of the resolved knots is 32 pc, with about 65\% of them smaller than 40 pc. This is significantly larger than the largest knots found in other systems, and several times larger than their mean knot radii (e.g., 10 pc in~\citealt{Whitmore99}). Given the distances of the galaxies in the sample (from about 65 Myr to more than 500 Myr) and the resolution limit, in general  the knots identified in our galaxies are likely aggregates of individual clusters, which would increase their (apparent) size. According to the simulations (see section 4.5), the apparent size of the knots changes from an average of 25 pc to 78 pc when galaxies at an average distance of 100 Mpc are moved at distances 2.6 times further away.

\begin{figure}[!t]
\vspace{-4cm}
   \includegraphics[trim = -3cm -1cm 0cm 0cm,clip=true,width=0.95\columnwidth]{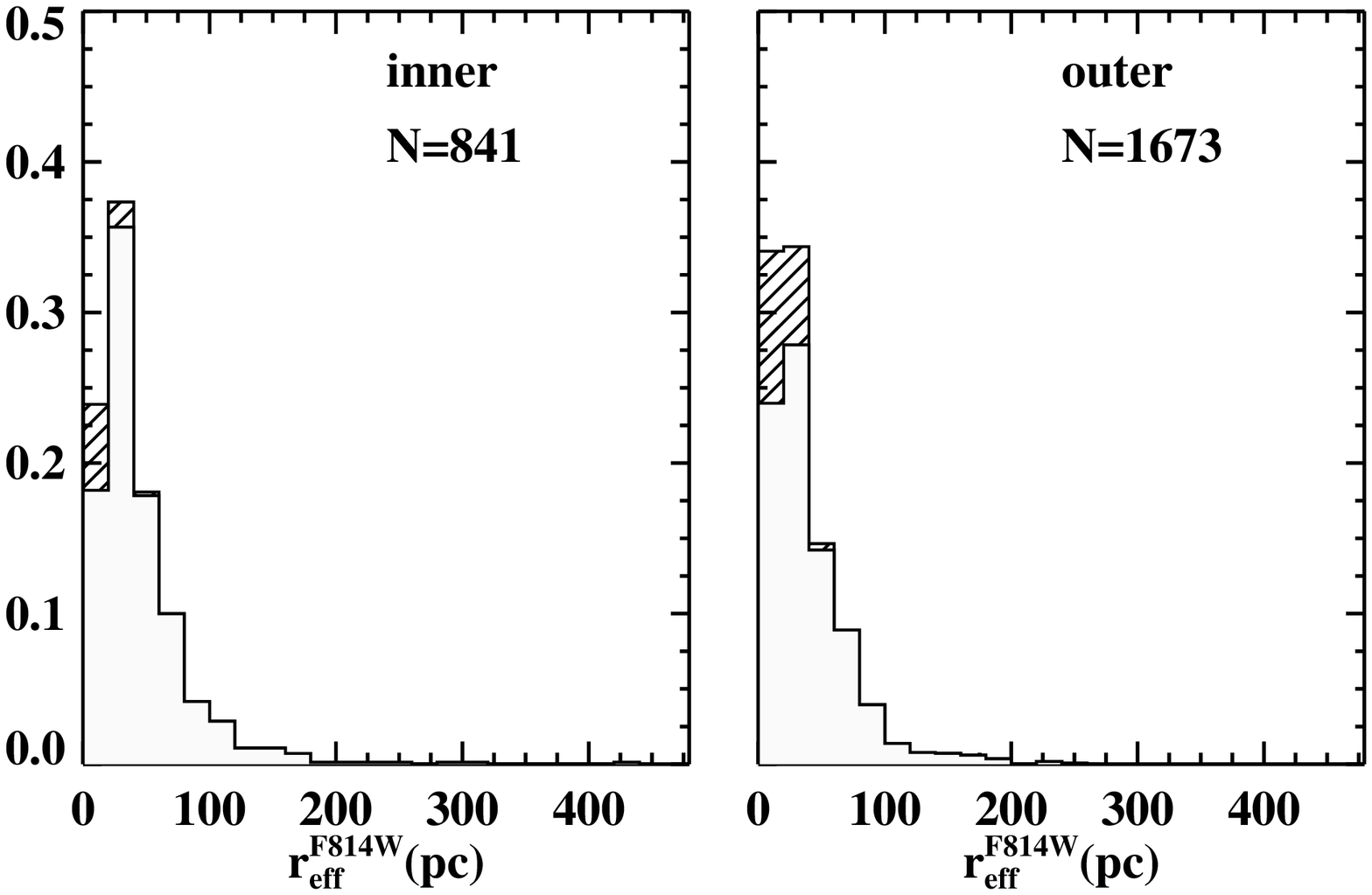}
   \caption{Effective radius for inner and outer knots. Throughout the rest of the paper the resolved sizes are grouped into blank bins. The knots with unresolved sizes are grouped into hatched bins, above the blank bins. \label{fig:reff1}}
              
 \end{figure}

If we consider the sample of inner and outer knots, their distributions are similar (Figure~\ref{fig:reff1}), so there is no evidence of a radial dependence of the sizes at the angular resolution of our data.

\subsubsection{The bluest knots. Age and mass estimates}
\label{sub:ages_masses}

The determination of ages and masses of stellar po\-pu\-la\-tions is highly degenerate when using a single color index. In addition, uncertainty (or lack of
knowledge) in the internal extinction in heavily obscured systems like the luminous infrared galaxies considered here, adds more degeneracy. However, a first-order estimate of the age and mass can be obtained for a specific range in magnitude and colors where degeneracy can be minimized. In order to perform this kind of analysis, two independent synthetic stellar population mo\-dels are considered in this study: the evolutionary synthesis code starburst99 (hereafter SB99) v5.1~\citep{Leitherer99,Vazquez05} and the code by~\cite{Maraston05}. The former is optimized for young population and also models the ionized gas whereas the latter gives a rigorous treatment of the thermally pulsing asymptotic giant branch (TP-AGB) phase, which is re\-le\-vant within the interval 0.1-1 Gyr. In both cases we assume instantaneous burst models with a Kroupa Initial Mass Function~\citep{Kroupa02} over the range 0.1-120 \msun and solar metallicity. These models are normalized to 10$^6$ \msun\onespace. Mass estimates must be multiplied by a factor of 1.56 if the Salpeter IMF is considered.

\begin{figure}[!t]
\hspace{0.3cm}
   \includegraphics[angle=90,width=0.95\columnwidth]{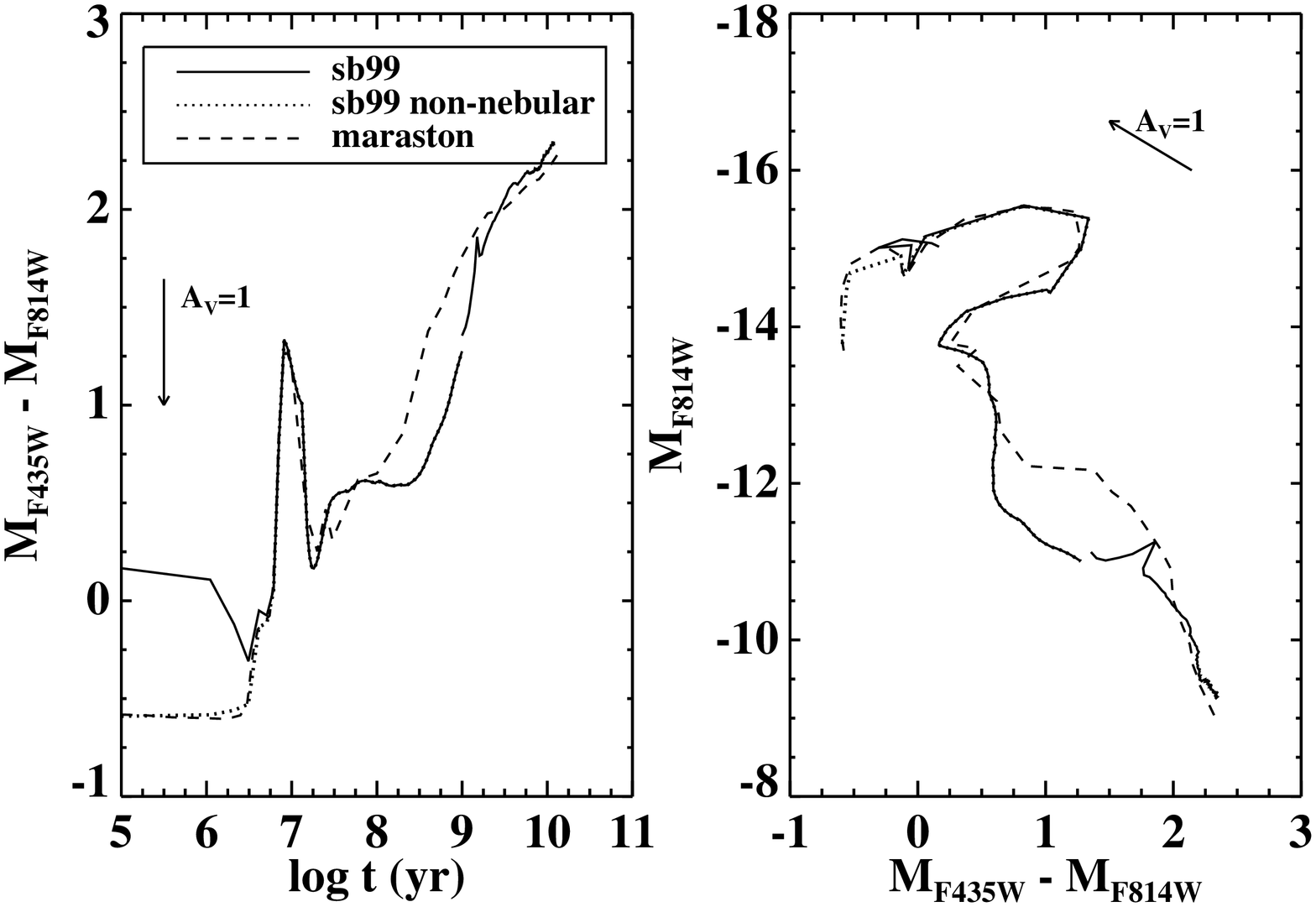}
    \vspace{-0.3cm}
   \caption{Color as a function of age (left) and color-magnitude track (right) according to the Stellar Synthesis Population SB99 models with and without the nebular continuum (solid and pointed line, respectively) and to the Maraston05 models (dashed line). The curves are normalized to 10$^{6}$ \msun\twospace. From \mbox{logt $\sim$ 6.8} on both SB99 curves are identical, since the inclusion of the nebular continuum is only relevant at younger ages. The de-reddening vector (A$_{V}$=1 mag) is shown in both plots. For the whole article it is computed using the curve of~\cite{Calzetti00}. \label{fig:spp_models}}
              
	\vspace{-0.3cm}
 \end{figure}

The evolutionary tracks using both SB99 and Maraston SEDs are presented in Figure~\ref{fig:spp_models}. The shift in color at 1 Gyr in the SB99 models is the result of using Geneva tracks (optimized for young ages) at ages younger than that and Padova tracks at older ages, since incompleteness sets in for Geneva models at \mbox{t $>$ 1 Gyr}~\citep{Vazquez05}. At very young ages (\mbox{log t $\lesssim$ 6.5}) the nebular continuum, which is considered in SB99, reddens the colors by up to 0.6 mag. From \mbox{t $\gtrsim$ }50 Myr (log t = 7.7) up to 1 Gyr the differences between both models also become significant. However, regardless of the model used, the de\-ge\-ne\-ra\-cy in color within the range \mbox{\mbi=[0.6,1.4]} is considerable for two rea\-sons: stellar populations with ages within the interval [50-500] Myr have the same \mbi color in the range 0.6-0.8; the Red Super Giant population reddens the tracks at \mbox{ages t$\sim$ 7-14 Myr}, when this population dominates the lu\-mi\-no\-si\-ty, hence making it indistinguishable from a population of up to \mbox{t$\sim$1 Gyr}. Considering this, there is
a clear age degeneracy for the observed color interval \mbox{\mbi=[0.6-1.4]} that translates into an uncertainty of about a factor of 100 in the mass estimates. Hence, although the ignorance of the extinction contributes as well, the age degeneracy represents the primary source of uncertainty.

However, as already mentioned in previous sections, a non negligible fraction (15\%) of all the knots in this study are blue, implying young stars and some reddening. In particular, knots with colors \mbi $\lesssim$ 0.5 (hereafter blue knots) must be young and should not be very much affected by degeneracy and  extinction, according to the stellar population models considered in this study. They are bright (\mbox{$<$\mb\onespace$>$=-11.5}) and in general found in the outer regions of the systems, with some exceptions such as IRAS 13536+1836 and IRAS 20550+1656 where they are also detected very close to the nuclei (see Figure~\ref{fig:spatial_dist}). Some first-order estimation of the age and mass can be achieved for the blue knots with colors \mbox{\mbi $\lesssim$ 0.5}. In that color interval the mass uncertainty reduces considerably down to a factor of 2-4 if a single stellar population is considered. That color can only indicate young population (if \mbox{\mbi$ \lesssim$0.5}, then \mbox{t$\lesssim$ 30} Myr), hence the SB99 tracks will be used for this estimation. In case of having some degeneracy (contribution from the RSG branch), the average age is taken, in order to minimize the uncertainty of the estimation.

Note that the embedded phase of the young population in any environment can last few Myr and that (U)LIRGs are known to have young population highly enshrouded in dust. Previous works have measured very high extinction values for clusters younger than about 3 Myr in different environments~\citep{Larsen10}. They have measured extinction values as low as \av= 0.5 mag for some clusters older than 3 Myr. Finally, it is thought that the embedded phase does not last more than 5 Myr, since they have observed clusters older than that age which are already extinction-free. Hence, the knots for which we are trying to make age and mass estimates, though young, are likely to be older than 3 Myr, probably even older than 5 Myr.

\begin{figure}[!t]
\includegraphics[angle=90,width=0.95\columnwidth]{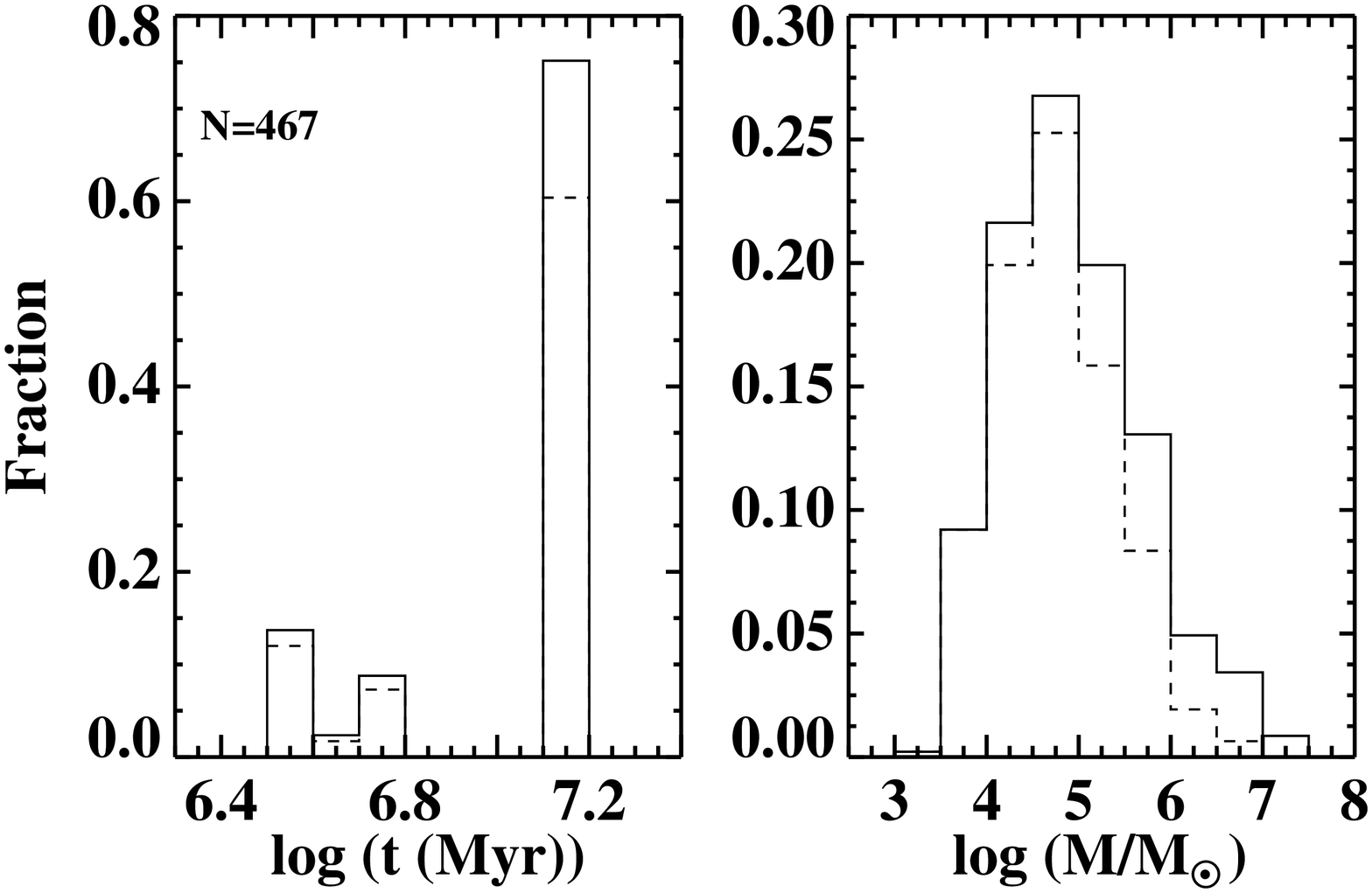}
\caption{Ages and masses estimated for the knots with color \mbi $\lesssim$ 0.5, where the degeneracy in the evolutionary tracks is less severe. The dashed line histograms refer to the external knots (projected distance $>$ 2.5 kpc). \label{fig:ages_mass}}

\vspace{-0.5cm}
\end{figure}

Under these assumptions and considering all caveats above, an estimate of the age and mass of the blue knots has been performed, assuming a single stellar population (Figure~\ref{fig:ages_mass}). If small extinction is present in these regions, the estimated mass would be increased by a small amount (up to a factor of 1.8 for A$_V$=1 mag). Some blue knots appear to be more massive than the most massive old globular clusters in the Milky Way, which have masses up to 10$^6$~\msun \citep{Harris01}. However, most of the blue knots have masses in the 10$^4$ to few 10$^6$ \msun range, similar to those of giant \hii regions and brightest Young Massive Clusters in other less luminous interacting galaxies (e.g., ~\citealt{Bastian05b};~\citealt{Whitmore99};~\citealt{Mengel08};~\citealt{Konstantopoulos09}). Complexes of less massive star clusters can also constitute these blue knots.

It is interesting to mention that the blue knots in the outermost regions (i.e., d $>$ 2.5 kpc) represent the 80\% of the total number, while most of the star formation is likely to be occurring within the central regions of the galaxies, as mentioned in section~\ref{sub:mag_col_rad}. This suggests that in general in the outermost regions the dust extinction is very low in (U)LIRGs. Hence, the most obscured knots must be located close to the nuclei, as mentioned in the previous section and in agreement with recent spectroscopic studies of these systems~\citep{Alonso-Herrero06,Garcia-Marin09b}. Part of the difference between the number of blue knots in the inner and outer regions may also be caused by a stronger disruption process in the central regions, as claimed by~\cite{Haas08}, as well as by different histories of star formation of the knots in both fields.

The number of blue knots with estimated masses higher than 10$^6$ \msun ($\sim$ 10\%) drops rapidly in the external areas (the dashed line histogram in Figure~\ref{fig:ages_mass}). Hence, the most massive young knots are also located close to the nuclei.

\subsubsection{Mass-radius relation of the bluest knots}
\label{sub:mass_rad}

\begin{figure}[!t]
   \includegraphics[angle=90,width=1\columnwidth]{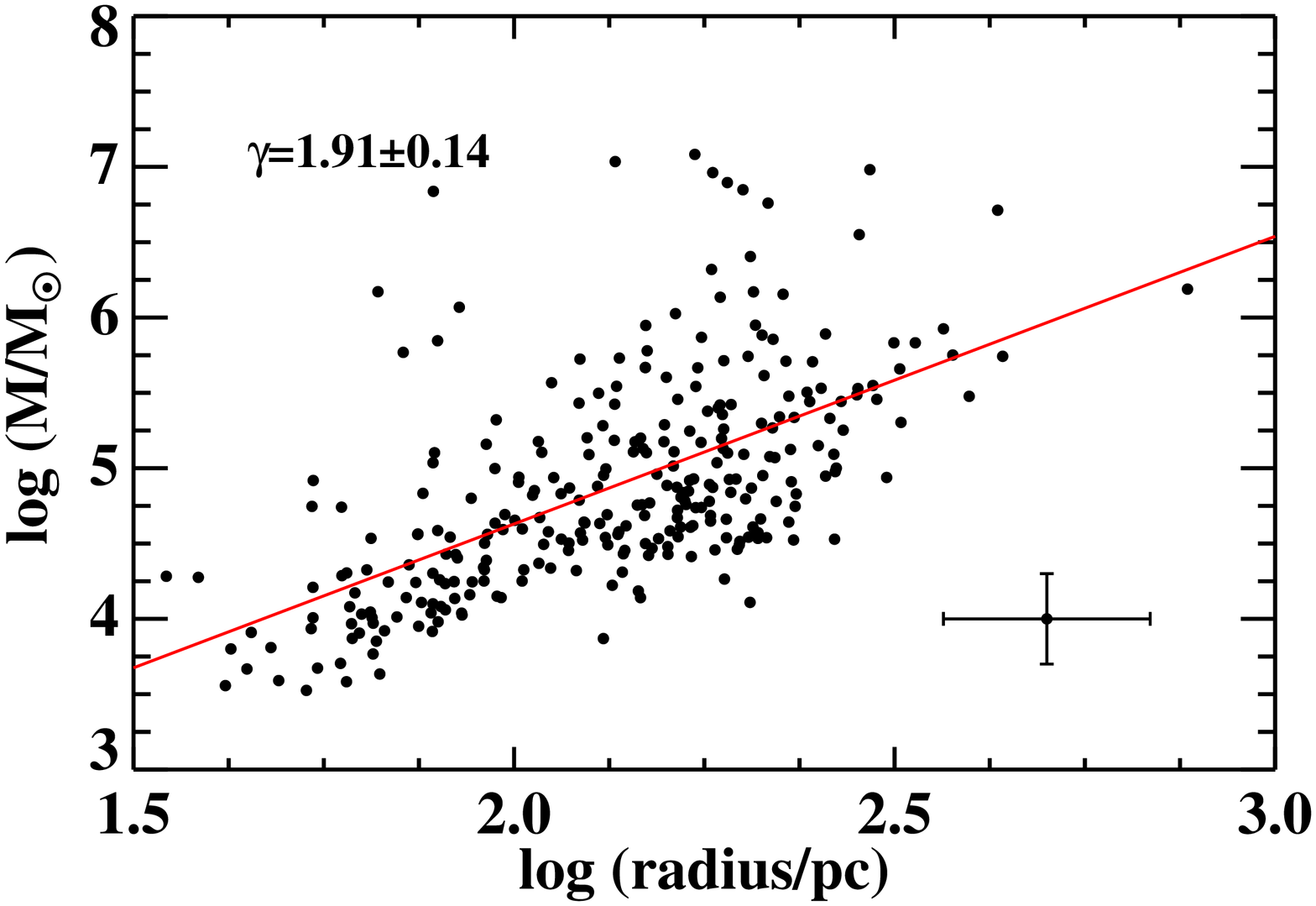}
   \caption{The mass vs. radius relation for knots in (U)LIRGs. The solid red line is a power-law fit to the data, with index $\gamma$. The typical uncertainty associated with the data is shown in the right bottom corner. \label{fig:size_mass_all}}
              
 \end{figure}

We know that in general the knots represent associations of star clusters created in clumps. A single giant molecular cloud (GMC) does not usually produce a single star cluster, but a complex of star clusters (\citealt{Elmegreen83};~\citealt{Bastian05a}). Thus, it is interesting to see how these complexes fit into the hierarchy of star formation by searching for a relation between the mass and the radius of the knots. This can be done for the bluest knots, for which an estimation of the mass with an uncertainty of less than a factor of 4 was achieved (section~\ref{sub:ages_masses}). Their radius was derived as explained in section~\ref{sub:tot_size}. We could obtain reliable measurements of both values for a total of 294 knots, which is about a 10\% of the total number of knots or about 2/3 of the total number of blue knots.

The surface brightness profile of the bluest knots is well fitted by a power-law with index $\beta$=0.37$\pm$0.33, which corresponds to (assuming sphericity) a three dimensional density profile of $\rho \propto r^{\beta}$, with $\beta$=1.37. This power-law profile is similar to that observed for GMCs associated with high-mass star-forming regions ($\beta$=1.6$\pm$0.3;~\citealt{Pirogov09}). The similarity between the profiles in GMCs and the subsample of bluest knots suggests that the amount of luminous material formed is proportional to the gas density. Obviously, measurements of the gas in molecular clouds in (U)LIRGs are needed to prove this.

Although with high dispersion, the bluest knots follow a mass-radius relation (see Figure~\ref{fig:size_mass_all}). We have fitted a function of the form \mbox{M $\propto$ R$^{\gamma}$}, with index $\gamma$=1.91$\pm$0.14. The size and the mass of GMCs in the Milky Way are also related, \mbox{M$_{cloud} \propto$ R$_{cloud}^{2}$}, as a consequence of virial equilibrium (\citealt{Solomon87}). This relation has also been found for extragalactic GMCs (e.g.,~\citealt{Ashman01};~\citealt{Bastian05a}). Our result is therefore consistent with the bluest knots having the same mass-radius relation as GMCs in general, contrary to what is measured for individual young star clusters (e.g.,~\citealt{Bastian05b}). An offset between both relations can be used to estimate the star formation efficiency within the cloud, as done in \cite{Bastian05a}. Data from the ALMA observatory, which will be able to reach sub-arcsecond resolution, would allow this estimation for (U)LIRG systems.

\subsubsection{Luminosity Function of the knots in the closest systems}
\label{sub:LF}

The large number of detected knots allows us to study their luminosity function and compare it with that for other less luminous and/or interacting galaxies. However, this comparison should be restricted to the closest (U)LIRGs, since given the sizes measured complexes of star clusters are detected only for galaxies located at distances larger than 100 Mpc with the resolution of the ACS. Although many knots detected in our closest systems may still consist of complexes of star clusters, in order to minimize distance effects (see section~\ref{sec:distance_effect}) we have considered here a subsample that comprises the seven systems closer to this distance (see specific distances in table~\ref{table:sample}).

\begin{figure}[!t]
    \centering
   \includegraphics[angle=90,width=0.85\columnwidth]{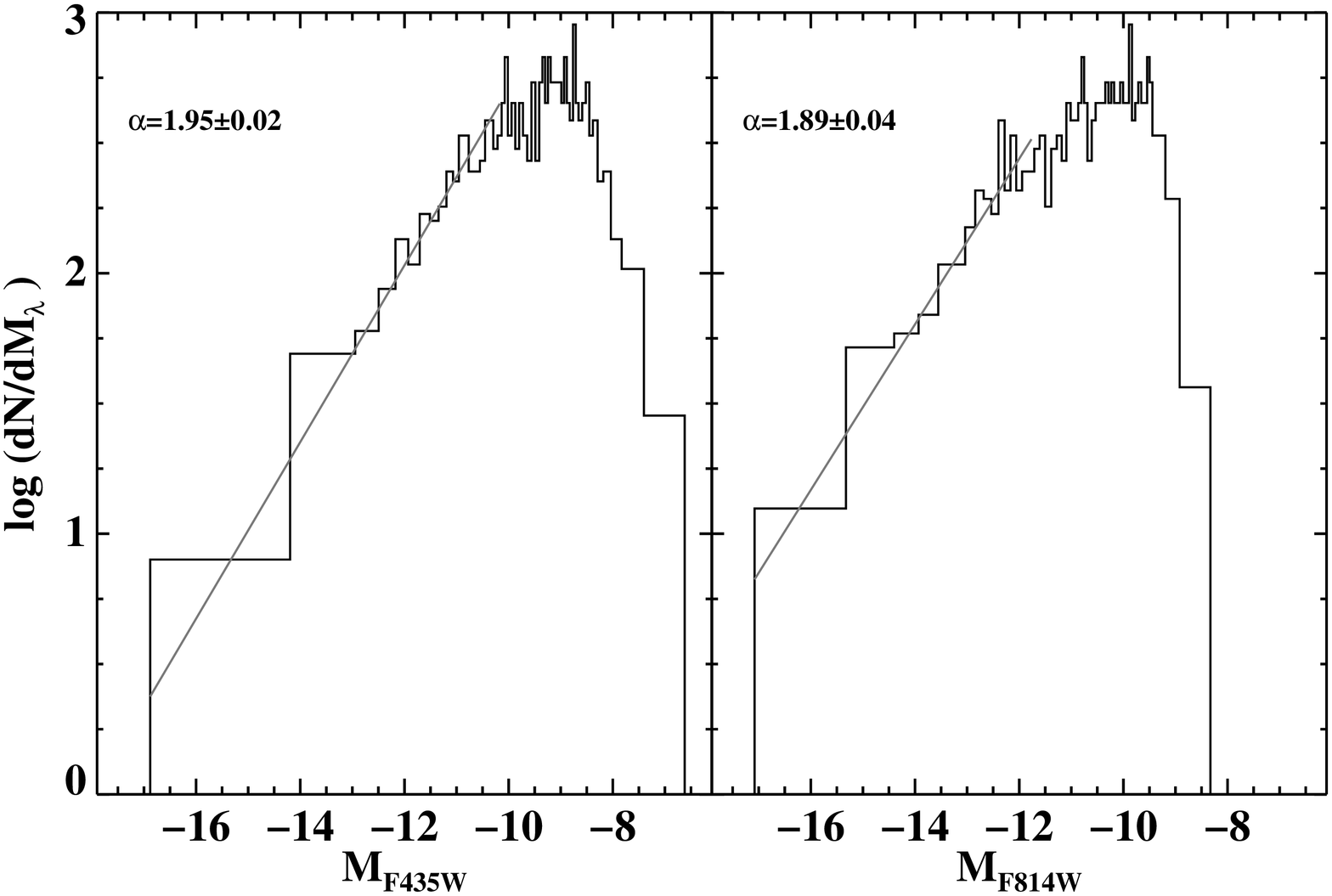}
   \caption{Luminosity functions for the knots identified in both the red and the blue filters for the subsample of (U)LIRGs closer than 100 Mpc. The line shows the fit up to the completeness limit computed for each filter. \label{fig:LFall}}
              
 \end{figure}

In order to compute the slope of the LF for the knots in both the blue and red filters (as explained in section~\ref{sec:LF}) we assigned a lower limit cutoff to the brightest value of the 90\% completeness limit in order to ensure the same level of completeness for the whole subsample. The single power-law fits to our blue and red luminosity functions are not significantly different, with slopes of $\alpha = 1.95$ and $\alpha = 1.89$ for the \filterb (blue) and the \filteri (red) filters, respectively (see Figure~\ref{fig:LFall}). 

The slopes in this study are in disagreement with the much flatter slopes previously measured in ULIRGs ~\citep{Surace98}. Surace and coworkers argued that the flattening of the slope is the artificial consequence of not resolving individual clusters, but associations of clusters (the bright end is overpopulated), the same bias that we encounter but more severe. In addition, other factors could contribute to the flattening of the LF in that study, as well as in other works (see table~\ref{table:LF}): (i) the slopes are not corrected for incompleteness (like in NGC 7252 in~\citealt{Miller97} and NGC 4038/39 in~\citealt{Whitmore99}), (ii) the statistics are more limited, since the distribution of less than 90 knots is fitted within their completeness limit interval, and (iii) ACS imaging is more sensitive than WFPC2, and since we are observing closer systems, the LF can be measured to intrinsically fainter magnitudes. 

%\textbf{We did not expected noticeable differences, since the mass function of molecular clouds in the Local Group (\citealt{Blitz07}) and in different ISM models (\citealt{Fleck96,Wada00,Elmegreen02}) is consistent with a slope of index 2, like the mass function for clusters in interacting systems (e.g~\citealt{Bik03,Gieles09}). This similarity can be reflected in the LF. Given the fact that several clusters can be formed from a single molecular cloud, if there is not much blending (like in our case of nearby systems), a similar slope of the LF is expected.}

The slopes measured for our subsample are however similar to those obtained for other less luminous interacting galaxies such as in NGC 4038/39, NGC 3921, NGC 7252 and Arp 284 (see table~\ref{table:LF}). As the studies start to resolve individual clusters and corrections for incompleteness are applied, the slopes tend to consolidate around a value of 2-2.2. For instance, \reff (between 5 or unresolved, and 50 pc) derived in early studies (\citealt{Whitmore95}) for clusters in NGC 4038/39, indicate that some blending existed and thus the slope of the LF is somewhat flatter than in more recent studies which detect only individual clusters and corrections for incompleteness are applied as well (\citealt{Whitmore99,Whitmore10}). Therefore, the result obtained in this study extends the universality of the slope of the luminosity function regardless of the intensity of the star formation in interacting galaxies, to extreme star-forming systems like the luminous and ultraluminous infrared galaxies, at least for systems closer than 100 Mpc.

\begin{deluxetable}{lcccccc@{}}
\tabletypesize{\small}
\tablewidth{0pt}
\tablecaption{Slopes of the LF computed for this work and for other interacting systems \label{table:LF}}

\tablehead{
\colhead{Galaxy}	&\colhead{\ld \tablenotemark{b} (Mpc)} & \colhead{Band}	&	\colhead{Slope LF ($\alpha$)\tablenotemark{a}}	&	\colhead{Uncertainty}	&	\colhead{Ref.}}
\startdata
ULIRGs Surace	&	422	&	I	&	1.39	&	0.08	&	[1]	\\
NGC 3921	&	84.5	&	V	&	2.10	&	0.30	&	[2]	\\
Nearby (U)LIRGs	&	72.8	&	B,I	&	1.95,1.89	&	0.02,0.04	&	This work*	\\
NCG 7252	&	62.2	&	V	&	1.84	&	0.06	&	[3]	\\
Arp 284	&	33.6	&	I	&	2.30	&	0.30	&	[4]	\\
NGC 4038/39	&	27.5	&	V	&	1.78	&	0.05	&	[5]	\\
NGC 4038/39\tablenotemark{c}	&	27.5	&	V	&	2.01	&	0.11	&	[6]	\\
NGC 4038/39	&	27.5	&	V	&	2.13	&	0.07	&	[7]*	\\
\enddata

\tablenotetext{a}{In some works they also fit a double power-law, but the slopes given in this table correspond to a single power-law fitting.}
\tablenotetext{b}{Distances taken from NED. For the (U)LIRG samples (this work and that by \citealt{Surace98})  the mean distance is given.}
\tablenotetext{c}{Computed for cluster-rich regions on the PC chip.}
\tablerefs{[1]~\cite{Surace98}, [2]~\cite{Schweizer96},   [3]~\cite{Miller97}, [4]~\cite{Peterson09}, [5]~\cite{Whitmore95}, [6]~\cite{Whitmore99} and [7]~\cite{Whitmore10}. In asterisk, studies that have used the prescription in~\cite{Maiz05}.}
\end{deluxetable}

%In order to investigate any possible dependence of the LF with the distance to the nucleus, we have computed the corresponding LF for the sample of inner and outer knots separately. Since the inner knots are in general located within regions of higher surface brightness than the outer knots, different limiting magnitudes for the LFs of the inner and outer knots have been considered. 

%The slopes for the inner (outer) knots are 1.91$\pm$0.03 (2.25$\pm$0.10) and 1.89$\pm$0.02 (2.16$\pm$0.09) for the F814W and F435W filter, respectively. There is a trend toward steeper slopes as the galactocentric distance increases.~\cite{Haas08} observes the same trend when fitting the LF with a double power-law for clusters in M51 at three different galactocentric distances. They conclude that this is an indication of the variation of the disruption time for star cluster with the galactocentric distance, being strongest in the central regions of the galaxy. However, given the fact that we are not observing individual clusters, this trend can also be due to the lack of resolution. Since the innermost regions are more crowded, if not enough resolution is available to separate them blending occurs, and this is more severe in the inner regions than in the outer parts of the galaxies. Besides, there is also more dilution with the diffuse background. As a result, we only detect bright objects in the innermost regions, which overpopulates the bright end of the LF, making it flatter. 

\subsection{Properties of the knots as a function of \lir}
\label{sub:lir}

\begin{figure*}
\centering
   \vspace{-2cm}
   % trim -> left bottom right top
   \includegraphics[trim = -1.5cm 12.5cm -1cm -4cm,clip=true,width=0.9\textwidth]{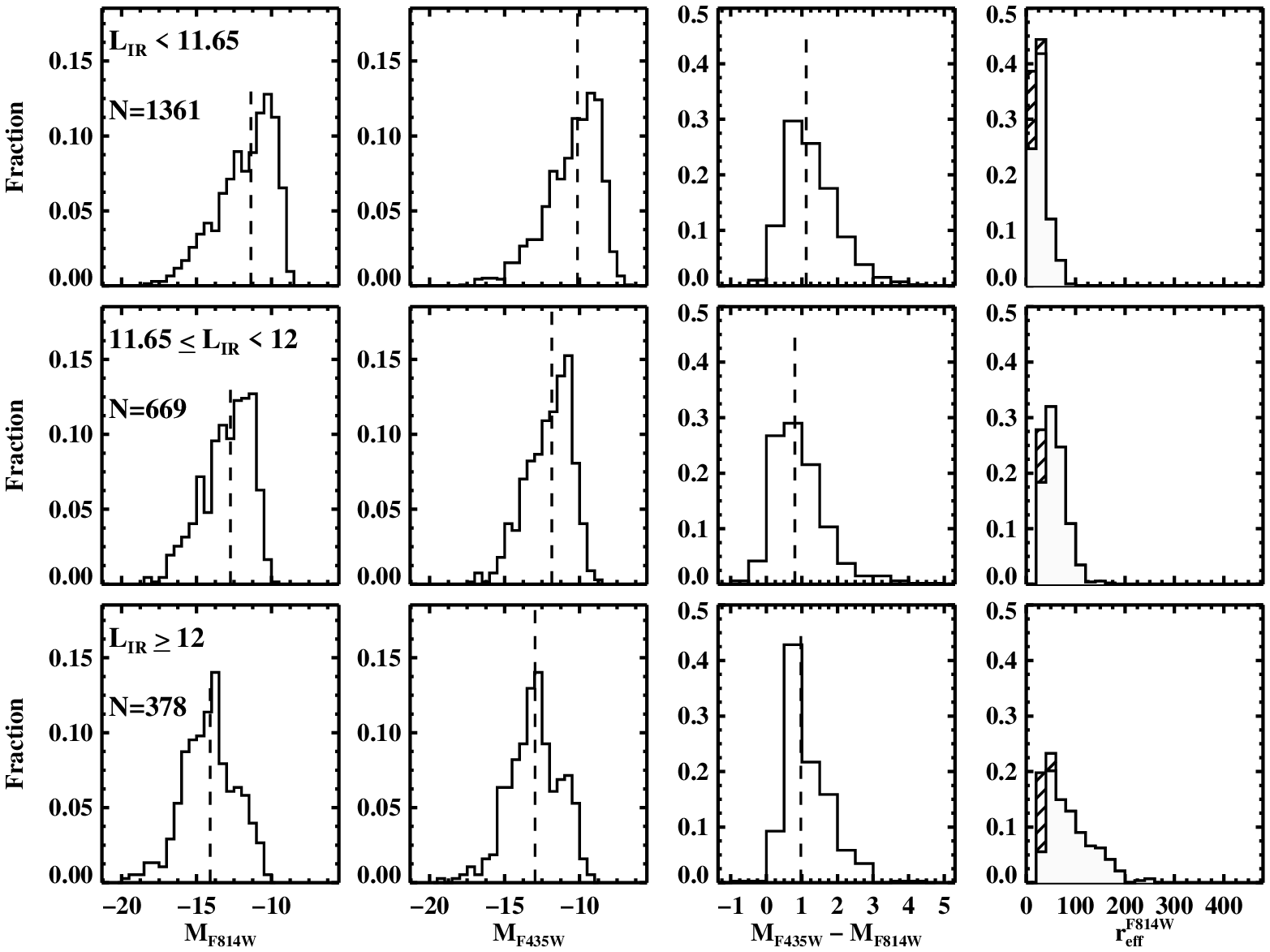}
   \vspace{-0.5cm}
   \caption{Photometric properties of the sample in three \lir intervals, excluding 3 systems in the sample, as explained in the text. The dashed vertical line indicates the median value of the distributions of magnitudes \mb and \mi\onespace, and color \mbi\onespace.\label{fig:prop_lir}}

 \end{figure*}

\begin{deluxetable}{l@{}c@{}c@{}c@{}c@{}c@{}c@{}c@{}c@{}c@{\hspace{0.2cm}}c@{\hspace{0.2cm}}c@{}}
%\tabletypesize{\tiny} %referee
\tabletypesize{\scriptsize}  %normal
\tablecaption{Photometric properties of the sample as a function of \lir \label{table:prop_lir}}
\tablewidth{0pt}
\tablecolumns{11}

\tablehead{
\colhead{System/s} &  \colhead{\lir} & \colhead{Number of} & \colhead{knots per} & \colhead{$<$ D$_{L}$ $>$} & \colhead{$<$ \mi $>$} & \colhead{$<$ \mb $>$} & \colhead{$<$ C $>$} & \colhead{$<$ $r_{\rm{eff}}^{F814W}$ $>$} & \colhead{$\alpha$ LF}  & \colhead{$\alpha$ LF}  \\
 & \colhead{(or interval)} & \colhead{systems} & \colhead{system} & \colhead{(Mpc)} &   &  &  & \colhead{(pc)} & \colhead{\filteri} &  \colhead{F435W}\\
 \colhead{(1)} & \colhead{(2)} & \colhead{(3)} & \colhead{(4)} & \colhead{(5)} & \colhead{(6)}  & \colhead{(7)} &  \colhead{(8)} & \colhead{(9)} & \colhead{(10)} &  \colhead{(11)}} 
\startdata
(U)LIRGs& \lir $<$ 11.65 (all)& 11& 113 $\pm$ 102& 99 $\pm$ 26& -11.36& -10.13& 1.13& 25& 1.80 $\pm$ 0.02& 1.77 $\pm$ 0.02\\
(this work)& 11.65 $\le$ \lir $<$ 12.0& 8& 84 $\pm$ 59& 178 $\pm$ 27& -12.74& -11.84& 0.81& 55& 1.73 $\pm$ 0.04& 1.79 $\pm$ 0.04\\
& \lir $\ge$ 12.0& 10& 38 $\pm$ 15& 246 $\pm$ 72& -14.09& -12.97& 0.97& 81& 1.83 $\pm$ 0.03& 1.78 $\pm$ 0.04\\
& 11.65 $\le$ \lir $<$ 12.0 (all)& 9& 115 $\pm$ 109& 166 $\pm$ 45& -12.11& -11.18& 0.82& 45& 1.78 $\pm$ 0.04& 1.78 $\pm$ 0.04\\
& \lir $\ge$ 12.0 (all)& 12& 47 $\pm$ 41& 258 $\pm$ 125& -13.39& -12.20& 1.10& 59& 1.78 $\pm$ 0.04& 1.78 $\pm$ 0.04\\
 \hline
   \noalign{\smallskip}
M51& 10& 1& 881& 10.6& -8.4& -7.6& 0.7& -& - & - \\
NGC 4038/4039& 10.7& 1& $\sim$ 10$^4$& 27.5 & -11.7& -11.0& 0.9& 16.8& - & - \\
ULIRGs (Surace98)& \lir $>$ 12.0& 9& 12 $\pm$ 10& 520 $\pm$ 210& -14.7& -13.7& 0.8& 65& - & - \\
\enddata
\tablecomments{\footnotesize{The first three rows show the statistics for the systems considered in the text (some few have been excluded in the intermediate and high infrared luminosity intervals; see beginning of Section~\ref{sub:lir}. The following two rows implement the statistics for all the systems that define the \lir interval. Below the horizontal middle line the values for clusters and knots from other works are shown (M51,~\citealt{Bik03}; NGC 4038/4039,~\citealt{Whitmore99}; and ULIRGs,~\citealt{Surace98}). Col (1): galaxy name or designation for a sample of galaxies. Col (2): infrared luminosity (or interval) of the galaxy/sample. Col (3): number of systems that comprises each sample. Col (4): average number of knots per system. Note that even though more than 10$^4$ clusters were detected, the values for NGC 4038/4039 refer to the 86 brightest clusters. Col (5): luminosity distance or average value. Col (6): median value of the \mi absolute magnitude distribution. Col (7): median value of the \mb absolute magnitude distribution. Col (8): median value of the \mbi color distribution. Col (9): median value of the effective radius distribution. Col (10): slope of the \textit{F814W} luminosity function distribution. The slopes of the LFs for interacting systems in other works are already showed in Table~\ref{table:LF}. Col (11): slope of the \textit{F435W} luminosity function distribution.}}
\end{deluxetable}

For most (U)LIRGs the infrared luminosity comes mainly from the re-emission by dust of light emitted in intense episodes of star formation. In this study we span  a range of a factor of 15 in infrared luminosities from \lir\twospace\footnote{We remind the reader that for simplicity, we identify the infrared luminosity as \mbox{\lir(\lsun) $\equiv$ log ($L_{IR}[8 - 1000 \mu m]$) }.}~= 11.39 to \mbox{\lir = 12.54}. In order to investigate whether the properties of the star-forming knots do show dependence on the infrared luminosity (i.e.,  star formation rate), we have divided the initial sample into three \lir intervals. We aim at covering the low, intermediate and high luminosity regimes while preserving a similar number of systems per luminosity bin for the subsequent statistical analysis. We have therefore chosen the following luminosity intervals: \lir $<$ 11.65 (low), \mbox{11.65 $\leq$ \lir $<$ 12.0} (intermediate) and \lir $\geq$ 12.0 (high). The number of systems that lie in each interval is 11, 9 and 12, being at an average distance of 99, 166 and 258 Mpc, respectively. A summary of the properties of the knots per luminosity bin is given in Table~\ref{table:prop_lir} and will be discussed in detail in subsequent subsections. 

As shown in Figure~\ref{fig:LirVsZ},  systems with higher infrared luminosity are in general intrinsically more distant. We have tried to get rid of any distance effects by selecting a subsample of systems located at a similar distance (those at 65-75 Mpc) and sampling all the \lir bins. This subsample, not homogeneously distributed, includes only one system in the intermediate and high luminosity bins (4,1 and 1 systems, respectively per luminosity bin). Besides, the results based only on the ULIRG Arp 220 (IRAS 15327+2340) are not reliable since it is known to suffer extreme obscuration from its silicate features~\citep{Spoon07}. In order to achieve more reliable statistics we need to include a larger number of systems in the defined luminosity bins and hence we inevitably have to deal with distance effects.

Finally, for each \lir bin, we have excluded the knots of systems located at the two extremes of the distance scale, in order to diminish distance effects within the intervals. Thus, the statistics that are shown first in Table~\ref{table:prop_lir} for the intermediate luminosity interval do not take into account IRAS 04315-0840, and for the high luminosity interval the knots in IRAS 15206+3342 and in IRAS 15327+2340 are not considered either. However, for completeness, the statistics is also provided considering all the sources in each \lir bin.

\subsubsection{Magnitudes and colors of the knots as a function of \lir} 
\label{sub:mc_lir}

\begin{figure*}[t]
\centering
    \includegraphics[angle=90,trim = 4cm 0cm 2cm 1.5cm,clip=true,width=1.05\textwidth]{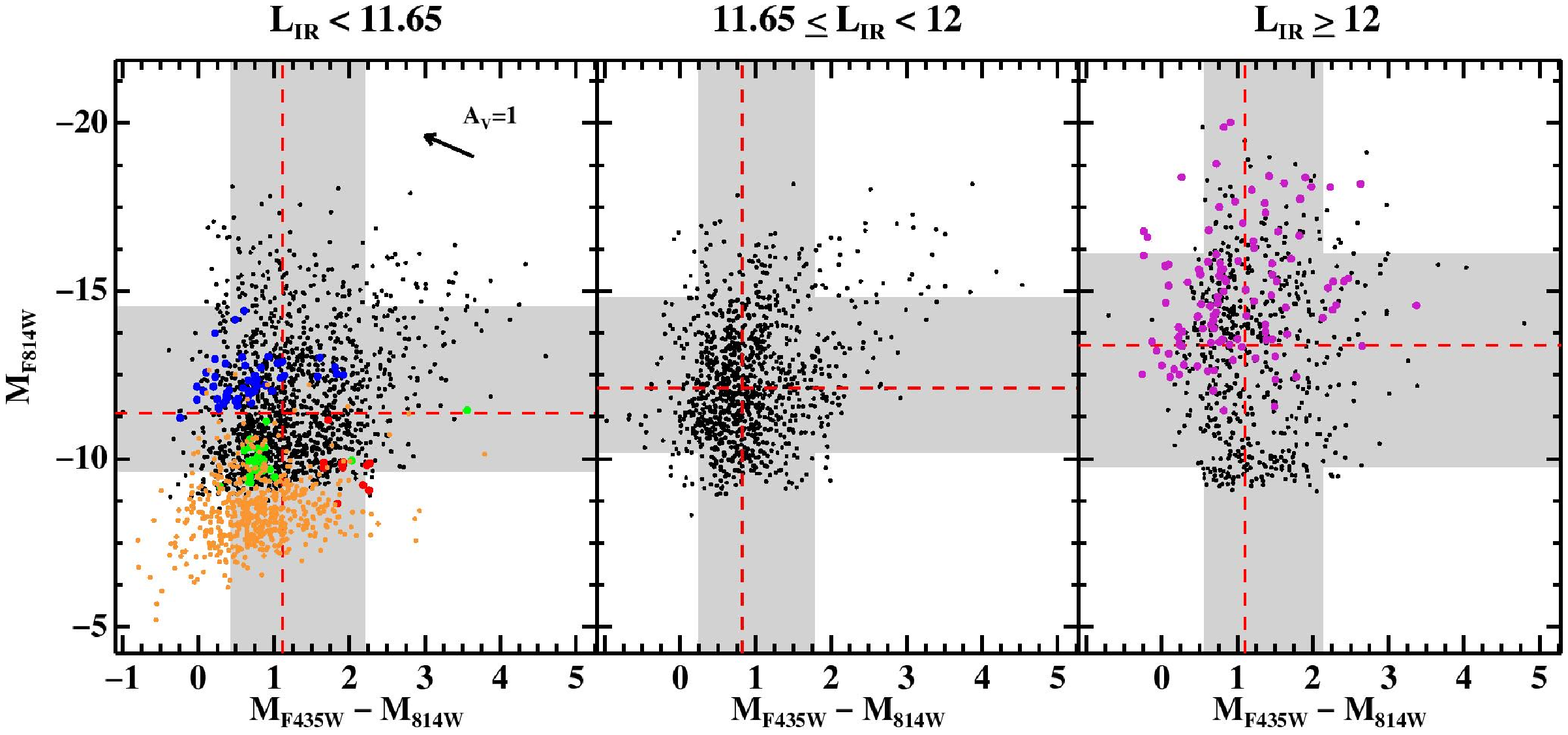}
   \caption{Color-magnitude diagrams of all the identified knots (black) at different \lir intervals, compared with other interacting systems from the literature. The gray band covers the 80\% of the total number of knots in each plot, as in Figure~\ref{fig:mag_col}. The colors refer to clusters/knots measured in M51 (orange) taken from~\cite{Bik03}, in the ULIRG sample of~\cite{Surace98} (pink) and in NGC 4038/4039 (blue, those younger than 10 Myr; green, those between 250 and 1000 Myr; and red, those older than 1 Gyr), taken from~\cite{Whitmore99}. \label{fig:mag_col_lira}}
              
\end{figure*}

Knots show a dramatic increase ($\times$ 12) in their blue and red luminosities (i.e., 2.7 mag) with the infrared luminosity of the entire galaxy while basically preserving their colors (see histograms in Figure~\ref{fig:prop_lir}, and average properties listed in Table~\ref{table:prop_lir}). Our simulations (see Figure~\ref{fig:simulation} and section~\ref{sec:distance_effect}) suggest that distance effects can explain only part of the differences revealed by the data. If the same type of galaxies are located at the average distance of the low, intermediate and high luminosity subsamples, the distance effect does not change significantly the colors while increasing the luminosity (absolute magnitude) by only a factor of 2.8 (1.1 mag). Therefore, we conclude that  knots in highly luminous galaxies are at least a factor of four more luminous than knots located in systems with lower infrared luminosity.

It is interesting to mention that when comparing these results with those independently obtained for 
nearby, less luminous systems (the weakly and strongly interacting galaxies M51 and NGC 4038/4039, respectively), and more distant ULIRGs (Surace et al. 1998), the same trend with infrared luminosity appears (see Figure~\ref{fig:mag_col_lira} and table~\ref{table:prop_lir}). In nearby systems like M51 and NGC 4038/4039, the angular resolution is such that individual clusters, instead of aggregates as in more distant galaxies, are detected, and therefore the difference in the luminosity of clusters and knots can be partly due to a distance effect. Knots in distant ULIRGs tend to be as luminous as knots detected in our closer high-luminosity galaxies (also ULIRGs) while having colors similar to those knots detected in our sample and clusters in M51 and NGC 4038/4039 (The Antennae). The most luminous clusters in the Antennae cover a luminosity range close  to the median value for knots detected in our low-luminosity systems. Finally, clusters in the very low luminosity, weakly interacting M51 galaxy are on average about 3 magnitudes fainter (i.e., x15 less luminous) than knots in our low-luminosity galaxies.   

If we assume a statistically similar extinction and age for the star-forming knots regardless of the \lir of the system, the most plausible explanation for the luminosity excess measured in the intermediate and high luminosity systems has to invoke a mass and/or density effects:

\begin{itemize}
\item \textit{Mass effect}. More high-mass knots are sampled in the most luminous systems.

\item \textit{Density effect}. The knots are aggregates of an intrinsically larger number of clusters as the infrared luminosity of the system increases. 
\end{itemize}

The star formation rate as well as the gas content in ULIRGs is higher than in less luminous systems and as a consequence, these systems form more clusters. Therefore, it is matter of simple statistics rather than a difference in physical formation mechanisms that in ULIRGs we find the brightest knots. Given the stability of the colors among the different infrared luminosity sample, they likely represent the most massive knots. This stochastic process, known as size-of-sample effect (\citealt{Whitmore07}), can explain the demographics of star cluster systems in the merging environment, where the star formation is enhanced. Therefore, although density effects cannot be discarded our result is compatible with a mass effect (which in turn corresponds to a size-of-sample effect), explained by statistics.

\subsubsection{Spatial distribution of the knots as a function of \lir}

\begin{figure*}[t]
\centering
\vspace{-3cm}
\includegraphics[angle=90,width=0.95\textwidth]{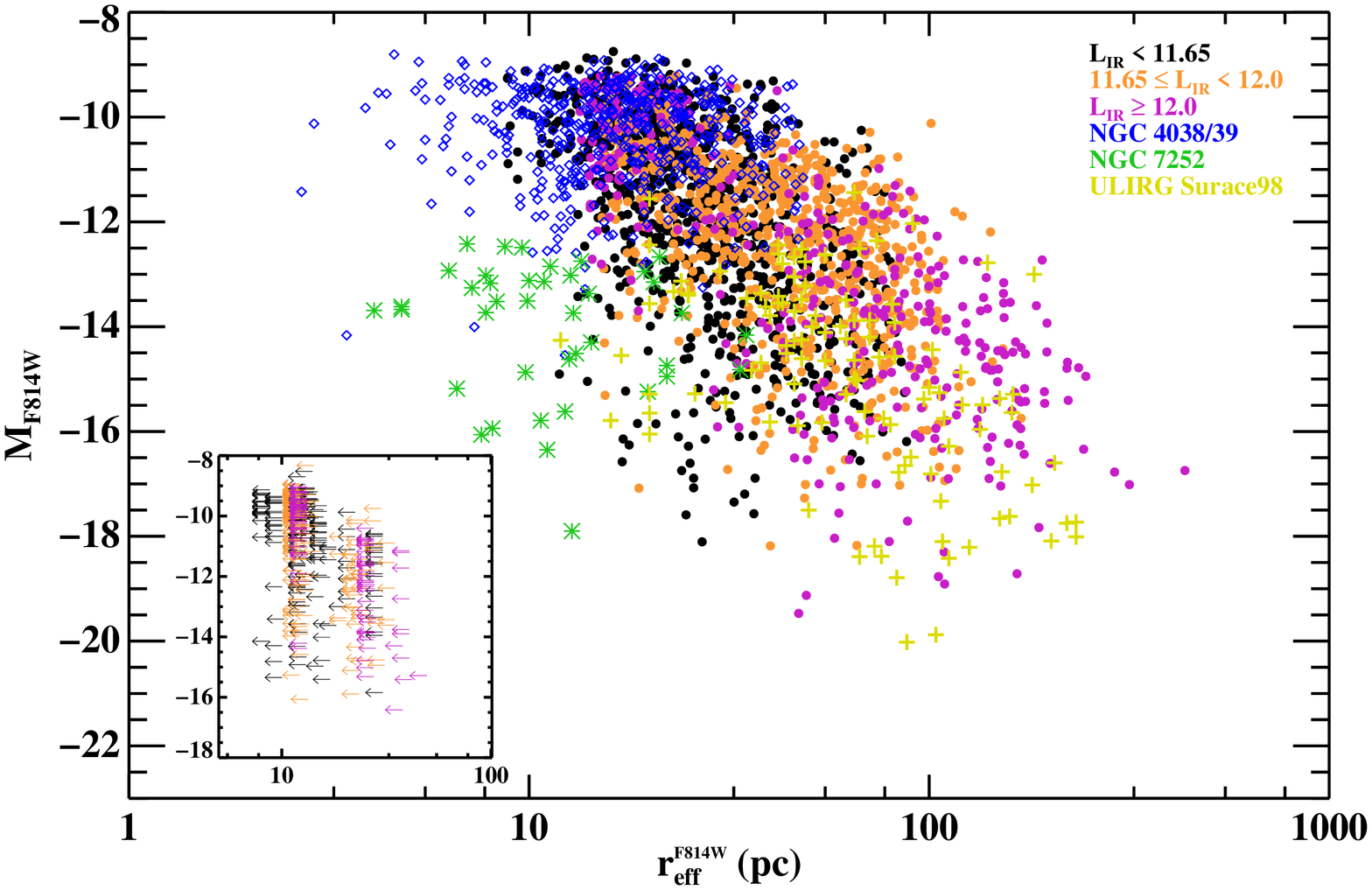}
   \vspace{-0.3cm}   
\caption{Absolute \filteri magnitude (uncorrected for internal extinction) as a function of size for clusters in interacting systems. Note that the x-axis is in log scale. We have divided the knots in our sample in three groups depending on their infrared luminosity. Positive values for clusters in the Antennae were taken from~\cite{Whitmore95}, in NGC 7252 (with \lir=10.70) from~\cite{Whitmore93} and in the ULIRG sample from~\cite{Surace98}. The inset plot shows the same for the unresolved knots in this study.\label{fig:reff2}}

 \end{figure*}

One important aspect to investigate is whether the spatial distribution of the knots shows dependence on the luminosity of the system. We measure a ratio of inner to outer knots (N$_{inner}$/N$_{outer}$) that changes with \lir from 0.68 to 0.49 and 0.31 for the three (low-, intermediate- and high-) luminosity intervals, respectively). According to our simulations, these differences are consistent with being purely due to a distance effect (see table~\ref{table:prop_lir} for the average distance of the different luminosity bins), and therefore no evidence for the dependence of the spatial distribution of the knots on the infrared luminosity of the systems is found.

\subsubsection{Effective radius of the knots as a function of \lir}

If there is a change in the luminosity of the knots with the bolometric luminosity of the system, their size could show a variation as well. At first look this seems to be the case (see Figure~\ref{fig:prop_lir} and~\ref{fig:reff2}). As the luminosity of the system increases, the distribution of \reff shows a broadening with a non negligible fraction of the knots having sizes larger than 100 pc. This is more clear for the high-luminosity systems (ULIRGs). However, as our simulations have shown (see Figure~\ref{fig:simulation}), this changes can be explained purely as a distance effect. On the one hand, \reff distribution of knots in ULIRGs is similar to the simulated \reff distribution. Moreover, the median value of \reff in ULIRGs is 81 pc (see table~\ref{table:prop_lir}), slightly larger, but consistent within the uncertainties,  with the 75 pc derived from the simulations. In fact, the sizes of the knots in the closest systems (i.e., \ld$<$100 Mpc), where individual clusters are almost resolved, are closer to the sizes of clusters measured in less luminous interacting galaxies (e.g., see clusters of NGC 4038/4039 and NGC 7252 in Fig~\ref{fig:reff2}). Therefore, with the present angular resolution there is no evidence for a variation in size with luminosity. 

\subsubsection{Mass-radius relation of the bluest knots as a function of \lir}

\begin{figure}[!t]
\vspace{1cm}
   \includegraphics[width=1\columnwidth]{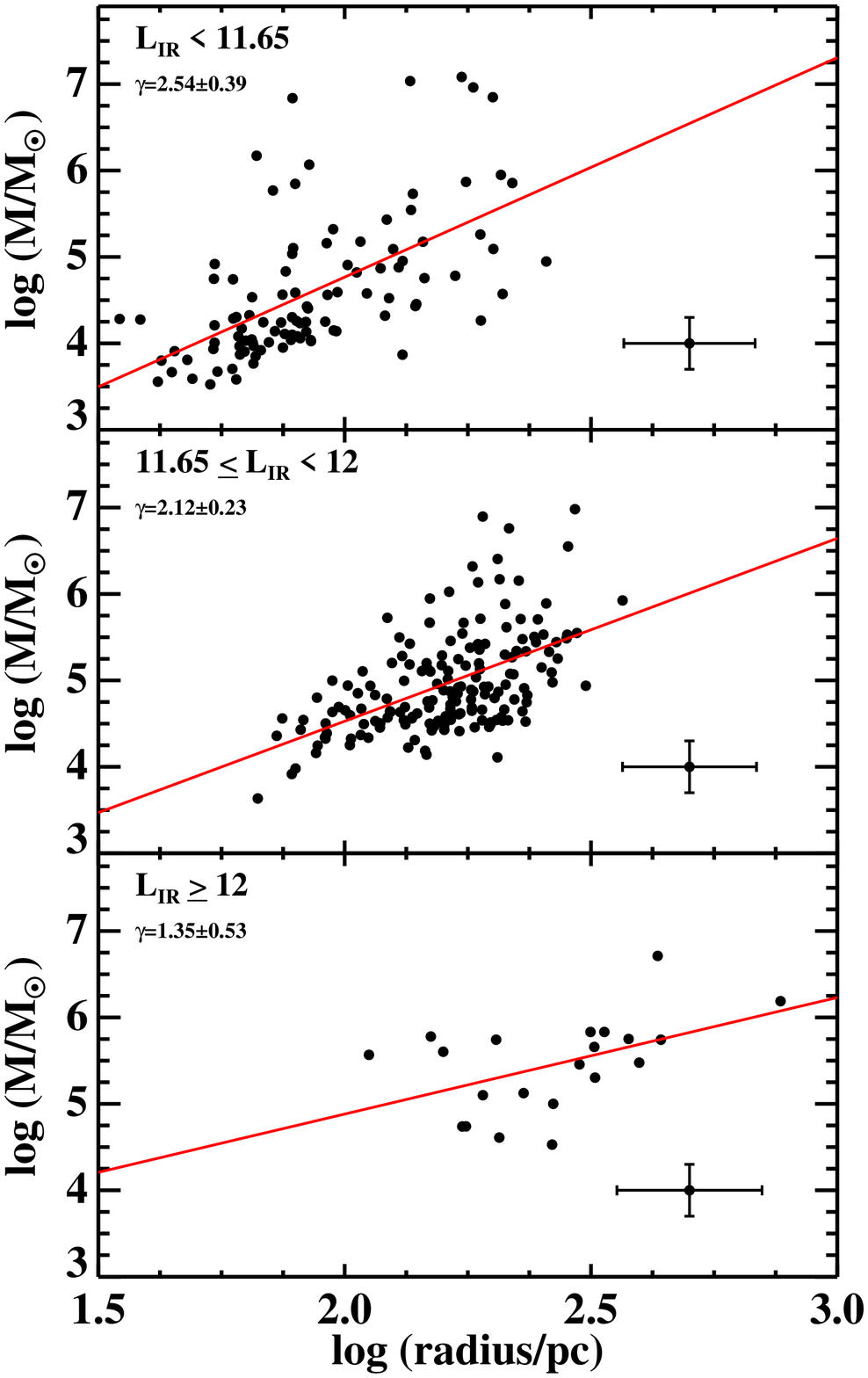}   
   \caption{The mass vs. radius relation for knots in (U)LIRGs as a function of \lir\twospace. The solid red line is a power-law fit to the data, with index $\gamma$. \label{fig:size_mass_lir}} 
 \end{figure}

We have seen that in general terms the bluest knots in our sample follow a mass-radius relation with index $\gamma \sim$2, similar to that in GMCs and complexes of clusters in less luminous interacting systems (see section~\ref{sub:mass_rad}). Here we have the opportunity to characterize this relation as a function of the infrared luminosity of the system.

Though with high dispersion, the correlation holds for the low and intermediate infrared luminosity intervals (see Figure~\ref{fig:size_mass_lir}). The fit for the bluest knots in ULIRGs gives a shallower index, though it could still be consistent with a value of $\gamma$=1.9 given the large dispersion on the index values. The fit is affected in any case by the small number of blue knots in ULIRGs and probably by distance effects. Therefore, according to our data, there is no clear variation of the mass-radius relation with the infrared luminosity of the system.

\subsubsection{Luminosity Function of the knots as a function of \lir}
\label{sub:LF_lir}

The derived slopes of the luminosity functions with values between 1.7 and 1.8 (see table~\ref{table:prop_lir})  do not show any clear trend of variation with the luminosity of the system. However,
our simulations of the distance effect show that the slope of the LF becomes shallower by about 0.15 dex  when moving the low-luminosity systems to the average distance of the high-luminosity subsample  (see Figure~\ref{fig:simulation}). This change in the slope is due to the reduction in the linear resolution such that knots that appear resolved in the closer, low-luminosity systems appear to be as
more luminous aggregates in the more distant luminous systems. Therefore the distance-corrected LF slopes for the intermediate and high-luminosity subsamples would be about 1.9-2, in the range of those measured in other low-z
interacting galaxies (see table~\ref{table:LF}).

\subsection{Properties of the knots as a function of interaction phase}
\label{sub:interacting_stage}

Strong interactions and mergers provide a natural laboratory for probing how star formation is affected by tidal forces and major rearrangements of the gaseous component during the different phases of the interaction. Most of the systems in our sample are interacting systems, hence we have divided the sample into four groups, depending on the morphology seen in the \filteri \hst images. Each group represents snapshots of the  different phases of the interaction/merger: first approach, pre-merger, merger and  post-merger. The morphology class is explained in section~\ref{sub:Morphology} and the selection is shown in Table~\ref{table:sample}. In order to know how the star formation evolves with the interaction process, in this section we study the photometric properties of the knots for each of the four morphology groups.

\subsubsection{Magnitudes and colors as a function of interaction phase}
\label{sub:mag_col_IS}

\begin{figure*}
   \centering
   \vspace{-0.5cm}
   \includegraphics[width=0.8\textwidth]{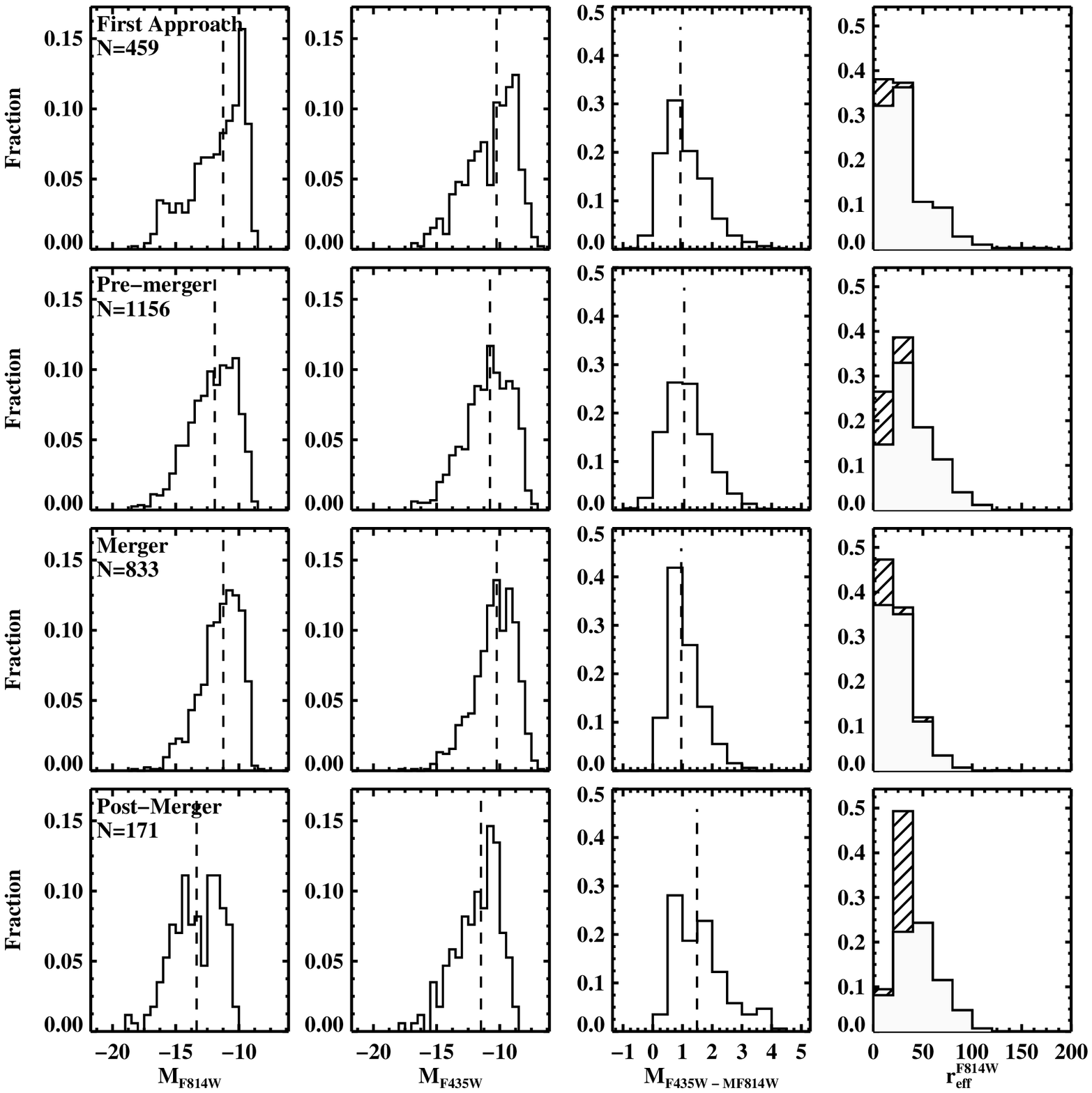}
   \vspace{-2.5cm} 
   \caption{Photometric properties as a function of merging phase for systems with similar D$_{L}$. Dashed vertical lines indicate the median value of the distributions of magnitudes \mi and \mb, and color \mbi, as in Figure~\ref{fig:prop_lir}. The relevant parameters of the plots are shown in the first four rows in Table~\ref{table:prop_IS}.\label{fig:propIS_sameD}}
              
 \end{figure*}

The magnitudes and colors of the knots in the first three interaction phases differ significantly from those of the knots in the post-merger phase (see results in Table~\ref{table:prop_IS}). 
Even if in a strictly statistical sense (Kolmogoroff-Smirnov test at a confidence level of at least 98\%) the population of knots in the first three phases are different, their median values and distributions are very similar suggesting the lack of strong differences. This is true in particular if, in order to mi\-ni\-mi\-ze the distance effect,  only galaxies at similar distances in each phase are considered (see table~\ref{table:prop_IS} and Figure~\ref{fig:propIS_sameD}). However, the population of knots in the post-merger phase are significantly more luminous (median I-band magnitude  difference of up to 2 mag), and significantly redder than in any of the three earlier phases. Besides, the population of knots in the post-merger phase shows in its color distribution an extension toward redder colors ($>$ 3 mag). The fact that knots in the post-merger phase have redder colors means that they likely represent either a more obscured or an older population. In the following we explore both alternatives.

\begin{deluxetable}{lccccccccc}
\tabletypesize{\scriptsize}
\tablecaption{Photometric properties of the sample as a function of interaction phase \label{table:prop_IS}}
\tablewidth{0pt}
\tablecolumns{10}

\tablehead{
\colhead{Interaction phase} & \colhead{Number of} & \colhead{knots per}  & \colhead{$<$ \ld  $>$} & \colhead{$<$ \mi $>$} & \colhead{$<$ \mb $>$} & \colhead{$<$ C $>$} & \colhead{$<$ $r_{\rm{eff}}^{F814W}$ $>$} & \colhead{$\alpha$ LF \filteri} & \colhead{$\alpha$ LF \filterb} \\
 & \colhead{systems} & \colhead{system}  & \colhead{(Mpc)} &  &  &  & \colhead{(pc)} &  &  }
\startdata
First Approach& 3& 153 $\pm$ 127& 121 $\pm$ 49& -11.25& -10.26& 0.93& 24& 1.50 $\pm$ 0.09& 1.63 $\pm$ 0.06\\
Pre-merger& 8& 145 $\pm$ 81& 126 $\pm$ 42& -11.91& -10.77& 1.06& 35& 1.84 $\pm$ 0.03& 1.86 $\pm$ 0.04\\
Merger& 6& 139 $\pm$ 123& 107 $\pm$ 37& -11.23& -10.24& 0.96& 25& 2.00 $\pm$ 0.04& 1.86 $\pm$ 0.03\\
Post-Merger& 5& 34 $\pm$ 19& 145 $\pm$ 50& -13.35& -11.49& 1.49& 44& 1.76 $\pm$ 0.12& 1.58 $\pm$ 0.17\\
 \hline
   \noalign{\smallskip}
First Approach& 4& 131 $\pm$ 113& 140 $\pm$ 55& -11.54& -10.66& 0.92& 27& 1.53 $\pm$ 0.10& 1.67 $\pm$ 0.02\\ 
Pre-merger& 13& 103 $\pm$ 83& 192 $\pm$ 98& -12.29& -11.12& 1.04& 40& 1.93 $\pm$ 0.03& 1.86 $\pm$ 0.03\\ 
Merger& 8& 112 $\pm$ 115& 142 $\pm$ 71& -11.41& -10.38& 0.96& 26& 1.83 $\pm$ 0.02& 1.78 $\pm$ 0.03\\ 
Post-Merger& 7& 29 $\pm$ 18& 213 $\pm$ 159& -13.51& -11.73& 1.55& 49& - & - \\
\enddata
\tablecomments{\footnotesize{The interaction groups above the horizontal middle line includes only systems having similar distance. Below the horizontal middle line all the systems have been considered, regardless their distance  (note the higher dispersion in the distance of the systems here). Apart from the first column, which refers to the different interaction phases defined for this study, the rest indicate the same as shown in Table~\ref{table:prop_lir}.}}
\end{deluxetable}

\begin{enumerate}
\item

In order to investigate whether the knots in the post-merger phase are intrinsically more obscured than in any of the  previous phases,  we define the infrared to ultraviolet luminosity ratio (IR/UV) as
\begin{center}
{\large $IR/UV =  \displaystyle\frac{L_{IR}}{\lambda L_{\lambda}(FUV)+\lambda L_{\lambda}(NUV)}$}
\end{center}

where \lir stands for the infrared luminosity, $\lambda$ to the pivot wavelength of a given filter and L$_{\lambda}$ to the flux in the near (NUV) and far (FUV) ultraviolet emission~\citep{Howell10}. This ratio is a useful empirical measurement of the dust absorption and emission in these heavily obscured systems. We searched for UV images in the Galaxy Evolution Explorer (GALEX) catalog, using the Multimission Archive at STScI (MAST\footnote{http://archive.stsci.edu/}) facility. With a typical angular resolution of about 4-5'' in the FUV-NUV ($\lambda$ = 1528-2271$\AA{}$) range, we can estimate an average value of IR/UV for each system assuming that the far infrared and ultraviolet flux is emitted in the same regions. Even if this is not necessarily true, taking into account that most of the nuclear star formation is obscured in (U)LIRGs, we can have a first-order estimate on which systems have a higher average extinction.

A total of 16 systems were detected in the GR5 data set, most of them in the All-Sky Imaging survey (ASI\footnote{Visit the GALEX website, http://www.galex.caltech.edu/, for more information}). The other systems were either not detected or not observed. In this dataset, the images are already reduced, the sources detected and the photometry performed. In general the photometry was performed using an aperture that covers the entire galaxy (if two galaxies form the system we add up the flux values). We then take the value of the flux provided by the data set facility.

\begin{deluxetable}{lccccc}
\tabletypesize{\small}
\tablewidth{0pt}
\tablecolumns{4}
\tablecaption{Infrared to ultraviolet luminosity ratio as a function of interaction phase \label{table:dust}}

\tablehead{
\colhead{Interaction phase}	&	\colhead{N$_{sys}$}	&	\colhead{$<$ IR/UV $>$}	&	\colhead{IR /UV}	& \colhead{$<$ IR$_{25}$/UV $>$} & \colhead{IR$_{25}$ /UV} \\
	&	&		&	\colhead{range} & & \colhead{range} \\
\colhead{(1)} & \colhead{(2)} & \colhead{(3)} &\colhead{(4)} &\colhead{(5)} & \colhead{(6)}
	
	}

\startdata
First Approach	&	2	&	-	&	[15,130]	&  -	&	[4,11]	\\
Pre-merger	&	6	&	48	&	[14,189]	&   7	&	[2,45]	\\
Merger	&	6	&	248	&	[114,1235]	& 32	&	[14,117]	\\
Post-merger	&	2	&	-	&	[385,895]	&  -	&	[86,93]	\\
\enddata
\tablecomments{Since the IR/UV ratio is not affected by distance effects (we compute it for the whole system) we give the statistics for all the systems per interaction phase with available UV image. Col (1): interaction phase. Col (2): number of systems with IR and UV data available. Col (3): median of the IR/UV ratio when more than two values are derived. Col (4): range covered of the ratio IR/UV for each interaction phase. Col (5): same as (3), but in this case only the contribution of the flux at 25$\mu$m is taken to compute \lir\twospace. Col (6): same as (4), but only the contribution of the flux at 25$\mu$m is taken to compute \lir\twospace.}
\end{deluxetable}

With the flux in the far and near UV and the infrared luminosity we have estimated an average value of the IR/UV ratio for each system. Although the assumption by~\cite{Kennicutt98} that the great majority of the luminosity from young stars would be absorbed by dust and re-radiated in the far-IR is likely to be correct, observationally there may be other contributions to the total IR luminosity from older populations of stars or other luminosity sources (e.g., see the review by~\citealt{Tuffs05}). For that reason we have also computed the IR/UV ratio taking only the contribution at 25$\mu$m in IRAS (IR$_{25}$), based on the good relation between the mid-IR 24$\mu$m-25$\mu$m (MIPS-IRAS) and the number of ionizing photons as derived from extinction corrected Pa$_ \alpha$ and \ha luminosities (\citealt{Wu05,Alonso-Herrero06,Calzetti07}). This way we select only hot dust, heated by the radiation of young population, and avoid any contribution from colder dust, heated by older stars.

Bearing in mind that we are dealing with a small number of statistics, there is a trend toward higher values of the IR/UV ratio in not only the post-merger but also the merger phase, with respect to early phases represented by the first approach and pre-merger (see values in Table~\ref{table:dust}). Although the absolute values are different and the refinement is better when using IR$_{25}$, the trend is practically the same if we use one infrared luminosity or the other. This is not surprising, as IRAS measurements poorly sample the output of the cold dust (the longest band is at 100$\mu$m). Since the IR/UV ratio is related to an average dust absorption, our result suggests that knots in the late phases of an interaction could be suffering on average intrinsically more extinction than in early phases.  For instance, an additional average extinction of \av\onespace=1 can already redden the colors by about 0.6, according to the SB99 models used in section~\ref{sub:ages_masses}. Were this the case, the population of clusters in the late phases of a merger would be more luminous than in previous phases. Therefore, we would need to invoke a physical process responsible for this significant increase in luminosity. Note however that dust absorption shows significant structure on scales of a kiloparsec or less~\citep{Garcia-Marin09b}, and therefore the measured IR/UV
ratios do not necessarily represent the intrinsic absorption toward the individual optically-selected clusters. A larger set of  high angular resolution UV and mid-IR imaging is required before any firm conclusion is derived. 

\item
An alternative scenario to explain the differences in color and luminosity of the knots identified in the post-merger phase with respect to early phases can invoke evolutionary effects. As proposed by ~\cite{Kroupa98}, many dozens or even hundreds of massive individual star clusters with masses of 10$^5$-10$^7$\msun  could merge into extremely massive super star clusters (SSCs) after a few hundred Myr. This hierarchical star formation evolution has also been suggested to be happening in complexes of clusters in M51 (\citealt{Bastian05a}). As the interaction proceeds, the smaller SSCs would be destroyed by disruption effects (e.g., by the mechanisms proposed in~\citealt{Whitmore07}) and a few massive individual clusters would still be able to form. Note that, although with a large dispersion, while the number of knots per system stays about the same (140-150) in the first three phases, this number drops by a factor of 4 to 5 in the post-merger phase (34 knots per system). Thus, in the post-merger phase systems we might be detecting only the evolved, extremely massive SSCs and a few massive individual clusters. The knots with the reddest colors could be evolved SSCs as massive as 10$^{7-9}$ \msun with an optical extinction (\av\twospace) of 3 mag, or massive young clusters with higher extinction.

\cite{Kroupa98} estimated that after 100 Myr or so an object with these characteristics can be formed. If the merging of clusters started during the early phases of the interaction we would have already detected some of them in the merger phase. Given the degeneracy of the \mbi color we could only set an upper-limit to the number of merged superclusters in this phase. Thus, with a color higher than \mbox{\mbi= 0.6} and a luminosity higher than \mbox{\mi= -15} an object could be more massive than 10$^7$ \msun\twospace, should it be around or older than 100 Myr. About three such knots per system are detected in our sample of (U)LIRGs. 

Obviously, photometry at other wavelengths is necessary to better constrain the ages and masses of these knots and thus to check the validity of this scenario for (U)LIRG systems. Likewise, simulations of galactic encounters help us know if this scenario can occur in this environment. A detailed comparison of the present observational results with state-of-the-art high-resolution numerical simulations (\citealt{Bournaud08a}) will be presented elsewhere (Miralles-Caballero et al. 2011, in preparation) to further investigate this scenario.
\end{enumerate}

\subsubsection{Spatial distribution as a function of interaction phase}

In order to know if major mergers have an impact on the spatial distribution of the knots,  we have investigated the relative fraction of inner (inside 2.5 kpc radius) and outer (outside 2.5 kpc radius) knots 
in the four interaction phases defined above. Although the number of systems in each phase is small (3, 8, 6 and 5 for the first approach, pre-merger, merger, and post-merger) the ratio N$_{inner}$/N$_{outer}$ is 0.50,0.65,0.52,0.68 from the first approach to the post-merger. 

We do not see any correlation toward a more dispersed or concentrated spatial distribution of the knots with interaction phase. However, in the pre- and post-merger phases, where the magnitudes are higher compared to the other phases, there is some evidence of greater concentration toward the nuclear regions. The greater the number of knots in the central regions, the harder it is to detect them, and therefore, only the brightest ones are identified given the local background conditions (bright and steep light profiles). Hence, this effect can contribute to a small increase in the luminosity distribution of the knots in these phases. Assuming that the knots in the pre-merger phase have a luminosity similar to that in the first contact and merger phases, this effect would increase the median luminosity by a factor of almost 2. Since the ratio N$_{inner}$/N$_{outer}$ in this phase is similar to the ratio in the post-merger phase, an additional increase of a factor of about 5 would still be unexplained by the first scenario explored in the previous section. A larger sample of galaxies would be required to establish a firmer result.

\subsubsection{Effective radius as a function of interaction phase}

All size distributions are statistically different with a high degree of confidence (higher than 99\%), yet the median value of \reff is similar during the three first interaction phases indicating that these differences must be rather small. However, knots in the post-merger phase tend to be larger by factors of 1.3 to 2 than in any of the previous phases (see table~\ref{table:prop_IS}). This result is not very robust, given the uncertainties in the determination of sizes, but if confirmed, it would support the hierarchical star formation scenario proposed by~\cite{Kroupa98}. According to this scenario, the merged super clusters would have a half-mass radius of about r$_{h}$=45-95 pc. Under the assumption of virial equilibrium and isotropic velocity distribution the effective and half-mass radius can be related as \reff=3/4r$_{h}$. Then, the merged extremely massive super clusters would have \reff=30-70 pc, similar to values measured for knots in the post-merger sample.

\subsubsection{Mass-radius relation of the bluest knots as a function of interaction phase}

\begin{figure*}
\includegraphics[trim = 0cm 3.5cm 5cm 0cm,angle=90,width=0.9\textwidth]{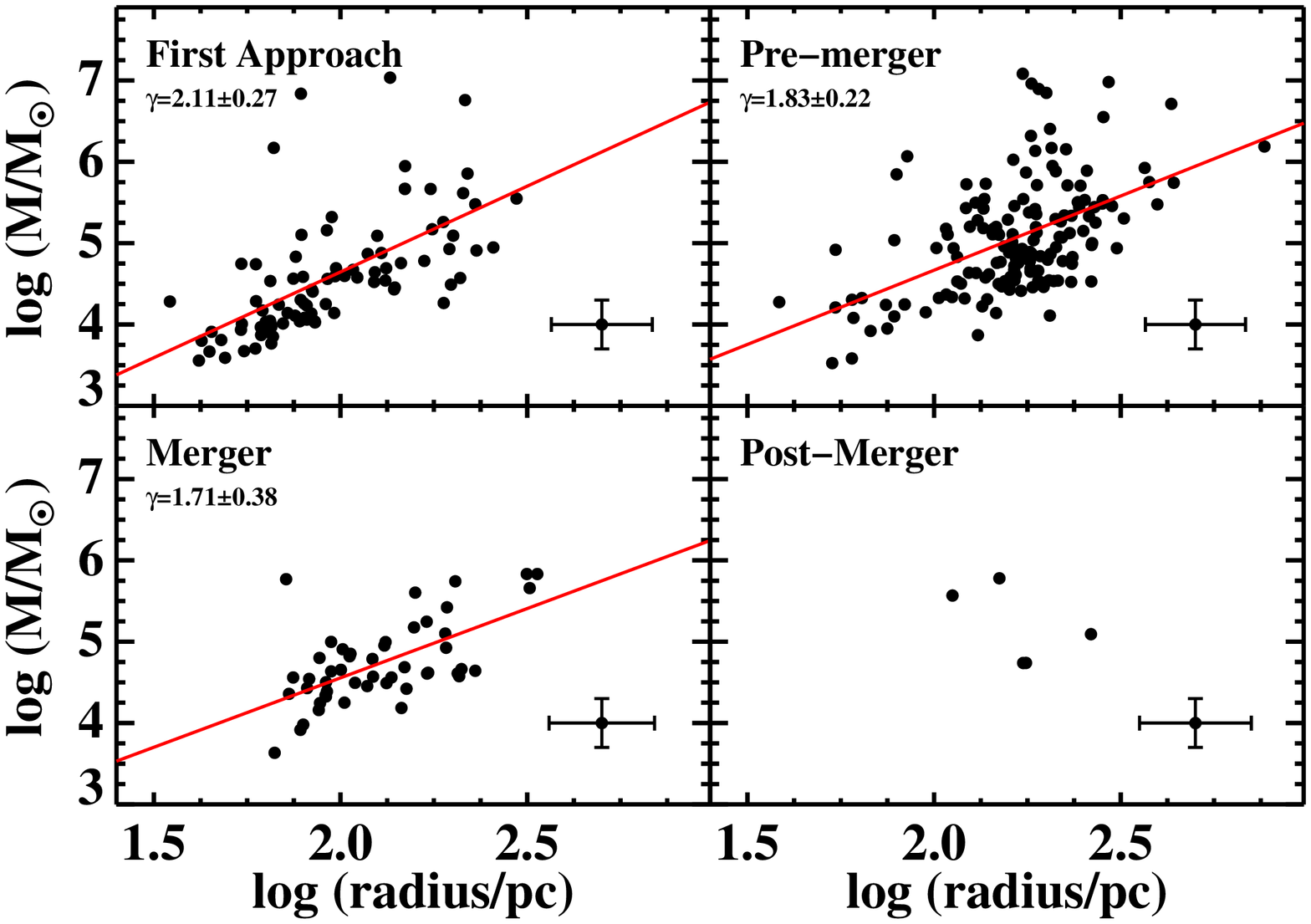}
  \caption{The mass vs. radius relation for knots in (U)LIRGs as a function of the interaction phase. The solid red line is a power-law fit to the data, with index $\gamma$. \label{fig:size_mass_is}}             
 \end{figure*}
 
The broad-band emission of the knots for which we have measured the mass is very much likely to come from young population (i.e., t$<$30 Myr; see section~\ref{sub:ages_masses}). Based on equivalent width measurements,~\cite{Bastian05a} determined the age of the complexes in M51 to be in the range 5-8 Myr. On the other hand, the phases of interaction are separated by hundreds of Myr. Thus, although we cannot study how the mass-radius relation of the knots behaves as they evolve, we can see if they all have a similar relation at a short time after they are born during the interaction process.

The mass-size relation of the bluest knots in the three interaction  phases for which a fit could be achieved is consistent with an index of $\gamma$=2 (Fig.~\ref{fig:size_mass_is}). Although the line-fit seems to become shallower as the interaction proceeds from the first approach up to the merger phase (from $\gamma$=2.11 to 1.71), the dispersion is too high to accept that it varies. Actually, the fit that we obtained using all the knots gives a slope inbetween both values (1.91; see section~\ref{sub:mass_rad}).

\subsubsection{Luminosity function as a function of interaction phase}

There appears to be a trend in the slopes of the LFs of the knots  with interaction phase such that they become steeper as the interaction evolves from first contact to pre-merger and up to merger and 
post-merger (see table~\ref{table:prop_IS}). This is particularly more visible when only systems at similar distances and \textit{I}-band LFs, less affected by extinction effects, are considered (slopes of 1.50, 1.84, 2.0 and 1.76, respectively). Note, however, that the slope of the LF for the post-merger phase is computed with less than 100 knots (brighter than the completeness limit), hence it is statistically less significant than the other slopes. Also note that the systematic uncertainties of the slope of the LF can be as high as about 0.1 dex (the tables give the uncertainty related only to the fit), as mentioned in section~\ref{sec:LF}. Thus, the observed trend is very weak.

If we select only systems closer than 100 Mpc (see section~\ref{sub:LF}), the phases are also sampled. We can try to further minimize distance effects by computing the LF only for these systems. While 400-600 measurements can be used in each of the other phases, on the post-merger phase we are instead restricted to only about 30 knots and thus, we will not include this phase in this analysis. The slopes derived for the closest systems are 1.68-1.68-1.92 (1.75-1.77-1.99)  for the \textit{I}- (\textit{B}-) band LF for the first contact, pre-merger and merger phases, respectively. It is interesting to note that in this case the LF still becomes steeper when the systems evolve from the pre-merger to the merger phases.

(U)LIRGs are known to have high SFRs at all phases of interaction. However, this can be concentrated at small scales or more dispersed and, at the same time, less or more hidden by dust. For instance, 80\% of the total infrared luminosity in IRAS 20550+1656 comes from an extremely compact, red source not associated with the nuclei of the merging galaxies but with a buried starburst (\citealt{Inami10}). If, on average, we detect more young knots in the optical at early phases of interaction than in more evolved phases, they can overpopulate the bright end of the luminosity function, and thus the measured slope would be shallower. A forthcoming study (Miralles-Caballero et al. 2011, in preparation), comparing the observed LFs with those computed from numerical simulations~\citep{Bournaud08a} (i.e., mass function) and stellar population synthesis models (i.e., mass to light ratios), will help us learn if the observed trend (change of the slope as the interaction evolves) is real. Should it be confirmed, then we will be able to investigate its origin.

\subsection{Star-forming knots in (U)LIRGs versus star-forming clumps in high-z
    galaxies}

\begin{deluxetable}{lcccccc}
\tabletypesize{\small} 
\tablewidth{0pt}
\tablecolumns{7}
\tablecaption{Comparison of knots in ULIGRs with clumps in the high-z Universe.\label{table:ULIRG_highz}}

\tablehead{
\colhead{System / field}	& \colhead{rest-frame} & \colhead{redshift} &	\colhead{N clumps}	&	\colhead{Mass range}	&	\colhead{Total size}	& \colhead{Ref.} \\
	&  band /line & 	& \colhead{per galaxy}	&	\colhead{(\msun\twospace)}	&	\colhead{(kpc)}	&	\\
\colhead{(1)} & \colhead{(2)} & \colhead{(3)} & \colhead{(4)} & \colhead{(5)} & \colhead{(6)} & \colhead{(7)}
	
	}
\startdata
UDF field	&	B	&	0.5 - 4	&	$\sim$ 5	&	10$^7$ - 10$^9$	&	1 - 2.5	&	[1], [2] 	\\
Lensed galaxies	&	\ha	&	1.7 - 3.1 	&	-	&	6$\times$10$^8$ - 3$\times$10$^9$	&	0.3 - 1 	&	[3]	\\
SF galaxies at z$\sim$2	&	optical g, UV, \ha	&	2.2 - 2.4	&	4.5	&	9$\times$10$^7$ - 9$\times$10$^9$	&	0.3 - 1.5	&	[4]	\\
GEMS \& GOODS fields	&	B	&	0.1 - 1.4	&	$\sim$ 4	&	5$\times$10$^6$ - 10$^8$	&	0.2 - 7	&	[5]	\\
Simulated ULIRGs at z=1	&	I	&	1	&	6.7	&	 2$\times$10$^5$ - 7$\times$10$^7$	&	0.4 - 3.5	&	This work	\\
TDG candidates in (U)LIRGs	&	I	&	0 - 0.1	&	$\sim$ 1	&	10$^6$ - 4$\times$10$^7$	& 	1 - 2 	&	[6] \\
\enddata
\tablecomments{Col (1): type of systems and deep fields. Col (2): rest-frame photometric band. Col (3): redshift interval of the galaxies. Col (4): average number of clumps per galaxy. Col (5): total stellar mass range of the clumps, except for the lensed galaxies, where the dynamical mass of the clumps is provided. Col (6): total diameter of the clumps. For star-forming galaxies at \mbox{z = 2} the full-width at half-maximum from the clump radial light profiles after subtraction of the local background is given. Col (7): references--~[1]~\cite{Elmegreen05};~[2]~\cite{Elmegreen07b};~[3]~\cite{Jones10};~[4]~\cite{Schreiber11b};~[5]~\cite{Elmegreen06};~[6]~Miralles-Caballero et al. (in prep)}
\end{deluxetable}

(U)LIRGs represent the extreme cases of low-redshift star-forming galaxies  associated with spirals, interacting galaxies and mergers. Deep cosmological imaging and spectroscopic surveys at redshifts (z$\sim$1-4) have identified massive star-forming galaxies with irregular structures where the star formation appears to proceed in large clumps,with sizes in the 0.1-1.5 kpc range. The most recent are the surveys conducted in the  GEMS and GOODS fields (\citealt{Elmegreen06,Elmegreen07a}), the Ultra Deep Field (UDF;~\citealt{Elmegreen05,Elmegreen07b,Elmegreen09}) and the Spectroscopic Imaging survey in the Near-infrared with SINFONI  (SINS;~\citealt{Schreiber11b}). While some of these galaxies show structural characteristics of being involved in mergers (\citealt{Elmegreen07a};~\citealt{Schreiber11a}), most systems appear as thick turbulent disks (\citealt{Elmegreen09}). Faint tidal features are difficult to discern at higher redshifts because of cosmological dimming. It is important therefore to make a direct comparison between the star-forming knots in (U)LIRGs and in clumpy high-z galaxies in order to understand the similarities and differences of the star formation in the present Universe and at cosmological distances.

The clumps in high-z galaxies are less numerous and intrinsically brighter and more massive (by a factor of 10-1000) than the largest star-forming complexes in local galaxies (\citealt{Efremov95}; \linebreak \citealt{Elmegreen05};~\citealt{Elmegreen07a,Elmegreen09};~\citealt{Schreiber11b}). The integrated luminosity of these clumps  usually accounts for about 10-25\% of the total luminosity. The larger masses in the high-redshift galaxies are thought to be the result of high turbulence in the disks. In fact, velocity dispersions of 40 km s$^{-1}$ have been derived from spectroscopic observations of UDF6462, a z=1.6 galaxy (\citealt{Bournaud08b}. This large-scale star formation can also be reproduced by means of numerical simulations which require velocity dispersions several times higher than in local quiescent galaxies (\citealt{Bournaud07}). Finally, these high velocity dispersions have already been observed in clumps in local ULIRGs (\citealt{Monreal07}), since gravitational torques in major mergers can also induce star formation in large condensations of gas. Yet, the last mention has to be taken with care since strong winds and shocks in mergers can also induce large gradients in the velocity fields (i.e., high velocity dispersions).

In the standard $\Lambda$CDM cosmology (\citealt{Spergel07}) the angular size is nearly constant over the redshift range \mbox{0.7 $<$ z $<$ 4}. At closer redshifts, the angular size decreases approximately linearly with increasing redshift. A kpc-sized clump  corresponds to 3-4 pixels in the ACS images at high redshift values. Thus, clumps much smaller than this could not be measured at high redshifts. On the other hand, if there were clumps significantly larger than this in nearby galaxies, they would be easily discernible. The lack of spatial resolution at high redshifts can make a complex of star formation look like a clump since the light from the different clusters combines, blurs and some contrast with the local background is lost. Whether the high-z clumps correspond to a single entity or to complexes of star formation, the largest complexes can be identified and compared in both local and high-redshift galaxies. The \mbox{high-z} galaxy clumps have masses typically between 10$^7$ and 10$^9$ \msun from \mbox{z = 1 to 4}, decreasing to about 10$^6$ \msun at \mbox{z = 0.1-0.2}. The star-forming regions in the few interacting clumpy galaxies observed have sizes similar to star complexes in local interacting systems (i.e., the Antennae or M51), and their mass is 10-1000 times higher. On the other hand, complexes in the Hickson Compact Group 31 (HCG 31), stellar complexes, which are sensitive to the magnitude of disk turbulence, have both sizes and masses more characteristic of z = 1-2 galaxies (\citealt{Gallagher10}).

In section~\ref{sub:mc_lir} we have seen that knots in ULIRGs are intrinsically brighter than knots in less luminous systems. This means that for complexes of similar sizes and similar population (table~\ref{table:prop_lir} shows that the colors of the knots in LIRGs and ULIRGs are similar) we are sampling knots in ULIRGs (they could also be understood as clumps or complexes of star clusters) that can be more massive than in our low luminous LIRGs by at least a factor of 4. Although the knots in our ULIRGs only account for 0.5-2\% of the total luminosity (much less than clumps at high redshift), we can compare them with the complexes in high-redshift galaxies in order to determine if the star formation is similar or not to that in clumpy galaxies . Already at \mbox{z $\simeq$ 0.05-0.1} a significant percentage of knots in ULIRGs are more luminous than  \mi\twospace=-15 mag (see Fig.~\ref{fig:mag_col_lira}). According to the models used in this paper, if we assume that these knots are formed by a single young (e.g., t$\sim$6 Myr) population  they can be as massive as \mbox{few 10$^4$ - few 10$^7$ }\msun\twospace, with a median value of 1.4$\times$10$^5$ \msun\onespace. An older population (e.g., t$\sim$50 Myr) implies a knot with higher mass by one order of magnitude. Correction for internal extinction makes the population even more massive. Thus, these values have to be understood as a lower limit. In order to compare the knots in these ULIRGs with the clumps at high-redshift galaxies, we applied a Gaussian blur to our images in such a way that we simulate their appearance at z=1 (the scale at this redshift corresponds to 7.69 kpc/''), using the same technique as explained in section~\ref{sec:distance_effect}.

%\textbf{The absolute \textit{B} magnitude (not corrected for internal extinction) of the ULIRGs in this study ranges \mb from -21.5 to about -23.5. In order to minimize size-of-sample effects, we can compare our results with the clumps belonging to galaxies in the UDF, which have similar rest-frame luminosities (table~\ref{table:ULIRG_highz}). 
The knots in the blurred images (hereafter high-z knots) were measured. Table~\ref{table:ULIRG_highz} show the results together with the values for clumps in galaxies at high redshifts and TDG candidates found in our sample of (U)LIRGs. Some of the high-z knots were not recognizable in the blurred images because they were too faint or too small; a few were blended with other knots that were too close or with the nucleus. Most of the lowest mass knots and those close to the nucleus disappear, although the remaining high-z knots range from 2$\times$10$^5$\msun to 7$\times$10$^7$ \msun\twospace, with a median value of \mbox{$\sim$ 2.5$\times$10$^6$\msun\onespace}. The effective radius averages about 700 pc and the total size is typically \mbox{$\gtrsim$ 1.5 kpc}. Taking into account that the derived masses correspond to lower limits, they are typically one or two orders of magnitude less massive than clumps in high-z galaxies, having similar sizes. If the population that dominates the light of the high-z knots were older (e.g., 50 Myr), the masses of the high-z knots would significantly approach those of  clumps in high-z galaxies.

\begin{figure}[!t]
\hspace{-0.8cm}
\includegraphics[angle=90,width=1.1\columnwidth]{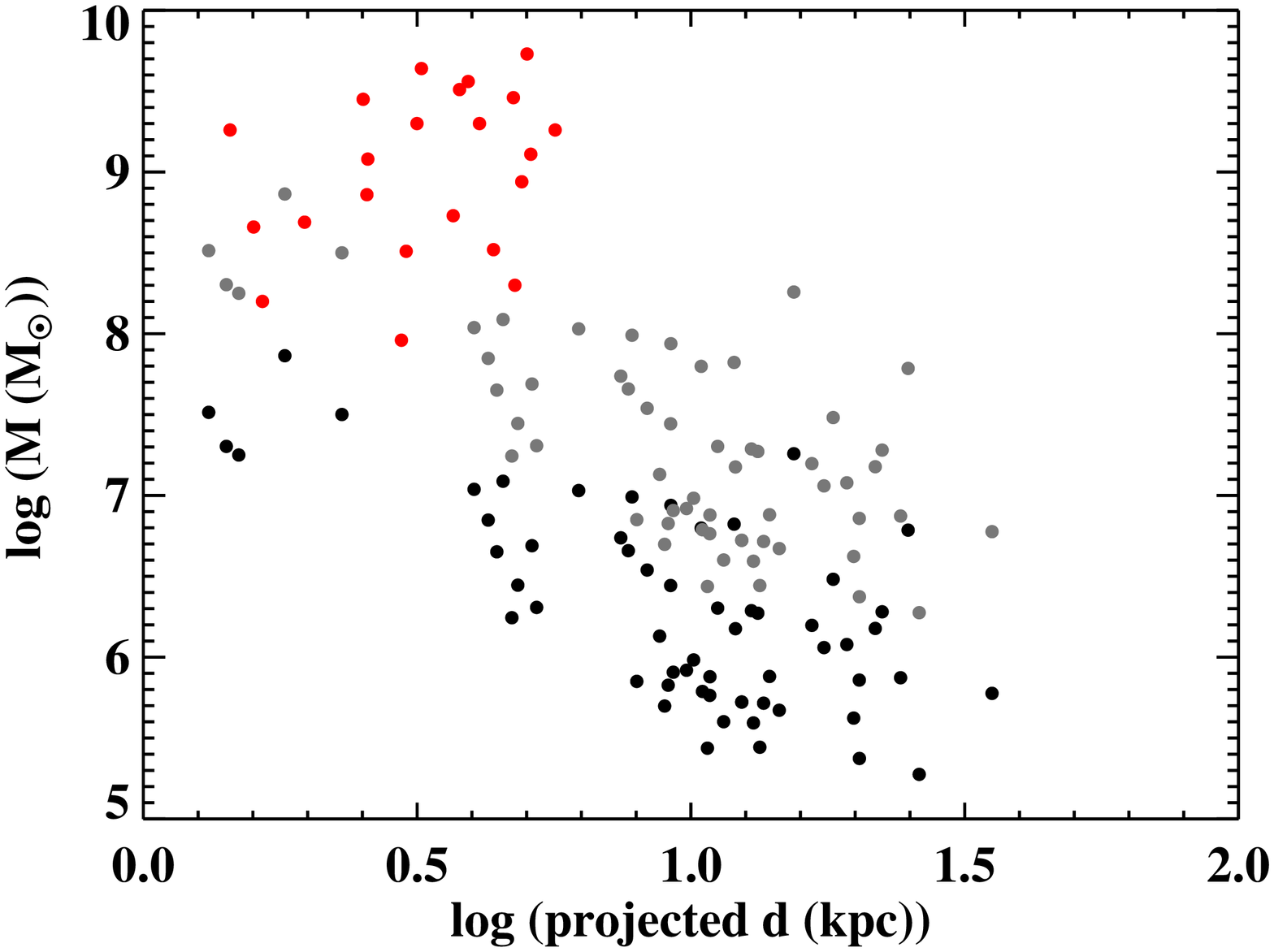}
      \caption{Mass vs projected distance for complexes in simulated ULIRGs at z=1 (in black for a population of 6 Myr and in gray for a population of 50 Myr) and clumps in z=2 star forming galaxies (red; values taken from~\citealt{Schreiber11b}) .\label{fig:deg_highz}
              }      
\end{figure}

Having set a lower limit for the mass of knots in ULIRGs, the mass of kpc-complexes in these systems is not likely to lay far from clumps in high-z galaxies with a similar size. Therefore, the extreme cases of low-redshift star-forming galaxies (ULIRGs) are likely to induce large-scale star formation similar to that observed in \mbox{high-z} disks (and some few mergers) with a high gas fraction.

We have also investigated how the projected distance of the high-z knots relates to their mass (see Fig.~\ref{fig:deg_highz}. While other works do not find any clear relation, we measure lower mass high-z knots with increasing distance. Although, according to our results it is clear that the mass is mainly concentrated at small distances in ULIRGs, we can only compare with one study of high-z galaxies and we explore larger distances by a factor of 10.

\section{Summary and conclusions}
\label{sec:summary}

A comprehensive study of bright compact knots (most of them apparent associations of clusters) in a representative sample of 32 (U)LIRGs as a function of infrared luminosity and interaction phase has been performed using \hst\twospace-ACS \mbox{B- and} I-band imaging, providing linear resolutions of 10 to 40 pc, depending on the distance of the galaxy. The sample has been divided into three infrared luminosity intervals (low, \mbox{\lir $<$ 11.65}; intermediate, \mbox{11.65 $\leq$ \lir $<$ 12.0}; and high, \mbox{\lir $\geq$ 12.0}) and in 4 interaction phases, based on the projected morphology (first approach, pre-merger, merger, post-merger). We have extended toward lower infrared luminosities previous stu\-dies focused only on ULIRGs (e.g.,~\citealt{Surace98};~\citealt{Surace00}), and have detected close to 3000 knots, larger than a factor of ten more than in the previous studies of ULIRGs. The main conclusions derived from this study are the following:

\begin{enumerate}[-]

\item The knots span a wide range in magnitudes \linebreak  (\mbox{-20 $\lesssim$ \mi $\lesssim$ -9} and \mbox{-19.5 $\lesssim$ \mb $\lesssim$ -7.5}), and color \mbox{(-1 $\lesssim$ \mbi $\lesssim$ 5)}. The median values are \mbox{\mb=-10.84}, \mbox{\mi=-11.96} and \mbi = 1.0, respectively. \\

\item The knots are in general compact, with a median \reff of 32 pc, being 12\% unresolved and a few very extended, up to 200-400 pc. With these sizes, in general (and particularly for galaxies located at 100 Mpc or more) the knots constitute complexes or aggregates of star clusters.  Within the resolution limits there is no evidence of a luminosity dependence on the size. A slight dependence on the interaction phase, in particular in the post-merger phase, needs confirmation with larger samples and better angular resolution.\\

\item A non-negligible fraction (15\%) of knots are blue and luminous. Given their color (\mbox{\mbi$<$0.5}) and luminosity (\mbox{$<$\mb\twospace$>$=-11.5}) they are likely to be young (ages of about 5 to 30 Myr), almost free of extinction, and with masses similar to and up to one order of magnitude higher than the Young Massive Clusters observed in other less luminous interacting systems. \\

\item Unlike for isolated young star clusters in other systems, we find a clear correlation between the knot mass and radius, M$\propto$R$^{2}$, similar to that found for complexes of star clusters in less luminous interacting galaxies (e.g., M51 and the Antennae) and Galactic and extragalactic giant molecular clouds. This relation does not seem to be dependent on the infrared luminosity of the system or on the interaction phase.

\item The star formation is characterized by B- and I-band luminosity functions (LFs) with slopes close to 2, extending therefore the universality of the LF measured in interacting galaxies at least for nearby systems (i.e., \ld$<$100 Mpc), regardless of the total luminosity (i.e., the strength of the global star formation). This result has to be taken with caution because the LF of our knots (which can be mainly formed of complexes of star clusters) may not reflect the same physics as the LF of individual star clusters. Nevertheless, there are slight indications that the LF evolves as the interaction progresses becoming steeper (from about 1.5 to 2 for the I-band) from first approach to merger and post-merger phases.\\

\item Taking into account distance effects, knots in high luminosity systems (ULIRGs) are intrinsically more luminous (a factor of about 4) than knots in less luminous systems. Given the star formation rate in ULIRGs is higher than in less luminous systems, size-of-sample effects are likely to be the natural explanation for this. \\

\item Knots in ULIRGs have both sizes and masses characteristic of stellar complexes or clumps detected in galaxies at high redshifts (z$\gtrsim$1), as long as their population is about 50 Myr or older. Thus, there is evidence that the larger-scale star formation structures are reminiscent of those seen during the epoch of morphological galaxy transformations in dense environments.

\item Knots in systems undergoing the post-merger phase are on average redder and more luminous than those in systems in earlier phases. Scenarios involving either a different spatial distribution of the obscuration or a different evolutionary phase in which small knots do not survive for long are considered. The evolutionary scenario could also explain the trend in size toward higher values in post-merger systems. A comparison with numerical simulations to further investigate these scenarios will be presented in a forthcoming paper (Miralles-Caballero et al. 2011, in preparation). \\

\end{enumerate}

\begin{acknowledgements}

The authors thank the anonymous referee for useful comments and suggestions which improved the quality of the paper. This work has been supported by the Spanish Ministry of Education and Science, under grant BES-2007-16198, projects ESP2005-01480 and ESP2007-65475-C02-01.
This research has made use of the NASA/IPAC Extragalactic Database (NED) which is operated by the Jet Propulsion Laboratory, California Institute of Technology, under contract with the National Aeronautics and Space Administration.
We have made use of observations made with the NASA/ESA Hubble Space Telescope, obtained from the data archive at the Space Telescope Science Institute (STScI ), which is operated by the Association of Universities for Research in Astronomy, Inc., under NASA contract NAS 5-26555. 
\end{acknowledgements}

\bibliographystyle{apj}
% \bibliography{../../my_bib}{}

\end{document}